\documentclass[
superscriptaddress,
twocolumn,
amsmath,amssymb,
aps,
prb,
10pt
]{revtex4-2}

\usepackage{silence}
\newcommand*{\beginsupplement}{%
        \counterwithout{equation}{section}
        \setcounter{section}{0}
        \setcounter{table}{0}
        \renewcommand{\thetable}{S\arabic{table}}%
        \setcounter{figure}{0}
        \renewcommand{\thefigure}{S\arabic{figure}}%
        \setcounter{equation}{0}
        \renewcommand{\theequation}{S\arabic{equation}}%
        \setcounter{page}{1}
        \renewcommand{\thepage}{S\arabic{page}}%
     }
     
\makeatletter
\newcommand*{\balancecolsandclearpage}{%
  \close@column@grid
  \clearpage
  \twocolumngrid
}
\makeatother

\usepackage{titletoc}

\usepackage{xcolor}
\usepackage{epsfig}
\usepackage[T1]{fontenc}
\usepackage{lineno}

\definecolor{melanzana}{RGB}{127,0,127}
\definecolor{verde_ale_chiaro}{RGB}{79, 163, 150}

  \definecolor{seashell}{rgb}{1.00, 0.96, 0.93}
  \definecolor{honeydew}{rgb}{0.94, 1.00, 0.94}
  \definecolor{mintcream}{rgb}{0.96, 1.00, 0.98}
  \definecolor{azure}{rgb}{0.94, 1.00, 1.00}
  \definecolor{aliceblue}{rgb}{0.94, 0.97, 1.00}
  \definecolor{lavender}{rgb}{0.90, 0.90, 0.98}
  \definecolor{lavenderblush}{rgb}{1.00, 0.94, 0.96}
  \definecolor{mistyrose}{rgb}{1.00, 0.89, 0.88}
  \definecolor{white}{rgb}{1.00, 1.00, 1.00}
  \definecolor{black}{rgb}{0.00, 0.00, 0.00}
  \definecolor{midnightblue}{rgb}{0.10, 0.10, 0.44}
  \definecolor{navy}{rgb}{0.00, 0.00, 0.50}
  \definecolor{navyblue}{rgb}{0.00, 0.00, 0.50}
  \definecolor{cornflowerblue}{rgb}{0.39, 0.58, 0.93}
  \definecolor{darkslateblue}{rgb}{0.28, 0.24, 0.55}
  \definecolor{slateblue}{rgb}{0.42, 0.35, 0.80}
  \definecolor{mediumslateblue}{rgb}{0.48, 0.41, 0.93}
  \definecolor{lightslateblue}{rgb}{0.52, 0.44, 1.00}
  \definecolor{mediumblue}{rgb}{0.00, 0.00,0.80}
  \definecolor{royalblue}{rgb}{0.25, 0.41, 0.88}
  \definecolor{blue}{rgb}{0.00, 0.00, 1.00}
  \definecolor{dodgerblue}{rgb}{0.12, 0.56, 1.00}
  \definecolor{deepskyblue}{rgb}{0.00, 0.75, 1.00}
  \definecolor{skyblue}{rgb}{0.53, 0.81, 0.92}
  \definecolor{lightskyblue}{rgb}{0.53, 0.81, 0.98}
  \definecolor{steelblue}{rgb}{0.27, 0.51, 0.71}
  \definecolor{lightsteelblue}{rgb}{0.69, 0.77, 0.87}
  \definecolor{lightblue}{rgb}{0.68, 0.85, 0.90}
  \definecolor{powderblue}{rgb}{0.69, 0.88, 0.90}
  \definecolor{paleturquoise}{rgb}{0.69, 0.93, 0.93}
  \definecolor{darkturquoise}{rgb}{0.00, 0.81, 0.82}
  \definecolor{mediumturquoise}{rgb}{0.28, 0.82, 0.80}
  \definecolor{turquoise}{rgb}{0.25, 0.88, 0.82}

  \definecolor{Turquoise}{rgb}{0.25, 0.88, 0.82}

  \definecolor{cyan}{rgb}{0.00, 1.00, 1.00}
  \definecolor{lightcyan}{rgb}{0.88, 1.00, 1.00}
  \definecolor{cadetblue}{rgb}{0.37, 0.62, 0.63}
  \definecolor{mediumaquamarine}{rgb}{0.40, 0.80, 0.67}
  \definecolor{aquamarine}{rgb}{0.50, 1.00, 0.83}
  \definecolor{darkgreen}{rgb}{0.00, 0.39, 0.00}
  \definecolor{darkolivegreen}{rgb}{0.33, 0.42, 0.18}
  \definecolor{darkseagreen}{rgb}{0.56, 0.74, 0.56}
  \definecolor{seagreen}{rgb}{0.18, 0.55, 0.34}
  \definecolor{mediumseagreen}{rgb}{0.24, 0.70, 0.44}
  \definecolor{lightseagreen}{rgb}{0.13, 0.70, 0.67}
  \definecolor{palegreen}{rgb}{0.60, 0.98, 0.60}
  \definecolor{springgreen}{rgb}{0.00, 1.00, 0.50}
  \definecolor{lawngreen}{rgb}{0.49, 0.99, 0.00}
  \definecolor{green}{rgb}{0.00, 1.00, 0.00}
  \definecolor{chartreuse}{rgb}{0.50, 1.00, 0.00}
  \definecolor{mediumspringgreen}{rgb}{0.00, 0.98, 0.60}
  \definecolor{greenyellow}{rgb}{0.68, 1.00, 0.18}
  \definecolor{limegreen}{rgb}{0.20, 0.80, 0.20}
  \definecolor{yellowgreen}{rgb}{0.60, 0.80, 0.20}
  \definecolor{forestgreen}{rgb}{0.13, 0.55, 0.13}
  
  \definecolor{ForestGreen}{rgb}{0.13, 0.55, 0.13}
  
  \definecolor{olivedrab}{rgb}{0.42, 0.56, 0.14}
  \definecolor{darkkhaki}{rgb}{0.74, 0.72, 0.42}
  \definecolor{khaki}{rgb}{0.94, 0.90, 0.55}
  \definecolor{palegoldenrod}{rgb}{0.93, 0.91, 0.67}
  \definecolor{lightgoldenrodyellow}{rgb}{0.98, 0.98, 0.82}
  \definecolor{lightyellow}{rgb}{1.00, 1.00, 0.88}
  \definecolor{yellow}{rgb}{1.00, 1.00 ,0.00}
  \definecolor{gold}{rgb}{1.00, 0.84, 0.00}
  \definecolor{lightgoldenrod}{rgb}{0.93, 0.87, 0.51}
  \definecolor{goldenrod}{rgb}{0.85, 0.65, 0.13}
  \definecolor{darkgoldenrod}{rgb}{0.72, 0.53, 0.04}
  \definecolor{rosybrown}{rgb}{0.74, 0.56, 0.56}
  \definecolor{indianred}{rgb}{0.80, 0.36, 0.36}
  \definecolor{saddlebrown}{rgb}{0.55, 0.27, 0.07}
  \definecolor{sienna}{rgb}{0.63, 0.32, 0.18}
  \definecolor{peru}{rgb}{0.80, 0.52, 0.25}
  \definecolor{burlywood}{rgb}{0.87, 0.72, 0.53}
  \definecolor{beige}{rgb}{0.96, 0.96, 0.86}
  \definecolor{wheat}{rgb}{0.96, 0.87, 0.70}
  \definecolor{sandybrown}{rgb}{0.96, 0.64, 0.38}
  \definecolor{tan}{rgb}{0.82, 0.71, 0.55}
  \definecolor{chocolate}{rgb}{0.82, 0.41, 0.12}
  \definecolor{firebrick}{rgb}{0.70, 0.13, 0.13}
  \definecolor{brown}{rgb}{0.65, 0.16, 0.16}
  \definecolor{darksalmon}{rgb}{0.91, 0.59, 0.48}
  \definecolor{salmon}{rgb}{0.98, 0.50, 0.45}
  \definecolor{lightsalmon}{rgb}{1.00, 0.63, 0.48}
  \definecolor{orange}{rgb}{1.00, 0.65, 0.00}
  \definecolor{darkorange}{rgb}{1.00, 0.55, 0.00}
  \definecolor{coral}{rgb}{1.00, 0.50, 0.31}
  \definecolor{lightcoral}{rgb}{0.94, 0.50, 0.50}
  \definecolor{tomato}{rgb}{1.00, 0.39, 0.28}
  \definecolor{orangered}{rgb}{1.00, 0.27, 0.00}
  \definecolor{red}{rgb}{1.00, 0.00, 0.00}
  \definecolor{hotpink}{rgb}{1.00, 0.41, 0.71}
  \definecolor{deeppink}{rgb}{1.00, 0.08, 0.58}
  \definecolor{pink}{rgb}{1.00, 0.75, 0.80}
  \definecolor{lightpink}{rgb}{1.00, 0.71, 0.76}
  \definecolor{palevioletred}{rgb}{0.86, 0.44, 0.58}
  \definecolor{maroon}{rgb}{0.69, 0.19, 0.38}
  \definecolor{mediumvioletred}{rgb}{0.78, 0.08, 0.52}
  \definecolor{violetred}{rgb}{0.82, 0.13, 0.56}
  \definecolor{magenta}{rgb}{1.00, 0.00, 1.00}
  \definecolor{violet}{rgb}{0.93, 0.51, 0.93}
  \definecolor{plum}{rgb}{0.87, 0.63, 0.87}
  \definecolor{orchid}{rgb}{0.85,0.44,0.84}
  \definecolor{mediumorchid}{rgb}{0.73,0.33,0.83}
  \definecolor{darkorchid}{rgb}{0.60,0.20,0.80}
  \definecolor{darkviolet}{rgb}{0.58,0.00,0.83}
  \definecolor{blueviolet}{rgb}{0.54,0.17,0.89}
  \definecolor{purple}{rgb}{0.63,0.13,0.94}
  \definecolor{mediumpurple}{rgb}{0.58,0.44,0.86}
  \definecolor{thistle}{rgb}{0.85,0.75,0.85}
\definecolor{snow}{rgb}{1.00,0.98,0.98}
\definecolor{ghostwhite}{rgb}{0.97,0.97,1.00}
\definecolor{whitesmoke}{rgb}{0.96,0.96,0.96} 
\definecolor{gainsboro}{rgb}{0.86, 0.86, 0.86}
\definecolor{floralwhite}{rgb}{1.00, 0.98, 0.94}
\definecolor{oldlace}{rgb}{0.99, 0.96, 0.90}
\definecolor{mistyroseen}{rgb}{0.98, 0.94, 0.90}
\definecolor{antiquewhite}{rgb}{0.98, 0.92, 0.84}
\definecolor{papayawhip}{rgb}{1.00, 0.94, 0.84}
\definecolor{blanchedalmond}{rgb}{1.00, 0.92, 0.80}
\definecolor{bisque}{rgb}{1.00, 0.89, 0.77}
\definecolor{peachpuff}{rgb}{1.00, 0.85, 0.73}
\definecolor{navajowhite}{rgb}{1.00, 0.87, 0.68}
\definecolor{moccasin}{rgb}{1.00, 0.89, 0.71}
\definecolor{cornsilk}{rgb}{1.00, 0.97, 0.86}
\definecolor{ivory}{rgb}{1.00, 1.00, 0.94}
\definecolor{lemonchiffon}{rgb}{1.00, 0.98, 0.80}

\makeatletter
\newcounter{savesection}
\newcounter{apdxsection}
\renewcommand\appendix{\par
  \setcounter{savesection}{\value{section}}%
  \setcounter{section}{\value{apdxsection}}%
  \setcounter{subsection}{0}%
  \gdef\thesection{\@Alph\c@section}}
\newcommand\unappendix{\par
  \setcounter{apdxsection}{\value{section}}%
  \setcounter{section}{\value{savesection}}%
  \setcounter{subsection}{0}%
  \gdef\thesection{\@arabic\c@section}}
\makeatother

\newcommand{\blue}[1]{\textcolor{blue}{#1}}
\newcommand{\mel}[1]{\textcolor{melanzana}{#1}}
\newcommand{\im}{{\operatorname{i}}}
\newcommand{\pfig}[1]{\parbox{18.3mm}{\epsfig{file=#1,height=15mm}}}
\usepackage{simpler-wick}

\newcommand{\re}{\operatorname{Re}}
\newcommand{\iu}{\mathrm{i}}
\newcommand{\nus}{\nu_\mathrm{s}}

\newcommand{\taufcp}{\tau_{\mathrm{fcp}}}
\newcommand{\vs}{v_{\mathrm{s}}}
\newcommand{\ie}{\textit{i.e.}}
\newcommand{\eg}{\textit{e.g.}}
\newcommand{\vsdp}{v_{\mathrm{sdp}}}
\newcommand{\uex}{u_\mathrm{ex}}
\newcommand{\taup}{\tau_{\mathrm{p}}}

\usepackage{graphicx,framed} 
\usepackage{dcolumn} 
\usepackage{bm} 
\usepackage{braket}
\usepackage{dsfont}
\usepackage{color,xcolor}
\usepackage{amsthm}
\usepackage[normalem]{ulem} 
\usepackage{qcircuit}       
\usepackage[export]{adjustbox}

\usepackage{amsthm}

\usepackage{comment}

\usepackage{algorithm}
\usepackage[noend]{algpseudocode}
\algdef{SE}[DOWHILE]{Do}{doWhile}{\algorithmicdo}[1]{\algorithmicwhile\ #1}
\makeatletter
\def\algbackskip{\hskip-\ALG@thistlm}
\makeatother

\definecolor{lightblue}{RGB}{73,151,208}
\definecolor{crimson}{RGB}{140,41,53}

\theoremstyle{definition}

\newtheorem{proposition}{Proposition}

\newtheorem{lemma}{Lemma}
\newtheorem{theorem}{Theorem}

\begin{document}

{\footnotesize \hfill \blue{\today}}
\vspace{5mm}

\preprint{}

\newcommand{\ourtitle}{Partition function for several Ising model interface structures} 

\title{\ourtitle}

\author{Alessio~Squarcini}
\email{alessio.squarcini@uni.lu}
\affiliation{Department of Physics and Materials Science, University of Luxembourg, 30 Avenue des Hauts-Fourneaux, L-4362 Esch-sur-Alzette, Luxembourg}

\author{Piotr~Nowakowski}
\email{Piotr.Nowakowski@irb.hr}
\affiliation{Group for Computational Life Sciences, Division of Physical Chemistry, Ru\dj{}er Bo\v skovi\' c Institute, Bijeni\v cka 54, 10000, Zagreb, Croatia}

\author{Douglas~B.~Abraham}
\affiliation{Rudolf Peierls Centre for Theoretical Physics, Clarendon Laboratory, Oxford OX1 3PU, UK}

\author{Anna~Macio\l ek}
\affiliation{Institute of Physical Chemistry, Polish Academy of Sciences, Kasprzaka 44/52, PL-01-224 Warsaw, Poland}
\affiliation{Max Planck Institute for Intelligent Systems, Heisenbergstr.~3, D-70569, Stuttgart, Germany}

\begin{abstract}
We employ a procedure that enables us to calculate the excess free energies for a finite Ising cylinder with domain walls analytically. This procedure transparently covers all possible configurations of the domain walls under given boundary conditions and allows for a physical interpretation in terms of coarse-grained quantities such as surface and point tensions.
The resulting integrals contain all the information about finite-size effects; we extract them by careful asymptotic analysis using the steepest descent method.
To this end, we exactly determine the steepest descent path and analyse its features.
For the general class of
integrals, which are usually found in the study of systems with inclined domain walls, knowledge of the steepest descent path is necessary to detect possible intersections with poles of the integrand in the complex plane.
\end{abstract}

\maketitle


\section{Introduction}
\label{sec:Intro}
Two crucial advances in the theory of phase transitions were made in 1936 and 1944. In the first, Peierls~\cite{peierls1936ising} demonstrated that taking the Ising model on a square lattice and fixing the boundary spins to be all $+$ would cause the average magnetisation of a central spin to be strictly positive for low enough temperatures, no matter how large the system is. There was a slight lacuna in this work which was fixed independently by Dobrushin~\cite{Dobrushin_1968} and by Griffiths \cite{Griffiths, Griffiths_1970}.

\begin{figure}[t]
\centering
\includegraphics[width=0.99\columnwidth]{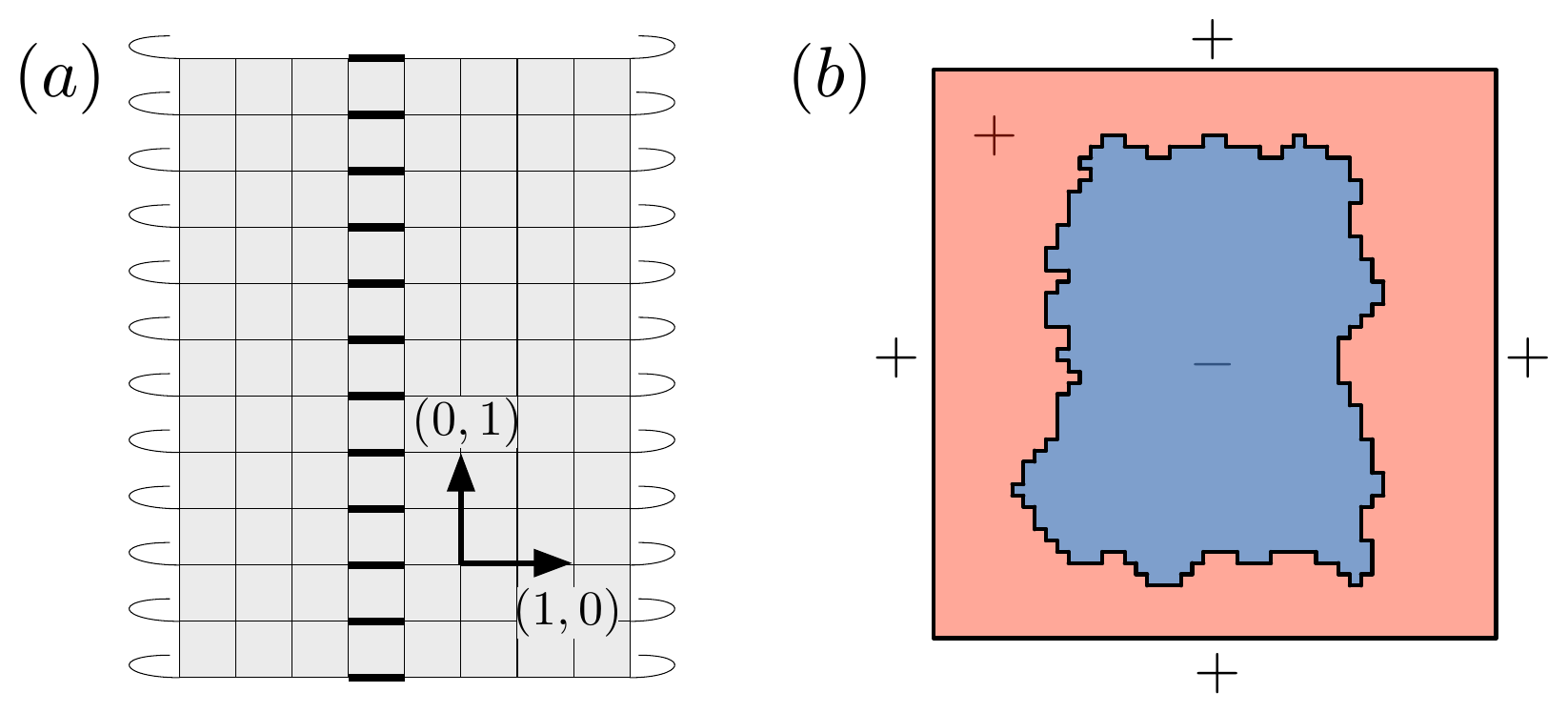}
\caption{Modification of the planar Ising model
 to introduce domain walls, or interfaces between pure phases: $(a)$~Lattice wrapped on a cylinder with a grain boundary of reversed bonds. $(b)$~Square domain with fixed total number of $+$ and $-$ spins.}
\label{fig_1new}
\end{figure}

The second advance was made in 1944, by Onsager~\cite{Onsager_44}. He showed that in the two-dimensional Ising model there is a critical temperature $T_{\mathrm{c}}$ at which the specific heat diverges as $\ln\left|(T_{\mathrm{c}}-T)/T_{\mathrm{c}}\right|$ (provided the lattice is infinite in extent). 
Further, Onsager gave a definition of surface tension which was carefully tailored to make it amenable to the techniques he had developed. It is based on a grain boundary of reversed bonds, as illustrated in Fig.~\ref{fig_1new}$(a)$, and assumes the existence of phase separation. The incremental free energy associated with this  is strictly non-zero per unit length for all $T<T_{\mathrm{c}}$ (the temperature at which the free energy is singular). 
Another idea for introducing phase separation lines was that of  Gallavotti, Martin-L\"of and Miracle-Sol\'e~\cite{GMM72}: 
suppose we take the Peierls setup and impose the additional restriction that $N_{+} + N_{-} = N$, where $N_{+}$ 
and  $N_{-}$ are the number of up 
and down spins, respectively, and these are fixed. Then we might hope for the picture as the one shown in Fig.~\ref{fig_1new}$(b)$. The $+$ phase is pinned to the boundary, the $-$ phase has a fluctuating boundary and ``floats around'' embedded in the $+$ phase as shown in Fig.~\ref{fig_1new}$(b)$. The boundary $\partial_{-}$ fluctuates as $N^{1/2}$ as shown in remarkable work by Gallavotti~\cite{Gallavotti_1972, Gallavotti}.

\begin{figure}[t]
\centering
\includegraphics[width=0.9999\columnwidth]{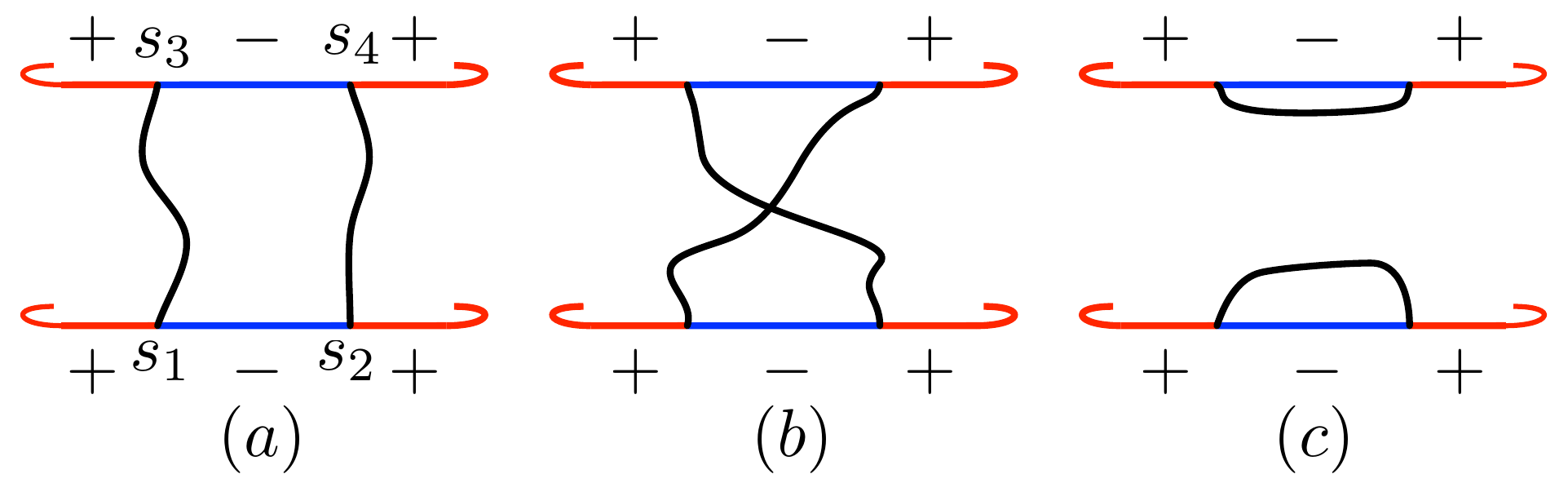}
\caption{Lattice with pattern of $+-+$ pattern of boundary conditions and corresponding long contour connections: a bridge of phase $-$ $(a)$, two crossing interfaces $(b)$, and two isolated droplets $(c)$.}
\label{fig_2}
\end{figure}

What is known as the ``Dobrushin program'' takes this matter in a different, but related, direction~\cite{Dobrushin_1968}; see also~\cite{Gallavotti}. Consider a lattice with boundary conditions of type $+ - +$ on the upper and lower edges and periodic boundary conditions along the horizontal direction. For this system there are three possible long contour connections, as shown in Fig.~\ref{fig_2}. In case (a), we will have two independent interfaces if $|s_{1}-s_{2}| \rightarrow \infty$ and $|s_{2}-s_{4}| \rightarrow \infty$ and configurations like $(b)$ and $(c)$ will be suppressed relative to $(a)$. On the other hand, for small values of $|s_{1}-s_{2}|$ and $|s_{2}-s_{4}|$ and  a sufficiently large separation between the two edges, the weight of the configuration $(c)$ can prevail on $(a)$ and $(b)$.
The leverage of Dobrushin boundary conditions enables geometrically controlled tuning of interface configurations, while preserving exact solvability. 

Closed-form expressions for the incremental free energies in systems in which the interface is pinned by $+-$ boundary conditions, as those illustrated in Fig.~\ref{fig_2},
can be obtained  by wrapping  the lattice into a cylinder and then employing the transfer matrix technique with the transfer direction along the cylinder axis, the direction $(0,1)$ in Fig.~\ref{fig_1new}$(a)$. This method has been successfully applied, among others, to study mass gaps in the Ising model with adjustable boundary conditions~\cite{AKS_1988, AKS_1989} or to study grain boundaries~\cite{AMW_2005} and their wetting effects~\cite{AMW_2004}. The calculation of incremental free energies can also be performed by taking the transfer matrix along the edge of the strip, the direction $(1,0)$ in Fig.~\ref{fig_1new}$(a)$. These two approaches  must be  equivalent and  have their own merits. In Ref.~\cite{AMSV_2017} we were able to demonstrate  that for a cylinder with free edges it is possible to calculate the point tension $\tau_{\mathrm{p}}$ using both approaches
and that they are consistent with each other despite the fact that they rely on completely different mathematical formulations.

Recently, interest in incremental free energies for certain interface configurations has gained new momentum, as they have proven essential for understanding the emergence of long-range order in the Ising  ``network'' model introduced in Refs~\cite{AMV_2014,AMSV_2017,AMS_2024}.

\begin{figure}[t]
\centering
\includegraphics[width=1.0\columnwidth]{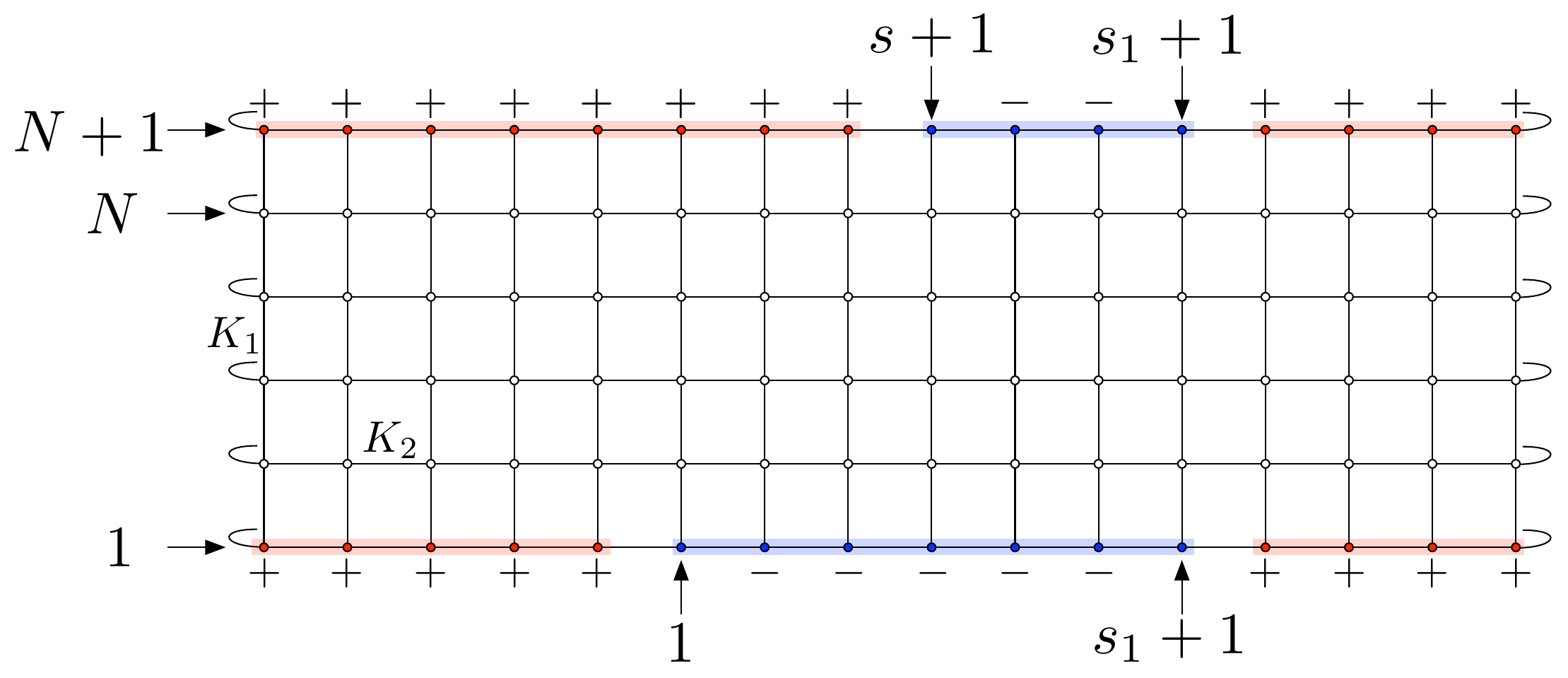}
\caption{Cylindrical lattice with boundary conditions leading to the formation of the interfacial structures shown in Fig.~\ref{fig_2}.}
\label{fig_3_new}
\end{figure}

The purpose of this paper is to provide in some detail the exact solution for incremental free energies associated with the interfacial structures shown in Fig.~\ref{fig_2}, which were not presented before. 
Specifically, we consider the edge-to-edge transfer matrix for the strip shown in Fig.~\ref{fig_2} and for a strip with the same lower edge, but whose upper edge has all spins fixed at $+$ (see Figs \ref{fig_3_new} and \ref{fig_3_new_bis}, respectively). 
The construction of an exact solution  in these cases is feasible because the block of spin rotation operators that flips the spins at the strip edge between positions $s_1$ and $s_2$ is quadratic in the lattice fermions.
The action of the transfer operator along the cylinder axis, propagates this boundary condition throughout the system; the direction of the propagation 
$(0,1)$  is equivalent  to the time direction for the two fermions emitted by the boundary. 
The resulting phase separation lines, due to topological constraints, cannot end in the middle of the system but at its boundary, as shown  in Fig.~\ref{fig_2}. Technically, the solution can be obtained  by employing a generalised version of Wick's theorem in which the in and out states are not the same~\cite{Abraham_2012}. Within this framework, the Wick contractions of lattice fermions can be handled by means of diagrammatic techniques and the classification of the resulting graphs is in one-to-one correspondence with the interfacial structures shown in Fig.~\ref{fig_2}. 
Anticipating some results, the Wick contraction for the cross-connected interfaces---not represented by Peierls contours---will acquire an overall minus sign due to the fermionic statistics under particle exchange.

The advantage in adopting the transfer along $(0,1)$ is that the final result for the incremental free energy can be expressed as a single integral involving a function that depends on the type of interfacial configuration, \eg, a single  interface as in Fig.~\ref{fig_2}$(a)$ or a single interface as in Fig.~\ref{fig_2}$(c)$. This is a significant simplification over the approach based on the transfer along $(1,0)$. The latter would yield the excess free energy in the form of a summation over a discrete set of momenta, which can ultimately be calculated by employing a summation kernel technique, as known for the Casimir effect~\cite{AM_2010,AM_2013}. In this sense, the generalised Wick theorem within the Dobrushin program allows to perform the mode summation in an implicit way. The resulting single integral contains all information about finite size effects that eventually can be extracted by a careful asymptotic analysis.

In order to compute a single integral giving a incremental free energy or to evaluate this integral asymptotically, we will need to perform saddle point calculations; the latter requires knowledge of the steepest descent path (SDP), which was never explored before. We will show that the SDP can be found exactly and in closed form.


\section{Formulation}
\label{sec_2}

We consider the planar  Ising model on a square lattice~\cite{Onsager_44} with the  Hamiltonian given by 
\begin{equation}
\label{30042021_1629}
E = - J_{1} \sum_{\langle i,j\rangle} \sigma_{i,j} \sigma_{i+1,j} - J_{2} \sum_{\langle i,j\rangle} \sigma_{i,j} \sigma_{i,j+1} ,
\end{equation}
with $J_{1}, J_{2}>0$ and spin variables $\sigma_{i,j}=\pm1$, where the sums are restricted to the nearest neighbouring pairs of sites in a square lattice.
We  wrap the lattice into a cylinder of circumference $M$ and  height $N$ with axis along the (vertical) lattice direction $(0,1)$. 
We introduce  domain walls into the system by setting the value of the boundary spins equal to $\sigma_{i,j}=+1$  everywhere, except for the segments where the spins are $\sigma_{i,j}=-1$, as shown in Fig.~\ref{fig_3_new}.


The incremental  free energy associated with the insertion of these  domain walls is given by   $-\ln (Z_{+-}/Z_{++})$, where $Z_{+-}(N,M \vert s,s_{1})$ is the partition function of the system depicted in Fig.~\ref{fig_3_new} and $Z_{++}(N,M)$ is the partition function for the cylinder with no reversed spins on both edges.
We show below that this quantity includes contributions from the interface configurations shown in Fig.~\ref{fig_2}. Taking the limit $M\rightarrow\infty$ followed by $s_{1} \rightarrow\infty$  eliminates possible interactions between the two domains and allows us to determine the incremental free energy for  only one of them. We denote
\begin{equation}
\label{30042021_1630}
\zeta_{\mathrm{A}}\left(N,s\right) = \lim_{s_{1} \to \infty} \lim_{M \to \infty}  \frac{ Z_{+-}\left(N,M \vert s,s_{1}\right)}{ Z_{++}\left(N,M\right) }.
\end{equation}

\begin{figure}[t]
\centering
\includegraphics[width=1.0\columnwidth]{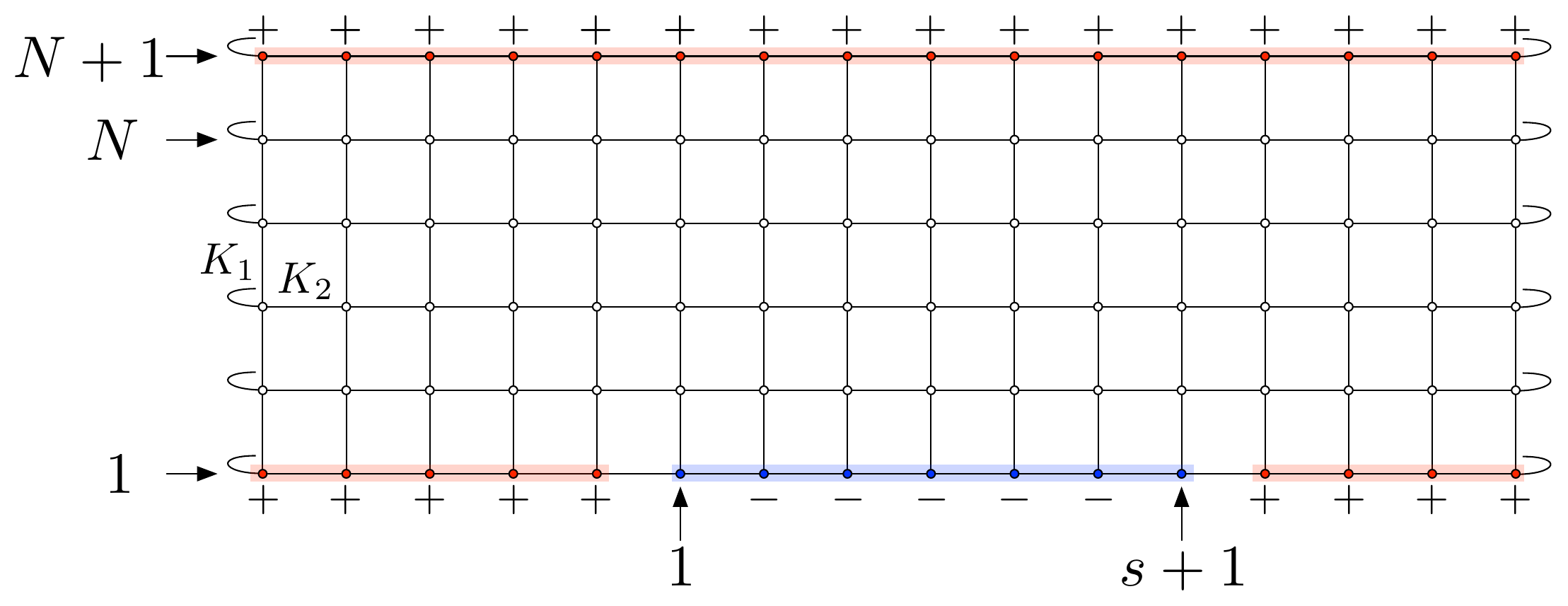}
\caption{Cylindrical lattice with boundary conditions leading to the formation of a droplet pinned on the lower edge.}
\label{fig_3_new_bis}
\end{figure}

A completely different type of boundary-induced effects can be expected when only the lower edge has a finite segment of reversed spins, as illustrated in Fig.~\ref{fig_3_new_bis}. This  induces a domain wall whose endpoints are attached to the lower edge and separated by $s$ lattice sites. We denote
\begin{equation}
\label{19112024_1131}
\zeta_{\mathrm{B}}\left(N,s\right) = \lim_{M \to \infty}  \frac{ Z_{+-}\left(N,M \vert s\right) }{ Z_{++}\left(N,M\right) },
\end{equation}
where $Z_{+-}(N,M \vert s)$ is the partition function for the lattice shown in  Fig.~\ref{fig_3_new_bis}, which gives us the associated incremental free energy.


Next, we use a transfer matrix formalism with cyclic boundary conditions in the $(1,0)$ direction in order to derive the quantities  $\zeta_{\mathrm{B}}$ and  $\zeta_{\mathrm{A}}$.


We introduce the transfer operator $\mathsf{V}_{1}$
\begin{equation}
\label{30042021_1406}
\mathsf{V}_{1} = \exp\left[ - K_{1}^{\star} \sum_{m=1}^{M} \sigma_{m}^{z} \right],
\end{equation}
which corresponds to the transfer between neighbouring rings of the cylindrical lattice~\cite{Abraham_SAM_1971}. The reduced coupling $K_{1}$ is defined by $K_{1}=\beta J_{1}$, $K_{1}^{\star}$ is the dual coupling defined by $\tanh(K_{1}^{\star}) = \exp(-2K_{1})$. The operator $\sigma_{m}^{\mu}$, defined by
\begin{equation}
\sigma_{m}^{\mu} = \underbrace{\mathsf{I} \otimes \cdots \otimes \mathsf{I}}_{m-1} \otimes \sigma^{\mu} \otimes \mathsf{I} \otimes \cdots \otimes \mathsf{I} \, , \qquad \mu=x,y,z \, ,
\end{equation}
acts as a Pauli spin operator $\sigma^{\mu}$ on site $m$ and as the identity elsewhere. Following the convention of \cite{SML}, we choose a representation for Pauli spin matrices in which $-\sigma^{z}$---for historical reasons---is identified with the spin flip operator and $\sigma^{x}$ is the magnetization operator. Strictly speaking, the transfer operator $\mathsf{V}_{1}$ is defined up to an overall factor, which however simplifies in the computation of partition function ratios, and for this reason such a factor is not shown here. The transfer within each horizontal row in the cylindrical lattice is taken into account by the operator $\mathsf{V}_{2}$,
\begin{equation}
\label{30042021_1408}
\mathsf{V}_{2} = \exp\left[ K_{2} \sum_{m=1}^{M-1} \sigma_{m}^{x}\sigma_{m+1}^{x} + K_{2} \sigma_{M}^{x}\sigma_{1}^{x} \right],
\end{equation}
with $K_{2}=\beta J_{2}$. The occurrence of the term $\sigma_{M}^{x}\sigma_{1}^{x}$ is signalling the fact that each horizontal row is closed, forming a ring of circumference $M$. The edge with spins locked all up or down, respectively, is indicated with $\vert \pm \rangle$. The latter is an eigenstate of $\sigma_{m}^{x}$ such that $\sigma_{m}^{x} \vert \pm \rangle = \pm \vert \pm \rangle$ for $1 \leqslant m \leqslant M$. The fixed spin states $\vert \pm \rangle$ are constructed as follows:
\begin{equation}
\vert \pm \rangle = 2^{-M/2} \prod_{m=1}^{M} \left( 1 \pm \sigma_{m}^{x} \right) \vert 0 \rangle,
\end{equation}
where $\vert 0 \rangle$ is an eigenstate of the spin flip operator with eigenvalue $+1$, \ie, $-\sigma_{m}^{z} \vert 0 \rangle = \vert 0 \rangle$ for $1 \leqslant m \leqslant M$.

A boundary configuration with pointing down spins between $s_{1} \geqslant 1$ and $s_{2} > s_{1}$, with $s_{2} \leqslant M$, is given by
\begin{equation}
\label{11062025_1058}
\vert s_{1}, s_{2} ; + \rangle = \prod_{j=s_{1}}^{s_{2}} (-\sigma_{j}^{z}) \vert + \rangle,
\end{equation}
which has the property
\begin{equation}
\sigma_{s}^{x} \vert s_{1}, s_{2} ; + \rangle
=
\begin{cases}
- \vert s_{1}, s_{2} ; + \rangle      & \text{if} \quad s_{1} \leqslant s \leqslant s_{2}, \\
+ \vert s_{1}, s_{2} ; + \rangle      & \text{otherwise} .
\end{cases}
\end{equation}

\subsubsection{Determination of $\zeta_{\mathrm{B}}$}
\label{Boltzmann_weight_B}

Using this notation we can write the partition function ratio for  the lattice in Fig.~\ref{fig_3_new_bis} as follows:
\begin{equation}
\label{19112024_1147}
\frac{Z_{+-}\left(M,N \vert s\right)}{Z_{++}\left(M,N\right)} = \frac{ \langle + \vert \left( \mathsf{V}_{1} \mathsf{V}_{2} \right)^{N-1} \mathsf{V}_{1} \vert 1,s+1 ; + \rangle }{ \langle + \vert \left( \mathsf{V}_{1} \mathsf{V}_{2} \right)^{N-1} \mathsf{V}_{1} \vert + \rangle }.
\end{equation}

The calculation proceeds by introducing the symmetrised product $\mathsf{V} = \mathsf{V}_{2}^{1/2} \mathsf{V}_{1}^{\vphantom{1/2}} \mathsf{V}_{2}^{1/2}$. It is known since the work of Kaufman~\cite{Kaufman_49} that, in order to diagonalize the symmetrized forms of the transfer operators, it is a good idea to introduce the Jordan--Wigner transformation and the lattice spinors $\Gamma_{m}$. These are introduced in the following together with Fermi lattice fields $f_{m}^{\vphantom{\dag}}$ and $f_{m}^{\dag}$ defined by
\begin{equation}
f_{m}^{\vphantom{\dag}} = \mathcal{P}_{m-1} \sigma_{m}^{-} , \qquad f_{m}^{\dag} = \mathcal{P}_{m-1} \sigma_{m}^{+},
\end{equation}
where $\sigma_{m}^{\pm}$ are the raising/lowering operators $\sigma_{m}^{\pm}=(\sigma_{m}^{x} \pm \im \sigma_{m}^{y})/2$, and
\begin{equation}
\mathcal{P}_{m} = \prod_{\ell=1}^{m}\left( -\sigma_{\ell}^{z} \right)
\end{equation}
is the Jordan--Wigner tail. We note that Jordan--Wigner tail operators satisfy $\mathcal{P}_{M} \vert + \rangle = \vert - \rangle$. It then follows that Fermi lattice fields satisfy the anti-commutation relations
\begin{equation}
\left[ f_{n}, f_{m} \right]_{+} = 0, \qquad \left[ f_{n}^{\vphantom{\dag}}, f_{m}^{\dag} \right]_{+} = \delta_{n,m},
\end{equation}
as required for fermions. The state $\vert 0 \rangle$ is a $f$-vacuum, namely $f_{m} \vert 0 \rangle=0$ for $1 \leqslant m \leqslant M$. Then, we introduce Kaufman spinors by means of
\begin{subequations}
\label{06052021_0658}
\begin{align}
\Gamma_{2m-1} & = f_{m}^{\dag} + f_{m}^{\vphantom{\dag}} = \mathcal{P}_{m-1} \sigma_{m}^{x}, \\
\Gamma_{2m} & = - \im \left( f_{m}^{\dag} - f_{m}^{\vphantom{\dag}} \right) = - \im\mathcal{P}_{m-1} \sigma_{m}^{y},
\end{align}
\end{subequations}
for $1 \leqslant m \leqslant M$ with $\mathcal{P}_{0} \equiv \mathsf{I}$. The lattice spinors satisfy the anti-commutation relation
\begin{equation}
\left[ \Gamma_{n}, \Gamma_{m} \right]_{+} = 2 \delta_{n,m}.
\end{equation}

Lattice spinors and lattice Fermi fields allow to write the boundary state with a patch of reversed bonds as  a block rotation operator acting on the hard edge state $\vert + \rangle$. The block rotation operator for the lower edge in Fig.~\ref{fig_3_new_bis} is explicitly constructed by using the algebra of Jordan--Wigner tail operators, which entails
\begin{equation}
\mathcal{P}_{s_{1}-1} \mathcal{P}_{s_{2}} = \prod_{j=s_{1}}^{s_{2}} (-\sigma_{j}^{z}).
\end{equation}
Therefore the state in Eq.~\eqref{11062025_1058}, describing the patch of reversed spins, can be written as
\begin{equation}
\label{11062025_1100}
\vert s_{1}, s_{2} ; + \rangle = \mathcal{P}_{s_{1}-1} \mathcal{P}_{s_{2}} \vert + \rangle.
\end{equation}
By bringing in Kaufman spinors, the above becomes
\begin{multline}
\label{11062025_1102}
\vert s_{1}, s_{2} ; + \rangle  = - \mathcal{P}_{s_{1}-1} \sigma_{s_{1}}^{x} \mathcal{P}_{s_{2}} \sigma_{s_{2}+1}^{x} \vert + \rangle \\
 = - \Gamma_{2s_{1}-1} \Gamma_{2s_{2}+1} \vert + \rangle \equiv R(s_{1},1+s_{2}) \vert + \rangle,
\end{multline}
which defines the block rotation operator $R(s_{1},1+s_{2})$ acting on the fixed spin state $\vert + \rangle$.
In terms of fermion operators, the block rotation operator becomes
\begin{equation}
\label{06052021_0659}
R\left(s_{1},1+s_{2}\right) = \mel{-} \left( f_{s_{1}}^{} + f_{s_{1}}^{\dag} \right) \left( f_{1+s_{2}}^{} + f_{1+s_{2}}^{\dag} \right).
\end{equation}

It is then important to note that $\vert + \rangle$ is an eigenstate of $\mathsf{V}_{2}$ such that $\mathsf{V}_{2} \vert \pm \rangle = \exp(MK_{2}) \vert \pm \rangle$. By using this property and the identity
\begin{equation}
\operatorname{e}^{ \im K \Gamma_{n}\Gamma_{m}} \Gamma_{m} \operatorname{e}^{ - \im K \Gamma_{n}\Gamma_{m} } = \operatorname{e}^{ 2 \im K \Gamma_{n} \Gamma_{m} } \Gamma_{m},
\end{equation}
Eq.~\eqref{19112024_1147} simplifies to
\begin{equation}
\label{19112024_1218}
\frac{Z_{+-}\left(M,N \vert s\right)}{Z_{++}\left(M,N\right)} = \frac{ \langle + \vert \mathsf{V}^{N} R\left(1,1+s\right) \vert + \rangle }{ \langle + \vert \mathsf{V}^{N} \vert + \rangle }.
\end{equation}

We then express the hard edge states $\vert \pm \rangle$ in terms of the (degenerate) maximal eigenvectors of $\mathsf{V}_{2}$, namely
\begin{equation}
\label{10062025_2141}
\vert \pm \rangle = \frac{1}{\sqrt{2}} \left( \vert \Phi_{+}^{0} \rangle \pm \vert \Phi_{-}^{0} \rangle \right) .
\end{equation}
By using projection operators
\begin{equation}
\mathcal{P}\left(\pm\right) = \frac{1}{2} \left( 1 \pm \mathcal{P}_{M} \right),
\end{equation}
we can write
\begin{equation}
\mathsf{V}_{2} = \mathcal{P}\left(+\right) \mathsf{V}_{2}\left(+\right) + \mathcal{P}\left(-\right) \mathsf{V}_{2}\left(-\right),
\end{equation}
where $\mathsf{V}_{2}(\pm)$ is
\begin{equation}
\label{06052021_1154}
\mathsf{V}_{2} = \exp\left[ \im K_{2} \sum_{m=1}^{M-1} \Gamma_{2m}\Gamma_{2m+1} - \im K_{2} \mathcal{P}_{M} \Gamma_{2M} \Gamma_{1} \right],
\end{equation}
with the appropriate eigenvalue of the parity operator $\mathcal{P}_{M}$, either $\pm1$. As a result, the ratio of partition functions (\ref{19112024_1218}) is given by the sum of two terms of the form $2^{-1} \langle \Phi_{\pm}^{0} \vert \mathsf{V}^{N}\left(\pm\right) R\left(1,1+s\right) \vert \Phi_{\pm}^{0} \rangle/\langle \Phi_{\pm}^{0} \vert \mathsf{V}^{N}\left(\pm\right) \vert \Phi_{\pm}^{0} \rangle$, where $\mathsf{V}\left(\pm\right)$ is the corresponding symmetrised transfer matrix. Since the limit $M\rightarrow\infty$ is performed, both the terms are asymptotically identical. We will therefore focus on one of the terms in the following analysis:
\begin{equation}
\begin{aligned}
\label{30052020_01}
\lim_{M\to\infty}\frac{Z_{+-}\left(M,N \vert s\right)}{Z_{++}\left(M,N\right)} & = \frac{ \langle \Phi_{+}^{0} \vert \left[ \mathsf{V}(+) \right]^{N} R\left(1,1+s\right) \vert \Phi_{+}^{0} \rangle }{ \langle \Phi_{+}^{0} \vert \left[ \mathsf{V}(+) \right]^{N} \vert \Phi_{+}^{0} \rangle } .
\end{aligned}
\end{equation}
The calculation of the matrix element shown in Eq.~\eqref{30052020_01} is carefully outlined in Appendix~\ref{app_g_w_t}. By taking the limit $M \to \infty$, the calculation yields for 
the droplet-like interface configuration:
\begin{multline}
\label{03052021_0909}
\zeta_{\mathrm{B}}\left(N,s\right) =\\ \int_{-\pi}^{\pi}\frac{\operatorname{d}\omega}{2\pi \im } \frac{ \sinh \left[N\gamma\left(\omega\right)\right] \sin\delta^{*}\left(\omega\right) \operatorname{e}^{\im s\omega } }{ \cosh\left[ N\gamma\left(\omega\right)\right] + \sinh \left[N\gamma\left(\omega\right)\right]  \cos\delta^{*}\left(\omega\right) }.
\end{multline}
Although this result contains an overall imaginary unit, $\zeta_{\mathrm{B}}\left(N,s\right)$ is purely real; it is because $\sin\delta^{*}\left(\omega\right)$ is an odd function of $\omega$. A careful analysis shows that $\zeta_{\mathrm{B}}\left(N,s\right)$ is actually positive, as it must be. Onsager function $\gamma\left(\omega\right)$ is defined as the positive solution of
\begin{equation}
\label{01022021_1752}
\cosh\gamma\left(\omega\right) = c_{1}^{\star}c_{2}^{\vphantom{\star}} - s_{1}^{\star}s_{2}^{\vphantom{\star}} \cos \omega ,
\end{equation}
where for the sake of brevity we use $s_{j}=\sinh(2K_{j})$, $c_{j}=\cosh(2K_{j})$, $s_{j}^{\star}=\sinh(2K_{j}^{\star})$, and $c_{j}^{\star}=\cosh(2K_{j}^{\star})$ with $j=1,2$. The angle $\delta^{*}(\omega)$ of Onsager hyperbolic triangle satisfies the hyperbolic sine law
\begin{equation}
\label{25112024_1235}
s_{1}^{\star} \sin \omega = \sinh\gamma\left(\omega\right) \sin \delta^{*}\left(\omega\right)
\end{equation}
and the identity
\begin{equation}
\sinh\gamma\left(\omega\right) \cos\delta^{*}\left(\omega\right) = c_{1}^{\star}s_{2}^{\vphantom{\star}} - s_{1}^{\star}c_{2}^{\vphantom{\star}} \cos \omega .
\end{equation}

\subsubsection{Determination of $\zeta_{\mathrm{A}}$}
\label{Boltzmann_weight_A}
In order to apply the formalism developed in the previous section to calculation of $\zeta_{\mathrm{A}}$, we need to insert another block rotation operator implementing the upper edge of the lattice shown in Fig.~\ref{fig_3_new}. The ket-state representing such an edge is $\vert 1+s, 1+s_{1} \rangle$; hence,
\begin{multline}
\label{19112024_1249}
\frac{Z_{+-}\left(M,N \vert s,s_{1}\right)}{Z_{++}\left(M,N\right)} = \\ \frac{ \langle 1+s,1+s_{1} \vert \left( \mathsf{V}_{1} \mathsf{V}_{2} \right)^{N-1} \mathsf{V}_{1} \vert 1,1+s \rangle }{ \langle + \vert \left( \mathsf{V}_{1} \mathsf{V}_{2} \right)^{N-1} \mathsf{V}_{1} \vert + \rangle }.
\end{multline}
By bringing in block rotation operators and the symmetrised transfer operator $\mathsf{V}$, the procedure described in the previous section naturally leads to a matrix element of the form
\begin{equation}
\label{19112024_1306}
\langle \Phi_{+}^{0} \vert R^{\dag}\left(1+s,1+s_{1}\right) \left[ \mathsf{V}(+) \right]^{N} R\left(1,1+s\right) \vert \Phi_{+}^{0} \rangle
\end{equation}
as the counterpart of Eq.~\eqref{30052020_01}. This matrix element contains three contributions stemming from certain contractions whose interpretation is in one-to-one correspondence with the three interfacial structures depicted in Fig.~\ref{fig_2}. Indeed, we can show that the partition function ratio admits the diagrammatic decomposition
\begin{widetext}
\begin{equation}
\label{19112024_1446}
\lim_{M \to \infty} \frac{Z_{+-}\left(M,N \vert s,s_{1}\right)}{Z_{++}\left(M,N\right)} = \pfig{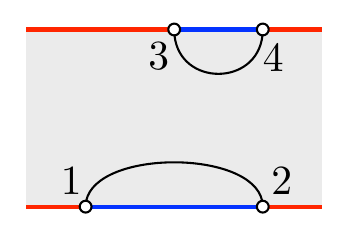} \quad + \pfig{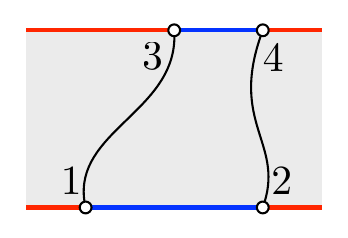} \quad + \pfig{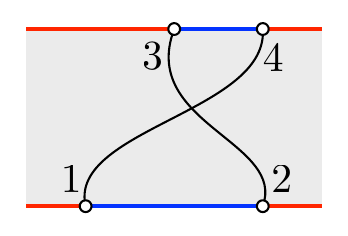} \quad.
\end{equation}
\end{widetext}
These diagrams, corresponding to certain Wick contractions, yield the incremental free energy for all possible configurations of domain walls given the boundary conditions of the lattice in Fig.~\ref{fig_3_new}. The first diagram describes the contribution due to two droplets, each with incremental free energy of type $\zeta_{\mathrm{B}}$. The calculation shows that
\begin{equation}
\label{27032025_1548}
\pfig{aa} \quad\!\! = \zeta_{\mathrm{B}}\left(N,r_{12}\right) \zeta_{\mathrm{B}}\left(N,r_{34}\right),
\end{equation}
where $r_{ij}$ are the distances between endpoints. In agreement with the lattice of Fig.~\ref{fig_3_new}, in our case $r_{12}=s_{1}$ and $r_{34}=s_{1}-s$. We note that the factorisation in the right hand side of Eq.~\eqref{27032025_1548} corresponds to two \emph{non-interacting} droplets. For the two droplets, we can write
\begin{equation}
\pfig{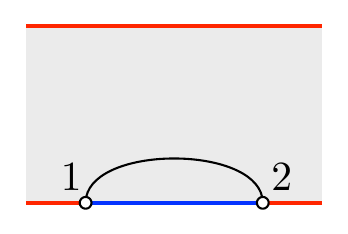} \quad \!\! = \zeta_{\mathrm{B}}\left(N,r_{12}\right), \pfig{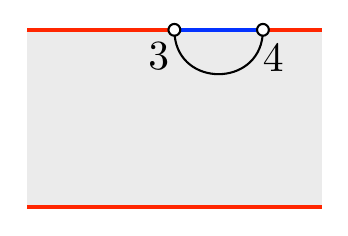} \quad\!\! = \zeta_{\mathrm{B}}\left(N,r_{34}\right).
\end{equation}
Then, the remaining diagrams are expressed as factorized forms involving the quantity $\zeta_{\mathrm{A}}$ for a domain wall with endpoints on opposite edges. More precisely:
\begin{equation}
\label{19112024_2017}
\pfig{bb} \quad\!\! = \zeta_{\mathrm{A}}\left(N,r_{13}\right) \zeta_{\mathrm{A}}\left(N,r_{24}\right),
\end{equation}
and
\begin{equation}
\label{20112024_0913}
\pfig{cc} \quad\!\! = - \zeta_{\mathrm{A}}\left(N,r_{14}\right) \zeta_{\mathrm{A}}\left(N,r_{23}\right).
\end{equation}
In agreement with the lattice of Fig.~\ref{fig_3_new}, the distances are $r_{13}=s$, $r_{24}=0$, $r_{14}=s_{1}$, and $r_{23}=s_{1}-s$. The occurrence of the factor $-1$ in the last diagram is due to the fermionic statistics accounting for the crossing of two interfaces. 
By performing the limit $s_{1} \to \infty$ the cross-connected diagram in Eq.~\eqref{20112024_0913}) vanishes as well as the contribution from the two droplets in Eq.~\eqref{27032025_1548}, therefore we are left with a single (left) interface in the  diagram (\ref{19112024_2017}), as expected.
Deferring to Appendix~\ref{app_g_w_t} for the details, the result of this procedure 
gives
\begin{multline}
\label{03052021_1247}
\zeta_{\mathrm{A}}\left(N,s\right) =\\
\int_{-\pi}^{\pi}\frac{\operatorname{d}\omega}{2\pi} \frac{ \operatorname{e}^{\im s\omega } }{ \cosh \left[N\gamma\left(\omega\right)\right] + \sinh\left[ N\gamma\left(\omega\right)\right] \cos\delta^{*}\left(\omega\right)}.
\end{multline}

\section{Steepest descent path}
\label{sec_sdp_v1}
Integrals of the type (\ref{03052021_0909}) and (\ref{03052021_1247}) belong to a much more general class that is often found in the calculation of correlation functions and excess free energies in the Ising model. Generally, these integrals can be brought to the form
\begin{equation}
\label{00001}
I\left(M,N\right) = \int_{-\pi}^{\pi} \frac{\operatorname{d}\omega}{2\pi}  g\left(\omega\right) \operatorname{e}^{-N\gamma\left(\omega\right) + \im M\omega},
\end{equation}
where $g\left(\omega\right)$ is some function whose actual expression depends on the problem at hand. The geometrical parameters $N$ and $M$ are usually considered to be large compared to the lattice spacing, and the aim is to extract the asymptotic behaviour of $I\left(M,N\right)$ when $N,M \gg 1$ with fixed ratio $M/N$. It is clear that this task brings us to the realm of the asymptotic analysis of integrals. In particular, the analysis of $I\left(M,N\right)$ requires the application of standard techniques such as the steepest descent path method and complex analysis \cite{Bleistein1987}. Here we address this problem in full generality without relying on any specific choice for the function $g\left(\omega\right)$. We only require that $g\left(\omega\right)$ is an analytic complex function for $\re \omega\geqslant0$ and $-\pi \leqslant\operatorname{Im}\omega\leqslant \pi $ that satisfies the conditions
\begin{subequations}\label{g_conditions}
\begin{align}
g\left(-\pi+\im v\right) &=g\left(\pi+\im v\right), &&\text{for any }v\geqslant 0,\\
g\left(\omega\right)&\to 0,  &&\text{for } \operatorname{Im}\omega\to+\infty.
\end{align}
\end{subequations}
The analysis will proceed through the following stages: First, we determine the location of the saddle point and the shape of steepest descent path. Second, we modify the integration contour to incorporate the steepest decent path, and third, we calculate asymptotic expansion of the integral for large size of the system $M$ and $N$. Finally, we will apply our approach to the specific cases of incremental free energies as provided by 
$\zeta_{\mathrm{A}}$ and $\zeta_{\mathrm{B}}$.

\subsection{Saddle point and steepest descent path}
\label{sec_sdp_v1a}
We begin the analysis of Eq.~\eqref{00001} by locating the saddle point of the function in the exponent. As anticipated, $M$ and $N$ are taken to be large with a finite ratio. Thus, we have $M,N\to\infty$ with the ratio parametrized by the angle $\vartheta$ as follows: $M=[N\tan\vartheta]$, $\vartheta \in [0,\pi/2)$, where the notation $[\cdot]$ stands for the integer part. The exponential part of the integral \eqref{00001} containing the large parameters can be written as $\exp\left[-N\gamma\left(\omega\right) + \im M\omega\right] \equiv \exp\left[-N L\left(\omega\right)\right]$ with $L\left(\omega\right) = \gamma\left(\omega\right) - \im \omega  \tan \vartheta$. The saddle point $\omega_\mathrm{s}$ is determined as the solution of $L^{(1)}\left(\omega\right)=0$, \ie,
\begin{equation}
\label{09022021_0712}
\gamma^{(1)}\left(\omega_{\mathrm{s}}\right) = \im M/N=\im \tan\vartheta,
\end{equation}
where the superscript denotes the first derivative with respect to $\omega$. It is straightforward to check (see Sec.~\ref{sec_3} of Supplementary Information (SI)) that the saddle point is \emph{unique} and it is located on the imaginary axis, $\omega_{\mathrm{s}}\left(\vartheta\right) = \im \vs\left(\vartheta\right)$ with positive $\vs\left(\vartheta\right)$. The function $\vs\left(\vartheta\right)$ is known for any $\vartheta\in \left[0,\pi/2\right)$ and arbitrary couplings $K_{1}$ and $K_{2}$, the expression is derived in SI, see Eq.~\eqref{01022021_1857}. In Sec.~\ref{sec_3} of SI we also prove that $\vs\left(\vartheta\right)$ is strictly monotonically increasing for $\vartheta \in \left(0,\pi/2\right)$ with the following limiting values: $\vs\left(0\right)=0$ and $\vs\left(\pi/2\right) = \hat{\gamma}\left(0\right)$, where $\hat{\gamma}\left(\omega\right)$ is $\gamma\left(\omega\right)$ given by Eq.~\eqref{01022021_1752} with $K_1$ and $K_2$ interchanged.

We are now in the position to discuss the steepest descent path (SDP) and its main properties. By definition, the imaginary part of $L\left(\omega\right)$ stays constant along the steepest descent path. Since $\operatorname{Im}\left(L\left(\omega\right)\right)$ is zero at the saddle point, along the SDP the following condition is satisfied
\begin{equation}
\label{00007}
\operatorname{Im}\left[N\gamma\left(\omega\right) - \im M\omega\right]=0,
\end{equation}
and consequently, the exponential factor in Eq.~\eqref{00001} is purely real along the SDP.

Finding the analytic expression for the SDP is a conceptually simple exercise, although in the case of arbitrary $\vartheta$ it requires cumbersome calculations. A detailed account on the analytic expression of the SDP, its derivation and properties, are provided in Sec.~\ref{sec_4} of SI. Here, we outline the main ideas underlying the exact determination of the SDP and, on the way, we will present its salient features.
We introduce the complex wave number $\omega = u+ \im v$ with $u$ and $v$ real numbers and parametrize a point on the SDP with the coordinates $\left(u,\vsdp\left(u\right)\right)$ for $u \in \left(-\pi, \pi\right)$. The Onsager function becomes
\begin{equation}
\label{00009}
\gamma\left(u+ \im v\right) = \gamma_{1}\left(u,v\right) + \im \gamma_{2}\left(u,v\right),
\end{equation}
where $\gamma_{1}\left(u,v\right)$ and $\gamma_{2}\left(u,v\right) \in \mathbb{R}$ are the real and imaginary parts of $\gamma\left(u+ \im v\right)$, respectively. By equating real and imaginary parts of $\cosh\gamma\left(u+ \iu v\right)$ obtained from
\begin{equation}
\label{00010}
\cosh\gamma\left(u+ \im v\right) = c_{1}^{\star}c_{2}^{\vphantom{\star}} - s_{1}^{\star}s_{2}^{\vphantom{\star}} \cos\left(u+\im v\right),
\end{equation}
we find the pair of equations
\begin{subequations}\label{00011}
\begin{align}
\cosh\gamma_{1}\cos\gamma_{2} & =  c_{1}^{\star} c_{2}^{\vphantom{\star}} - s_{1}^{\star} s_{2}^{\vphantom{\star}} \cos u \cosh v \equiv \mathcal{A}\left(u,v\right), \\
\sinh\gamma_{1}\sin\gamma_{2} & =  s_{1}^{\star} s_{2}^{\vphantom{\star}} \sin u \sinh v \equiv \mathcal{B}\left(u,v\right).
\end{align}
\end{subequations}
Since, along the SDP the imaginary part of the function $L\left(\omega\right) = \gamma\left(\omega\right) - \im \omega \tan\vartheta$ vanishes,
\begin{equation}
\label{00008}
\gamma_{2}\left(u,v\right) = u \tan\vartheta .
\end{equation}
Thus, the SDP is implicitly given via
\begin{equation}
\label{00014}
\frac{ \mathcal{A}^{2}\left(u,\vsdp\left(u\right)\right)}{\cos^{2}\left(u\tan\vartheta\right) } - \frac{ \mathcal{B}^{2}\left(u,\vsdp\left(u\right)\right) }{ \sin^{2}\left(u\tan\vartheta\right)} = 1.
\end{equation}
In Sec.~\ref{sec_4} of SI we solve this equation in a closed form and provide analytical expression for the SDP. Here, we only analyse the simple and instructive case of isotropic lattice ($K_1=K_2\equiv K$) and $\vartheta=\pi/4$.
%
In this situation, the function $\vsdp\left(u\right)$ is given by the solution of
\begin{equation}
\label{28102024_1135}
\cos u \cosh \vsdp\left(u\right) = cc^{*}/2,
\end{equation}
where $c=\cosh 2K$, $c^{*}=\cosh 2K^{*}$,  $-\pi/2 < u < \pi/2$, and the saddle point is located at $u=0$ and $\cosh \vsdp\left(0\right)=\cosh\vs\left(\pi/4\right) = cc^{*}/2$. From the exact solution \eqref{28102024_1135} we observe that the SDP is symmetric around the imaginary axis. Moreover, the SDP is convex and it diverges to infinity for $u\to\pm\pi/4$.

In SI we show that the mirror symmetry, $\vsdp\left(u\right) = \vsdp\left(-u\right)$, is a general property valid for any $\vartheta$. Furthermore, the function $\vsdp\left(u\right)$ diverges for $u\to \pm u_{\rm ex}\left(\vartheta\right)$ with $u_{\rm ex}\left(\vartheta\right) = \pi/(1+\tan \vartheta) < \pi$. This means that SDP does not intersect the side lines $u= \pm \pi$. Additionally, it is possible to verify numerically that $\operatorname{d} \vsdp\left(u\right)/\operatorname{d} u>0$ for $u>0$---this property implies that the saddle point is located in the minimum of $\vsdp\left(u\right)$.

\begin{figure}[t]
\centering
\includegraphics[width=0.79\columnwidth]{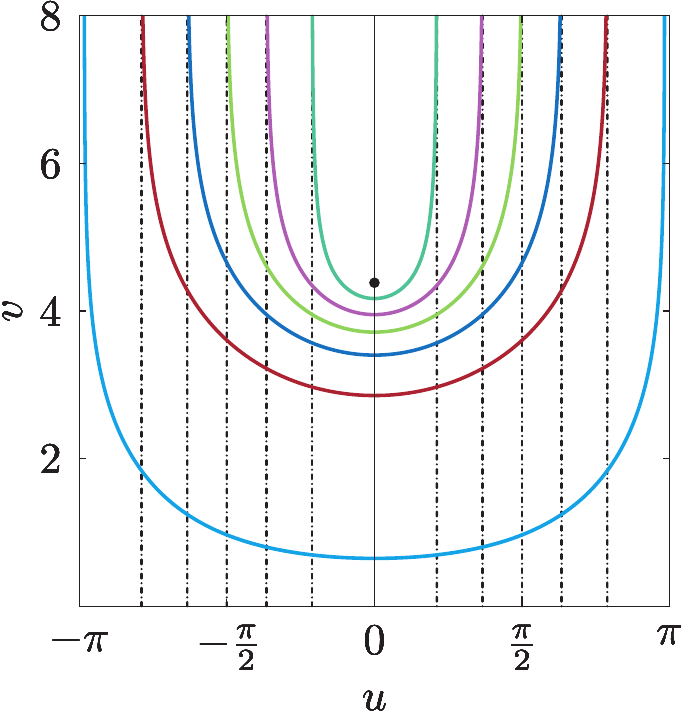}\\[4mm]
\includegraphics[width=0.79\columnwidth]{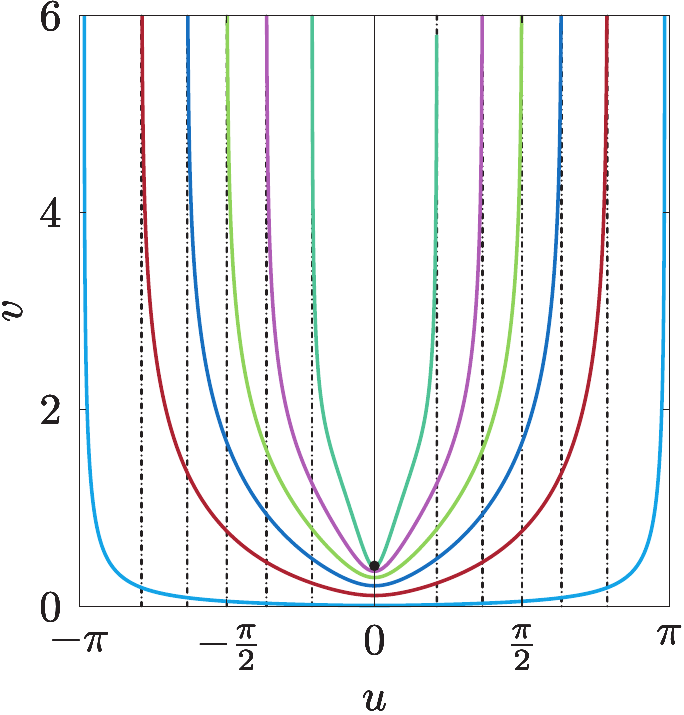}
\caption{SDPs for $T=0.2\, T_{\rm c}$ (upper panel) and for $T=0.8\, T_{\rm c}$ (lower panel) and $K_1=K_2$. The angle $\vartheta$ takes the values $1^{\circ}$, $15^{\circ}$, $30^{\circ}$, $45^{\circ}$, $60^{\circ}$, and $75^{\circ}$, respectively in cyan, red, blue, light green, violet and green. The dot-dashed lines $u=u_{ex}(\vartheta)$ are the vertical asymptotes corresponding to the exit points of the SDP. A black bullet indicates the point $v=\hat{\gamma}(0)$.}
\label{fig_7_sdp}
\end{figure}

In Appendix~\ref{app:B}, we present a typical behaviour of the function $L\left(\omega\right)$ on the complex plane.
In Fig.~\ref{fig_7_sdp} we show typical shape of SDP for a fixed temperature and various angles $\vartheta$. The figure illustrates the location of the saddle point and the divergence of SDP upon approaching the angle-dependent exit point $u_{\mathrm{ex}}\left(\vartheta\right)$.

\subsection{Evaluation of the integral}
\begin{figure}[t]
\centering
\includegraphics[width=0.99\columnwidth]{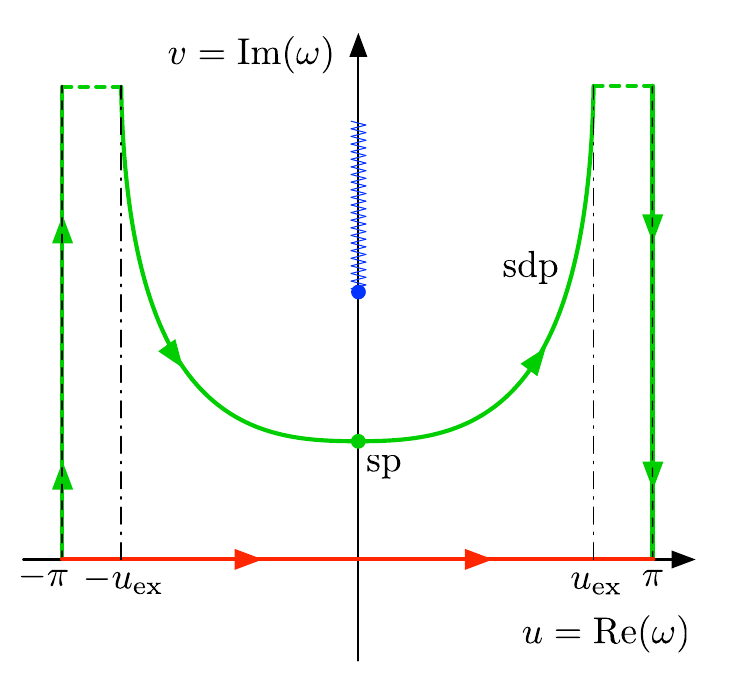}
\caption{The contour used for calculation of integral $I\left(M,N\right)$ (see Eq.~\eqref{00001}). Initially, the integral is calculated from $-\pi$ to $\pi$ along the real axis (red contour). For the calculation, we modify it to the contour denoted with green curve. For $-u_\mathrm{ex}<u<u_\mathrm{ex}$ this contour follows steepest descent path of the function $L\left(\omega\right)$ (denoted with ``sdp'' in the figure). In the limit of $N,M\to\infty$ the value of the integral is dominated by the contribution stemming from the vicinity of the saddle point (denoted with ``sp''). The blue zigzag denotes a branch cut of the function $L\left(\omega\right)$.}
\label{fig_contours}
\end{figure}

The calculation of the integral \eqref{00001} can be performed by using the residue theorem. If there are no singularities of $g(\omega)$ in the region enclosed between the real axis and the SDP, then it is possible to deform the integration contour from the straight line into a line integral along the SDP. In such a case, we get
\begin{equation}
\label{28102024_1200}
I\left(M,N\right) = \int_{\text{sdp}} \frac{\operatorname{d}\omega}{2\pi}  g\left(\omega\right) \operatorname{e}^{-N\gamma\left(\omega\right) + \im M\omega}.
\end{equation}
This result follows by completing the integration over the segment on the real axis shown in Fig.~\ref{fig_contours} with a closed contour $\mathcal{C}$. The contour $\mathcal{C}$ is composed by the segment $\left(-\pi,\pi\right)$, the vertical lines $u = \pm\pi$, the SDP, and two horizontal segments necessary in order to close the loop for $v\to\infty$. The contributions from the vertical side lines as well as those from the horizontal segments for $v\to\infty$ vanish due to the properties \eqref{g_conditions} of the function $g\left(\omega\right)$. 
Fig.~\ref{fig_contours} shows a qualitative sketch of the SDP together with the saddle point.

In order to calculate the integral \eqref{28102024_1200} we note that the exponent in the integrand $-N L\left(\omega\right)$ is real along the SDP and it has a maximum at the saddle point. Therefore, in the limit $N\to\infty$ the integral is dominated by the vicinity of the saddle point. By using Laplace's method~\cite{Bleistein1987}, 
we expand the expression in the exponential to quadratic order in the wave number $\omega=u$. The expansion yields $L(u) = L\left( \im \vs\left(\vartheta\right)\right) + L^{(2)}\left(\im \vs(\vartheta)\right) u^{2}/2! + O(u^{3})$ for $u \to 0$. 
We note that the linear term $L^{(1)}\left(\im \vs\left(\vartheta\right)\right) = \gamma^{(1)}\left(\im \vs\left(\vartheta\right)\right) - \im \tan \vartheta$ vanishes by virtue of the saddle point condition \eqref{09022021_0712}.

Now we introduce the physical interpretation by recalling  the notion of \emph{angle-dependent surface tension} $\tau\left(\vartheta\right)$~\cite{AR74,AU88}:
\begin{equation}
\label{01022021_1750}
\tau\left(\vartheta\right) = \gamma\left( \im \vs\left(\vartheta\right)\right)\cos\vartheta + \vs\left(\vartheta\right) \sin\vartheta,
\end{equation}
where the location of saddle point $\vs\left(\vartheta\right)$ is defined as the solution of
\begin{equation}
\label{01022021_1751}
\gamma^{(1)}( \im \vs(\vartheta)) = \im \tan\vartheta.
\end{equation}
The angle $\vartheta$ is the angle measured from the principal lattice direction $(0,1)$. Within the notations of Fig.~\ref{fig_3_new}, the direction $(0,1)$ is the one along the $K_{1}$ bonds. By using the properties of the saddle point it follows that $\tau(0) = \gamma(0)$ and $\tau(\pi/2) = \hat{\gamma}(0)$. We recall that below the critical temperature $\gamma(0) = 2K_{2}-2K_{1}^{\star}$ and that $\hat{\gamma}(0) = 2K_{1}-2K_{2}^{\star}$ are both positive and they approach zero for $T\to T_\mathrm{c}$. 

By using the definition of angle-dependent surface tension, the $u$-independent constant in the expansion of $L(\omega)$ simplifies to
\begin{equation}
N L\left( \im v_{s}\left(\vartheta\right)\right) = \ell \tau\left(\vartheta\right),
\end{equation}
where $\ell = \sqrt{N^{2}+M^{2}}$ is the Euclidean distance between the pinning points of the domain wall. The meaning of this term is the energetic cost of a straight domain wall of length $\ell$ between the pinning points, \ie, in the direction that forms angle $\vartheta$ with $\left(0,1\right)$ direction of the lattice. Of course, the information about the fluctuations about this straight line is encoded in the curvature of $L(\omega)$ evaluated at the saddle point, which is $L^{(2)}\left(\im \vs\left(\vartheta\right)\right) = \gamma^{(2)}\left(\im \vs\left(\vartheta\right)\right)$. 
Formally, $\tau\left(\vartheta\right)$ is defined only for $\vartheta\in\left[0,\pi/2\right)$. However, its interpretation as an angle-dependent surface tension allows to extend its domain by requiring that $\tau$ is symmetric around $\vartheta=0$ and $\vartheta=\pi/2$, \ie, by enforcing the symmetries of the lattice. The details of this extension, as well as other properties of $\tau\left(\vartheta\right)$ are discussed in Sec.~\ref{SI:tau} of SI.

We can now evaluate the integral by using Laplace's method, for $M,N\to\infty$
\begin{equation}
I\left(M,N\right) \sim g\left(\im \vs\left(\vartheta\right)\right) \operatorname{e}^{-\ell \tau\left(\vartheta\right)} \int_{-\infty}^{\infty} \frac{\operatorname{d}u}{2\pi} \operatorname{e}^{-N\gamma^{(2)}\left(\im \vs\left(\vartheta\right)\right)u^{2}/2},
\end{equation}
where we use the symbol ``$\sim$'' to denote the leading order term in the asymptotic expansion of the expression for $N\to\infty$.
In the above, we further assumed that $g\left(\im \vs\left(\vartheta\right)\right) \neq 0$. It is clear that if this is not the case, the function $g\left(\omega\right)$ needs to be Taylor-expanded. The desired large-$N$ asymptotic result can be obtained by carrying out the Gaussian integration
\begin{equation}
\label{28102024_1307}
I\left(M,N\right) \sim g\left(\im \vs\left(\vartheta\right)\right) \frac{ \operatorname{e}^{-\ell \tau\left(\vartheta\right)} }{ \left[ 2 \pi N \gamma^{(2)}\left(\im \vs\left(\vartheta\right)\right) \right]^{1/2} }.
\end{equation}
By using the definition of \emph{interface stiffness}~\cite{AR74, AU88, Fisher_Fisher_Weeks}
\begin{equation}
\begin{aligned}
\label{zigzag13}
\Sigma\left(\vartheta\right) & = \left[ \gamma^{(2)}\left(\im \vs\left(\vartheta\right)\right) \cos^{3}\vartheta \right]^{-1},
\end{aligned}
\end{equation}
the result (\ref{28102024_1307}) can be rewritten as
\begin{equation}
\label{28102024_1342}
I\left(M,N\right) \sim g\left(\im \vs\left(\vartheta\right)\right) \cos\vartheta \left[ \Sigma\left(\vartheta\right)/2\pi\ell \right]^{1/2} \operatorname{e}^{-\ell \tau\left(\vartheta\right)},
\end{equation}
from which it follows that in the limit $M,N\to\infty$
\begin{equation}
\label{29102024_1005}
- \frac{1}{\ell} \log I\left(M,N\right) = \tau\left(\vartheta\right) + \frac{\ln \ell}{2\ell} + \frac{c\left(\vartheta\right)}{\ell},
\end{equation}
where $c\left(\vartheta\right)$ is some $\ell$-independent constant. The first term in Eq.~\eqref{29102024_1005} is the only one that remains in the limit $\ell \to \infty$, the latter yields the surface tension $\tau\left(\vartheta\right)$ of the inclined domain wall. The logarithmic term accounts for the fluctuations around the straight domain wall. This type of term has already been encountered in the study of floating domain walls in finite-size cylinders~\cite{Abraham_Svrakic_1991}.

For the sake of completeness, in Appendix~\ref{app:B} we provide the polar charts showing the comparison between the angle-dependent surface tension $\tau(\vartheta)$ and interface stiffness, the latter can also be written in the form $\Sigma(\vartheta) = \tau(\vartheta) + \tau^{(2)}(\vartheta)$.

Finally, we note that the result \eqref{29102024_1005} is also valid when the function $g\left(\omega\right)$ has a branch cut along the imaginary axis for $v\geqslant \hat{\gamma}\left(0\right)$. Moreover, if $g\left(\omega\right)$ has a pole in the region between the real axis and SDP, the residue at such a pole must be taken into account. This situation is not happening for the integrals considered in this manuscript, however, it can arise in the study of wetting properties in systems with surface fields~\cite{AM_2002,AM_2005} due to the presence of a wetting pole, something that is known also in integrable field theories \cite{Squarcini_Tinti_2023}. An analogous singular structure occurs in the study of the behaviour of correlation functions in the planar symmetric eight-vertex, or Baxter, model, where the pole is due to a single-particle bound state from the two-particle continua \cite{Abraham_Owczarek}.



\subsection{Incremental free energies: leading asymptotics}
Let us see how the above framework can be effectively used to evaluate certain partition functions in finite geometries. Strictly speaking, the expressions \eqref{03052021_1247} and \eqref{03052021_0909} for $\zeta_{\mathrm{A}}$ and $\zeta_{\mathrm{B}}$, respectively, are not yet written in a form that allows for a straightforward usage of the technique we just illustrated. The reason is that for both the weights, the corresponding function $g\left(\omega\right)$ is actually $N$-dependent. This feature can be better noticed by writing the hyperbolic functions in terms of exponentials. A simple manipulation yields the alternative representation
\begin{equation}
\label{28102024_1355}
\zeta_{\mathrm{A}}\left(N,s\right) = \int_{-\pi}^{\pi}\frac{\operatorname{d}\omega}{2\pi} g_{\mathrm{A},N} \left(\omega\right) \operatorname{e}^{ - N \gamma\left(\omega\right) + \im s\omega },
\end{equation}
with
\begin{equation}
\label{28102024_1542}
g_{\mathrm{A},N}\left(\omega\right) = \frac{ 2 }{ 1+\cos\delta^{*}\left(\omega\right) } \frac{ 1 }{ 1 + \tan^{2}\left(\delta^{*}\left(\omega\right)/2\right) \operatorname{e}^{-2N\gamma\left(\omega\right)} }.
\end{equation}
It is thus clear that besides the exponential part, the parameter $N$ appears also at the denominator.

However, since we are interested in the leading asymptotic behaviour of $\zeta_{\mathrm{A}}\left(N,s\right)$ we can replace $g_{\mathrm{A},N}\left(\omega\right)$ with its limit for large $N$ by, \textit{de facto}, setting to zero the exponential term in the denominator. This means that the second factor in the function $g_{\mathrm{A},N}\left(\omega\right)$ in Eq.~\eqref{28102024_1542} can be approximated by unity up to corrections of the order $\exp\left(-2N\gamma\left(\omega\right)\right)$. A systematic treatment of these exponentially small corrections will be discussed in Sec.~\ref{sec_rattling}. 

By using Eq.~\eqref{28102024_1342} the expression for $\zeta_{\mathrm{A}}$ at leading order for large $N$ is
\begin{equation}
\label{zeta_A_approx}
\zeta_{\mathrm{A}}\left(N,s\right) \approx g_{\mathrm{A}}\left(\im \vs\left(\vartheta\right)\right) \cos\vartheta \left[ \Sigma\left(\vartheta\right)/2\pi\ell \right]^{1/2} \operatorname{e}^{-\ell \tau\left(\vartheta\right)},
\end{equation}
with $g_{\mathrm{A}}\left(\omega\right) \equiv \lim_{N \to \infty} g_{\mathrm{A},N}\left(\omega\right) = 1/\cos^{2}\left(\delta^{*}\left(\omega\right)/2\right)$ and $\ell=\sqrt{N^{2}+s^{2}}$. It is instructive to consider the case $s=0$ corresponding to an interface running along the vertical axis. Recalling that $\vs\left(0\right)=0$, we find
\begin{equation}
\label{zeta_A_approx2}
\zeta_{\mathrm{A}}\left(N,0\right) \approx \left[ \Sigma\left(0\right)/2\pi \ell \right]^{1/2} \operatorname{e}^{-\ell \gamma\left(0\right)},
\end{equation}
where $\tau(0) = \gamma(0)$ is the surface tension along the principal lattice direction $\left(0,1\right)$.

\begin{figure}[t]
\centering
\includegraphics[width=0.99\columnwidth]{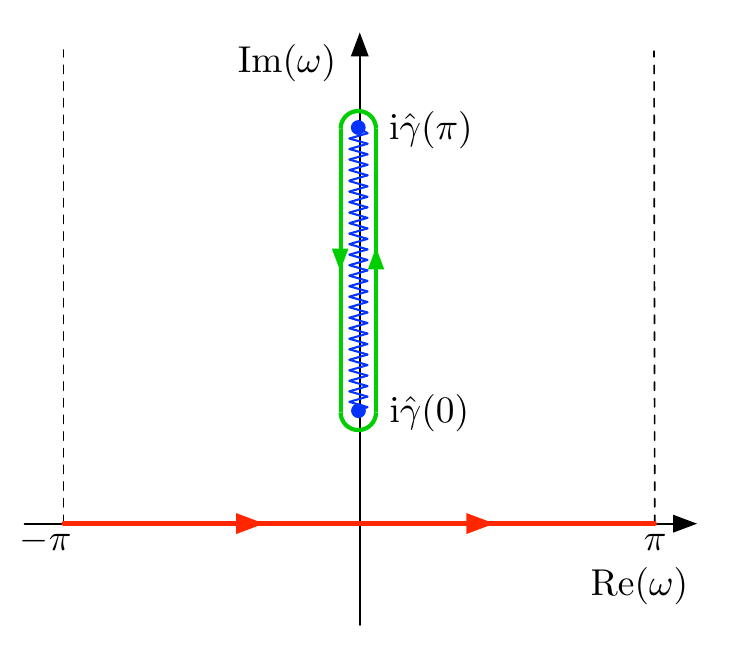}
\caption{Modification of the integration from $-\pi$ to $\pi$ to the contour surrounding the branch cut.}
\label{fig_branch_cut}
\end{figure}

The incremental free energy for the droplet brings new features. The counterpart of Eq.~\eqref{28102024_1355} for $\zeta_{\mathrm{B}}$ is
\begin{equation}
\label{28102024_1354_1}
\zeta_{\mathrm{B}}\left(N,s\right) = \int_{-\pi}^{\pi}\frac{\operatorname{d}\omega}{2\pi\im} g_{\mathrm{B}, N}\left(\omega\right) \operatorname{e}^{\im s\omega },
\end{equation}
with
\begin{equation}
\label{28102024_1642}
g_{\mathrm{B}, N}\left(\omega\right) = \frac{\sin\delta^{*}\left(\omega\right)}{ 1 + \cos\delta^{*}\left(\omega\right) } \frac{ 1 - \operatorname{e}^{-2N\gamma\left(\omega\right)}  }{ 1 + \tan^{2}\left(\delta^{*}\left(\omega\right)/2\right) \operatorname{e}^{-2N\gamma\left(\omega\right)}}.
\end{equation}
The limit $N \to \infty$ for the configuration $\mathrm{B}$ removes the influence of the upper wall on the fluctuations of the droplet, and the $\zeta_{\mathrm{B}}$ reduces to
\begin{equation}
\label{28102024_1354}
\zeta_{\mathrm{B}}\left(\infty,s\right) = \int_{-\pi}^{\pi}\frac{\operatorname{d}\omega}{2\pi\im} \tan\left[\delta^{*}\left(\omega\right)/2\right] \operatorname{e}^{ \im s\omega }.
\end{equation}
This integral can be computed by deforming the straight integration path to the one around the branch cut, as shown in Fig.~\ref{fig_branch_cut}.

This procedure is possible because the function $H\left(\omega\right) = \tan\left[\delta^{*}\left(\omega\right)/2\right]$ does not possess isolated poles. The point $\omega = \im \hat\gamma\left(0\right)$ corresponds to a branch point with branch cut extending along the imaginary axis $\operatorname{Im}\omega > \hat\gamma\left(0\right)$, as shown in Fig.~\ref{fig_branch_cut}. The calculation can be performed along the green path passing aside the branch cut, as sketched in Fig.~\ref{fig_branch_cut}. What is necessary to perform this calculation is the jump of the imaginary part of the function $H\left(\omega\right)$ across the branch cut, $\Delta H\left(\omega\right) = \lim_{\epsilon \to 0}H\left(\im v + \epsilon\right)-H\left(\im v - \epsilon\right)$. Alternatively, it is also possible to employ the trick of the \emph{isospectrality substitution}, \ie, change of variables $\omega = \im \hat{\gamma}\left(u\right)$. The calculation yields the large-$s$ asymptotic behaviour
\begin{equation}
\label{10062025_2217}
\zeta_{\mathrm{B}}\left(\infty,s\right) \propto s^{-3/2} \operatorname{e}^{-s\hat{\gamma}\left(0\right)},
\end{equation}
which gives the partition function for the droplet up to $s$-independent proportionality constants. We note the appearance of $\hat{\gamma}\left(0\right)$, the surface tension along the lattice direction $(1,0)$, \ie, $\tau(\pi/2) = \hat{\gamma}(0)$, as consistency requires. The factor $s^{-3/2}$ is the characteristic signature of the entropic repulsion experienced by the interface that is forced to fluctuate above the wall. In contrast to this case, it should be observed that the interface pinned, as in case $A$, is characterized by an overall different power law, $\zeta_{\mathrm{A}}\left(N,0\right) \propto N^{-1/2} \exp\left[-N\gamma\left(0\right)\right]$ \footnote{See \cite{DS_wetting} and \cite{DV, Squarcini_Multipoint}, respectively, for a derivation of \eqref{10062025_2217} and the latter result in field theory.}.

\section{Point tension}
\label{sec_point_tension}
We can now provide an interpretation of the quantity \eqref{zeta_A_approx}. To this end, we rewrite the function $g_{\mathrm{A}}$  evaluated at the saddle point as
\begin{equation}
g_{\mathrm{A}}\left( \im \vs\left(\vartheta\right)\right) \equiv \exp\left[ - 2\taup\left(\vartheta\right) \right],
\end{equation}
which defines the point tension $\taup$ that we discuss in detail below. For the remaining factor in Eq.~\eqref{zeta_A_approx} we introduce a new symbol 
\begin{equation}
\zeta_{\mathrm{f}}\left(\ell, \vartheta\right) \equiv \left( \frac{ \Sigma\left(\vartheta\right) }{ 2\pi \ell } \right)^{1/2} \cos\vartheta.
\end{equation}
Hence,  $\zeta_{\mathrm{B}}$ is decomposed as the product of three factors:
\begin{equation}
\label{pt:decompositionA}
\zeta_{\mathrm{A}}\left(N,s\right) = \zeta_{\mathrm{f}}\left(\ell, \vartheta\right)  \operatorname{e}^{ - 2\taup\left(\vartheta\right) - \ell \tau\left(\vartheta\right)}.
\end{equation}
To the above we can associate an incremental free energy $\mathcal{F}$ such that $\zeta_{\mathrm{A}} = \exp\left(-\mathcal{F}\right)$. 
With this identification, it is clear that $\ell \tau\left(\vartheta\right)$ is the energetic cost of a straight interface with inclination angle $\vartheta$ and length $\ell$. The point tension $\taup\left(\vartheta\right)$ is the energy stemming from each pinning point where the interface touches the wall. Lastly, the factor $\zeta_{\rm f}$  takes into account the thermal fluctuations around the straight interface connecting the pinning points. This term can be interpreted as an entropic contribution.

We conclude this section with some further comments on the point tension,
\begin{equation}
\label{25112024_1118_1}
\exp\left[ -2\taup\left(\vartheta\right) \right] = \frac{2}{1+\cos\left[\delta^{*}\left(\im \vs\left(\vartheta\right)\right)\right]}.
\end{equation}
By using the properties of the Onsager angle $\delta^{*}\left(\omega\right)$ at imaginary wave numbers $\omega$, we can prove that $\cos\left[\delta^{*}\left(\im v\right)\right] > 1$ for $0<v<\hat{\gamma}\left(0\right)$, which are the possible values of $\vs\left(\vartheta\right)$ for $\vartheta\in\left(0,\pi/2\right)$. It thus follows that the right-hand side of Eq.~\eqref{25112024_1118_1} is positive and bounded from below by $1$. By using the trigonometric identity $1+\cos\left(x\right)=2\cos^{2}\left(x/2\right)$, we find
\begin{equation}
\label{25112024_1133}
\exp\left[ -\taup\left(\vartheta\right) \right] = \frac{1}{\cos\left[\delta^{*}\left(\im \vs\left(\vartheta\right)\right)/2\right]},
\end{equation}
or equivalently
\begin{equation}
\label{25112024_1134}
\taup\left(\vartheta\right) = \ln\cos\left[\delta^{*}\left(\im \vs\left(\vartheta\right)\right)/2\right].
\end{equation}
The point tension $\taup$ is positive because the argument of the logarithm in Eq.~\eqref{25112024_1134} is larger than one. It is interesting to observe that $\cos\left[\delta^{*}\left(\im v\right)\right]$ diverges to $+\infty$ for $v \to \hat{\gamma}\left(0\right)^{-}$. This property implies that $\taup$ diverges as $\vartheta \to \pi/2$. This limit is never reached provided $\vartheta < \pi/2$.

\section{Rattling effect for incremental free energies}
\label{sec_rattling}
In Section~\ref{sec_sdp_v1} we have obtained asymptotic expansions of the quantities $\zeta_\mathrm{A}$ and $\zeta_\mathrm{B}$ for large distance $\ell$ between pinning points. These results were obtained by ignoring the exponentially small term $\exp\left(-2N\gamma\left(\omega\right)\right)$ in the integrand (see Eqs.~\eqref{28102024_1355} and \eqref{28102024_1354_1}). In this section we use the general approach developed in Section~\ref{sec_sdp_v1} for a systematic treatment of such integrals without neglecting any terms. We also provide a physical interpretation of the results in terms of coarse-grained quantities such as surface and point tensions.

\subsection{Inclined interface}
\label{sec_type_A}
\begin{figure}[t]
\centering
\includegraphics[width=0.8\columnwidth]{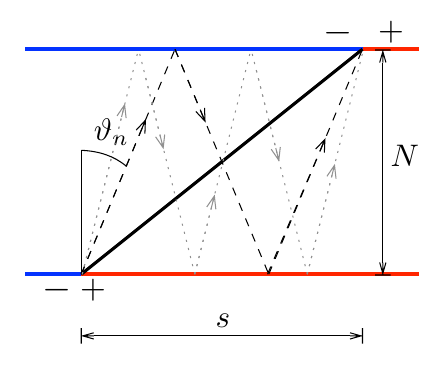}
\caption{Geometric construction explaining the angle $\vartheta_{n}$.}
\label{fig_zeta_A}
\end{figure}

Let us first consider the expression for $\zeta_{\mathrm{A}}$ given by Eq.~\eqref{28102024_1355}. For large $N$ it is legitimate to perform an expansion of the function $g_{\mathrm{A},N}\left(\omega\right)$ in terms of a geometric series in the exponentially small term $\exp\left(-2N\gamma\left(\omega\right)\right)$. With this in mind, Eq.~\eqref{28102024_1542} becomes
\begin{equation}
\label{28102024_1543}
g_{\mathrm{A},N}\left(\omega\right) = \sum_{n=0}^{\infty} g_{\mathrm{A}}^{(n)}\left(\omega\right) \operatorname{e}^{-2nN\gamma\left(\omega\right)},
\end{equation}
with
\begin{equation}
\label{28102024_1603}
g_{\mathrm{A}}^{(n)}\left(\omega\right) = \frac{ 2 }{ 1+\cos\delta^{*}\left(\omega\right) } \left[-\tan^{2}\left(\delta^{*}\left(\omega\right)/2\right)\right]^{n}.
\end{equation}
Each term in the above series is an integral that is of the form given in Eq.~\eqref{00001}. Therefore, the framework of Section~\ref{sec_sdp_v1} can be applied term by term to the series.
Absolute convergence of series in Eq.~\eqref{28102024_1543} yields the following series
for an inclined interface
\begin{equation}
\label{zigzag4}
\zeta_{\mathrm{A}}(N,s) = \sum_{n=0}^{\infty} \zeta_{\mathrm{A}}^{(n)}\left(N,s\right),
\end{equation}
where
\begin{equation}
\label{zigzag5}
\zeta_{\mathrm{A}}^{(n)}\left(N,s\right) = \int_{-\pi}^{\pi} \frac{\operatorname{d}\omega}{2\pi} g_{\mathrm{A}}^{(n)}\left(\omega\right) \operatorname{e}^{-\left(2n+1\right)N\gamma\left(\omega\right) + \im s\omega } 
\end{equation}
are integrals of the general type given by Eq.~\eqref{00001}. Since we are interested in the regime of large $N$ and $s$, we can apply the steepest descent method outlined in Sec.~\ref{sec_sdp_v1a} for each term in the series \eqref{zigzag4}. It is clear that now we have a \emph{sequence of saddle points} parametrized by the odd integer $2n+1$, with $n=0,1,2,\dots$. The $n$-th saddle point $\omega_n^{\mathrm{A}}$ satisfies
\begin{equation}
\label{zigzag6}
\gamma^{(1)}\left(\omega_n^{\mathrm{A}}\right) = \frac{\im s}{(2n+1)N}.
\end{equation}
Comparing the above relation with Eq.~\eqref{09022021_0712} we identify $\omega_n^{\mathrm{A}}=\vs\left(\vartheta_n^{\mathrm{A}}\right)$ with
\begin{equation}
\label{03052021_0818}
\tan\vartheta_{n}^{\mathrm{A}} = \frac{ s}{\left(2n+1\right)N}  = \frac{\tan\vartheta}{2n+1},
\end{equation}
where $\tan\vartheta=s/N$.
The angle $\vartheta_{n}^{\mathrm{A}}$ has a simple geometrical interpretation as shown in Fig.~\ref{fig_zeta_A}.

\begin{figure}[t]
\centering
\includegraphics[width=0.6\columnwidth]{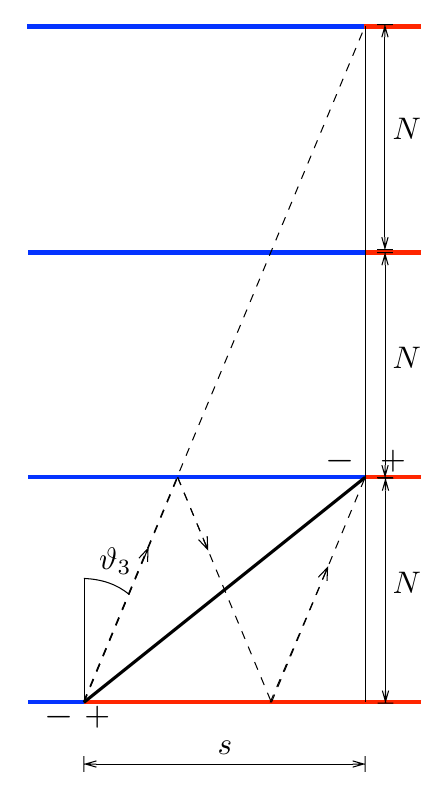}
\caption{Unfolding construction for $2n+1=3$.}
\label{fig_zeta_A_unfold}
\end{figure}

In the geometrical construction illustrated in Fig.~\ref{fig_zeta_A} the integer $2n+1$ is the number of broken straight segments connecting the pinning points. The resulting path is analogous to the trajectory of a billiard ball undergoing $2n$ perfectly elastic reflections. As shown in Fig.~\ref{fig_zeta_A_unfold}, the broken trajectory can be unfolded $2n+1$ times to a perfectly straight one of length $L_{n}$ given by $L_{n}^{2} = s^{2}+\left(2n+1\right)^{2}N^{2}$, with $L_{0} =\ell= \sqrt{s^{2}+N^{2}}$ the Euclidean distance between the pinning points.

Upon increasing $n$ the right-hand side of Eq.~\eqref{03052021_0818} decreases, and thus, also $\vartheta_{n}^{\mathrm{A}}$ decreases. In Sec.~\ref{sec_3} of SI we show that $\vs\left(\vartheta\right)$ is strictly increasing and, hence, we have the following ordering of saddle points:
\begin{equation}
\label{zigzag7}
\hat{\gamma}\left(0\right) > \vs\left(\vartheta\right) > \vs\left(\vartheta_{1}^{\mathrm{A}}\right) > \vs\left(\vartheta_{2}^{\mathrm{A}}\right) > \ldots>0.
\end{equation}

The evaluation of the integral in Eq.~\eqref{zigzag5} proceeds along the same steps as for the leading term corresponding to $n=0$. The definition \eqref{01022021_1750} of angle-dependent surface tension $\tau\left(\vartheta\right)$ allows us to write the exponential part in Eq.~\eqref{zigzag5} as
\begin{equation}
\label{zigzag11}
\left(2n+1\right)N\gamma\left(\im \vs\left(\vartheta_{n}^{\mathrm{A}}\right)\right)+s \vs\left(\vartheta_{n}^{\mathrm{A}}\right) = L_{n} \tau\left(\vartheta_{n}^{\mathrm{A}}\right),
\end{equation}
which reveals that it corresponds to the energetic cost of a single domain wall that bounces back and forth between the upper and lower boundaries, as shown in Fig.~\ref{fig_zeta_A}. By applying the result \eqref{28102024_1342}, we obtain
\begin{equation}
\label{25112024_1054}
\zeta_{\mathrm{A}}^{(n)}\left(N,s\right) = \zeta_{\mathrm{f}}^{(n)} g_{\mathrm{A}}^{(n)}\left(\im \vs\left(\vartheta_{n}^{\mathrm{A}}\right)\right) \operatorname{e}^{-L_{n}\tau\left(\vartheta_{n}^{\mathrm{A}}\right)},
\end{equation}
with
\begin{equation}
\zeta_{\mathrm{f}}^{(n)} = \left( \frac{\Sigma\left(\vartheta_{n}^{\mathrm{A}}\right) }{ 2\pi L_{n} } \right)^{1/2} \cos\vartheta_{n}^{\mathrm{A}}.
\end{equation}
Similar to the case of $n=0$, which was discussed in Section~\ref{sec_point_tension} (see Eq.~\eqref{pt:decompositionA}), the expression \eqref{25112024_1054} is decomposed into a product of three factors. The last one, $\exp\left[ - L_{n} \tau\left(\vartheta_{n}^{\mathrm{A}}\right)\right]$, can be explained by assuming that the fugacity for an inclined domain wall is of the form $\exp\left(-\mathcal{F}\right)$, where $\mathcal{F}=L_{n} \tau\left(\vartheta_{n}^{\mathrm{A}}\right)$ is the free energy associated to a line---of length $L_{n}$ and inclination $\vartheta_{n}^{\mathrm{A}}$---separating coexisting phases. The factor $g_{\mathrm{A}}^{(n)}\left(\im \vs\left(\vartheta_{n}^{\mathrm{A}}\right)\right)$ can be associated to a refinement of the above picture in which the incremental free energy $\mathcal{F}$ receives pointwise contributions stemming from each point in which the interface touches the wall.
Lastly, the prefactor $\zeta_{\mathrm{f}}^{(n)}$, originating from the integral \eqref{28102024_1307}, takes into account the thermal fluctuations around configuration with a simple broken line bouncing between pinning points (see Fig.~\ref{fig_zeta_A}).

Concerning pointwise contributions, we observe that any zigzag interface composed of $2n+1$ segments is scattered off the walls $2n$ times. On top of this, there are always two pinning points. It is, thus, straightforward to postulate a decomposition
\begin{equation}\label{rattling:taufcp_def}
    g_{\mathrm{A}}^{(n)}\left(\im \vs\left(\vartheta_{n}^{\mathrm{A}}\right)\right)=\exp\left[-2\taup\left(\vartheta_n^{\mathrm{A}}\right)-2n \taufcp\left(\vartheta_n^{\mathrm{A}}\right)\right],
\end{equation}
where $\taup\left(\vartheta_n^{\mathrm{A}}\right)$ is a point tension of pinned interface, discussed in Sec.~\ref{sec_point_tension}. The new parameter $\taufcp\left(\vartheta_n^{\mathrm{A}}\right)$, introduced in the above equation, is the point tension originating from scattering of interface on each bending of zigzag line. Using Eqs.~\eqref{rattling:taufcp_def}, \eqref{28102024_1603}, and \eqref{25112024_1118_1}, after short derivation, we get
\begin{equation}
\label{25112024_1119}
\exp\left[ -2\taufcp\left(\vartheta_{n}^{\mathrm{A}}\right) \right] =  - \tan^{2}\left[ \delta^{*}\left(\im \vs\left(\vartheta_{n}^{\mathrm{A}}\right)\right)/2\right].
\end{equation}
We note that the $\taufcp$ does not depend explicitly on number of scatterings $2n$ which confirms the proposed physical interpretation.
With this definition, the expression \eqref{25112024_1054} takes the suggestive form
\begin{equation}
\label{25112024_1120}
\zeta_{\mathrm{A}}^{(n)}\left(N,s\right) = \zeta_{\mathrm{f}}^{(n)} \operatorname{e}^{ -2\taup\left(\vartheta_{n}^{\mathrm{A}}\right) -2n\taufcp\left(\vartheta_{n}^{\mathrm{A}}\right) -L_{n}\tau\left(\vartheta_{n}^{\mathrm{A}}\right)}.
\end{equation}

The definition \eqref{25112024_1119} suggests that $\taufcp$ might be a complex number. We address this problem in Appendix~\ref{app:C}, where we show that $\taufcp$ is real and positive. This way the $n$-th quantity $\zeta_{\mathrm{A}}^{(n)}\left(N,s\right)$ is always positive. 

We note that the point tension $\taufcp$ diverges for $\vartheta \to 0$. This limit is analogous to the divergence of the point tension $\taup$ for $\vartheta \to \pi/2$; however, these regimes are never attained for finite shear $s$. Lastly, we report the following relation between the point tensions $\tau_{p}$ and $\taufcp$:
\begin{equation}
\operatorname{e}^{-2\taup\left(\vartheta\right)} + \operatorname{e}^{-2\taufcp\left(\vartheta\right)} = 1.
\end{equation}
From the latter it follows that $\taufcp$ diverges when $\taup$ vanishes, and \textit{vice versa}.
\subsection{Droplet}
\label{sec_type_B}
\begin{figure}[t]
\centering
\includegraphics[width=0.8\columnwidth]{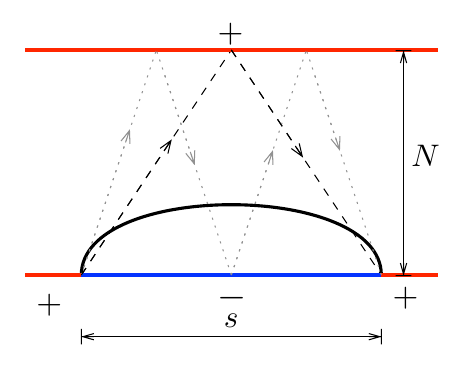}
\caption{Geometrical construction for the large-$N$ asymptotics of the droplet partition function $\zeta_{\mathrm{B}}$.}
\label{fig_zeta_B}
\end{figure}

The analysis outlined for the study of the inclined interface can be repeated, \textit{mutatis mutandis}, for the case of the droplet. The systematic treatment of the term $\exp\left[-2N\gamma\left(\omega\right)\right]$ entering the quantity $\zeta_{\mathrm{B}}\left(N,s\right)$ (and the partition function) follows from expanding Eq.~\eqref{28102024_1642} in a power series of $\exp\left[-2N\gamma\left(\omega\right)\right]$. After some trigonometric transformations we get
\begin{equation}
\label{28102024_1648}
g_{\mathrm{B}, N}\left(\omega\right) = \frac{\sin\delta^{*}\left(\omega\right)}{ 1 + \cos\delta^{*}\left(\omega\right)} - \sum_{n=1}^{\infty} g_{\mathrm{B}}^{(n)}\left(\omega\right) \operatorname{e}^{-2nN\gamma\left(\omega\right)},
\end{equation}
with
\begin{equation}
\label{rattling:gB}
g_{\mathrm{B}}^{(n)}\left(\omega\right) = \frac{2\im}{\sin\delta^{*}\left(\omega\right)} \left[ - \tan^{2}\left(\delta^{*}\left(\omega\right)/2\right)\right]^{n}.
\end{equation}
We note that the first term in the decomposition \eqref{28102024_1648} is the only one that persists in the limit $N\to\infty$, leading thus to the non-vanishing 
$\zeta_{\mathrm{B}}\left(\infty,s\right)$ given by Eq.~\eqref{28102024_1354}. In contrast with $\mathrm{A}$ type interface (Eq.~\eqref{zigzag5}), in the full formula for the $\zeta_{\mathrm{B}}$
\begin{subequations}
\begin{align}
\zeta_{\mathrm{B}}\left(N,s\right)&=\zeta_{\mathrm{B}}\left(\infty,s\right)-\sum_{n=1}^\infty \zeta_{\mathrm{B}}^{(n)}\left(N,s\right),\\
\zeta_{\mathrm{B}}^{(n)}\left(N,s\right) &= \int_{-\pi}^{\pi} \frac{\operatorname{d}\omega}{2\pi} g_{\mathrm{B}}^{(n)}\left(\omega\right) \operatorname{e}^{-2n\,N\gamma\left(\omega\right) + \im s\omega },
\end{align}
\end{subequations}
in the exponent $N$ is multiplied by the even integer number $2n$. This is in agreement with the number of segments of the scattering interface going between two pining points, as illustrated in Fig.~\ref{fig_zeta_B}.

Each term appearing in the geometric series can be handled as already done in Sec.~\ref{sec_type_A} for the inclined domain wall. In this case, the saddle-point equation \eqref{zigzag6} takes the form
\begin{equation}
\label{03052021_0811}
\gamma^{(1)}\left(\im \vs\left(\vartheta_{n}^{\mathrm{B}}\right)\right) = \frac{{\im} s}{2nN}, \qquad n \geqslant 1,
\end{equation}
and therefore
\begin{equation}
    \tan\vartheta_n^{\mathrm{B}}=\frac{\tan\vartheta}{2n}.
\end{equation}

We can now evaluate $g_{\mathrm{B}}^{(n)}$ in the saddle points. After some transformations, from Eq.~\eqref{rattling:gB} we get
\begin{multline}
    g_{\mathrm{B}}^{(n)}\left(\im \vs\left(\vartheta_n^{\mathrm{B}}\right)\right) =\\
    \frac{1}{\cos^2\left[\delta^{*}\left(\im \vs\left(\vartheta_n^{\mathrm{B}}\right)\right)/2\right]}\left[-\im \tan\left(\frac{\delta^{*}\left(\im \vs\left(\vartheta_n^{\mathrm{B}}\right)\right)}{2}\right)\right]^{2n-1}\\
    =\exp\left[-2\taup\left(\vartheta_n^{\mathrm{B}}\right)-\left(2n-1\right)\taufcp\left(\vartheta_n^{\mathrm{B}}\right)\right],
\end{multline}
where we have used relations \eqref{25112024_1133} and \eqref{25112024_1119}. This shows that the incremental free energy for the case $\mathrm{B}$ can be written using the same point tensions $\taup$ and $\taufcp$. As a result, the interpretation of the various factors entering the zigzag contributions follows along the same lines outlined in Sec.~\ref{sec_type_A}.

Finally, we note that the series expansions studied in this section does not cover all $N$-dependent terms of the incremental free energies. In fact, in the Laplace's method, the exponent in the integral \eqref{00001} is expanded around a saddle point only up to quadratic terms. Neglected, higher order terms produce corrections which typically dominate over the lower order terms in the series.

\section{Conclusions}
\label{sec_conclusions}

In this paper, we describe in detail the exact calculation of the incremental free energies associated with domain walls induced by Dobrushin boundary conditions in a ferromagnetic planar Ising lattice.

We have given closed-form expressions for the incremental free energies and presented their coarse-grained interpretation in terms of the ``rattling phenomenon'' in which the domain wall undergoes multiple reflections from the walls, an analogous effect has been pointed out in the context of the Lorentz gas or Ehrenfest wind-tree model (see \cite{DBA_Gates} and references therein). This analysis shows that the configurations presented in Fig. 2(a) and (c) are topologically distinct, since they involve an even and odd number of reflections, respectively.

The asymptotic forms of the incremental free energies obtained using the steepest descent method are given in terms of the angle-dependent surface tension $\tau(\vartheta)$, the point tension $\taup\left(\vartheta\right)$, the point tension for the fluctuating points of contact $\taufcp\left(\vartheta\right)$, and an entropic term. We have studied the main properties of $\tau(\vartheta)$, such as its maximum, minimum, and inflection points for the general case of lattices with anisotropic nearest neighbour interactions. Among the various results, we found a relationship between $\taup$ and $\taufcp$, showing that when one of the two diverges the other one must vanish. 

This paper also provides a comprehensive description of the steepest descent path method previously used in the analysis of the asymptotic correlation function for the planar Ising model. We then specialize this method for the correlation functions of boundary operators involved in the calculation of incremental free energies. In particular, we illustrate how the generalized Wick theorem in Dobrushin's program allows us to obtain the incremental free energies in the form of a single integral whose saddle point and asymptotic behaviour are handled exactly, as shown in the illustration. This approach offers a valuable perspective.


\section*{Acknowledgments}

This research was partially funded by the Polish National Science Center (Opus Grant No.\ 2022/45/B/ST3/00936).

\appendix
\numberwithin{equation}{section}
\renewcommand\thesubsection{\arabic{subsection}}

\section{Generalised Wick theorem}
\label{app_g_w_t}

\subsection{Type-B interfaces}
In this section, we will apply the generalised Wick theorem in order to compute incremental free energies. We refer the interested reader to \cite{Abraham_2012} for a detailed account on this technique. We begin our analysis by considering the incremental free energy for the droplet, namely
\begin{equation}
\label{11062025_1512}
Z_{+-}/Z_{++} = \langle + \vert \mathsf{V}^{N} \Gamma_{1} \Gamma_{2s_{1}+1} \vert + \rangle / \langle + \vert \mathsf{V}^{N} \vert + \rangle \, .
\end{equation}
We recall the definition of Fermi fields $F(\beta)$ and $F^{\dag}(\beta)$
\begin{equation}
\label{13062025_0922}
F^{\dag}(\beta) = M^{-1/2} \sum_{m=1}^{M} \operatorname{e}^{i k m} f_{m}^{\dag}
\end{equation}
with anti-periodic momenta $\beta$ defined by $\exp(\im \beta M)=-1$, i.e., $\beta=(2j-1)\pi/M$ with $j=1,\dots,M$. These fields allow us to express the block rotation operator $R(1,1+s_{1}) = \Gamma_{1}\Gamma_{2s_{1}+1}$ in terms of the lattice Fermi fields $f_{m}^{}, f_{m}^{\dag}$. To this end, we invert (\ref{13062025_0922}) and express Fermi fields in terms of symmetry-adapted ones; hence, 
\begin{equation}
f_{m}^{\dag} = M^{-1/2} \sum_{k} \operatorname{e}^{-i k m} F^{\dag}(k),
\end{equation}
and therefore the block rotation operator is
\begin{multline} 
\label{30052020_02}
R(1,1+s_{1})  =  \mel{-} M^{-1} \sum_{ \beta_{1}, \beta_{2} } \left( F(\beta_{1}) + F^{\dag}(-\beta_{1}) \right) \\
 \times  \left( F(\beta_{2}) + F^{\dag}(-\beta_{2}) \right) \operatorname{e}^{\mel{-} \im \beta_{1}\mel{-}\im (1+s_{1})\beta_{2}}.
\end{multline}
The next conceptual step brings in ideas of Lieb-Mattis-Schultz and the notion of Anderson-Nambu pairing (see, e.g., \cite{Abraham_2012}). The Anderson-Nambu (AN) spinor $\mathbb{G}_{0}(\beta)$ defined by
\begin{equation}
\label{30052020_08}
\mathbb{G}_{0}(\beta) =
\left(\begin{array}{c}
G_{0}(\beta) \\
G_{0}^{\dag}(-\beta)
\end{array}\right)
\end{equation}
is obtained through the Bogoliubov--Valatin transformation from the Fermi fields, which relates the two AN spinors via $\mathbb{G}_{0}(\beta) = \mathsf{U}(\theta_{0}(\beta)) \mathbb{F}(\beta)$, where  $\mathsf{U}(\theta)$ is the $2\times2$ matrix explicitly given by
\begin{equation}
\label{03}
\left(\begin{array}{c}G_{0}(\beta) \\G_{0}^{\dag}(-\beta)\end{array}\right) =
\left(\begin{array}{cc}\cos\theta_{0} & -\im \sin\theta_{0} \\ -\im \sin\theta_{0} & \cos\theta_{0}\end{array}\right) \left(\begin{array}{c}F_{0}(\beta) \\F_{0}^{\dag}(-\beta)\end{array}\right),
\end{equation}
see Eq.~(9) of Ref.~\cite{AMW_2005}. The unitary matrix can be written as
\begin{equation}
\label{04}
\textsf{U}(\theta_{0}) = \cos\theta_{0} \mel{-} \im \sin \theta_{0} \tau^{x},
\end{equation}
where $\tau^{x}$ is a Pauli matrix in the following representation
\begin{equation}
\label{02}
\tau^{x} = \left(\begin{array}{cc}0 & 1 \\1 & 0\end{array}\right), \quad \tau^{y} = \left(\begin{array}{cc}0 & - \im \\ \im & 0\end{array}\right), \quad \tau^{z} = \left(\begin{array}{cc}1 & 0 \\0 & -1\end{array}\right), 
\end{equation}
and the spinor $\mathbb{F}(\beta)$ is defined in analogy with (\ref{30052020_08}). The operators in (\ref{30052020_02}) become
\begin{equation}
\label{30052020_03}
F(\beta) + F^{\dag}(-\beta) = \operatorname{e}^{ \im \theta_{0}(\beta)} \left( G_{0}^{\vphantom{}}(\beta) + G_{0}^{\dag}(-\beta) \right),
\end{equation}
where $\theta_{0}(\beta)$ is the parameter implementing the transformation. The partition function ratio becomes
\begin{equation}
\label{30052020_04}
\frac{Z_{+-}(M,N \vert s)}{Z_{++}(M,N)} = -M^{-1} \sum_{ (\beta)_{2} } \mathcal{A}( (\beta)_{2} ) \mathbb{M}((\beta)_{2}) \, ;
\end{equation}
where $(\beta)_{2} = (\beta_{1},\beta_{2})$ denotes a pair of momenta, $\mathcal{A}( (\beta)_{2} )$ is given by
\begin{equation}
\mathcal{A}( (\beta)_{2} ) = \operatorname{e}^{-\im \beta_{1}} \operatorname{e}^{-\im(1+s)\beta_{2}} \prod_{j=1}^{2} \operatorname{e}^{\im \theta_{0}(\beta_{j})},
\end{equation}
and $\mathbb{M}((\beta)_{2})$ is the matrix element
\begin{equation}
\begin{aligned}
\label{21112024_2135}
\mathbb{M}((\beta)_{2}) & = \langle \Phi_{+}^{0} \vert \bigl[ \mathsf{V}(+) \bigr]^{N} \bigl[ G_{0}(\beta_{1}) + G_{0}^{\dag}(-\beta_{1}) \bigr] \\
& \bigl[ G_{0}(\beta_{2}) + G_{0}^{\dag}(-\beta_{2}) \bigr] \vert \Phi_{+}^{0} \rangle / Z_{0}
\end{aligned}
\end{equation}
where $Z_{0}=\langle \Phi_{+}^{0} \vert \bigl[ \mathsf{V}(+) \bigr]^{N} \vert \Phi_{+}^{0} \rangle$.
\textit{Sensu stricto}, Eq.~\eqref{21112024_2135} contains two terms that are obtained by expressing the state $\vert + \rangle$ in \eqref{11062025_1512} in terms of $\vert \Phi_{\pm}^{0}\rangle$. However, in the limit $M \to \infty$ these terms are asymptotically identical; see the discussion before Eq.~\eqref{30052020_01}.

The matrix element (\ref{21112024_2135}) can be computed by using the anti-commutation relation satisfied by the $G_{0}$-operators and by using the contraction function defined by
\begin{equation}
\begin{aligned}
\label{30052020_09}
\kappa_{0}(\beta_{1},\beta_{2}) & = \frac{1}{Z_{0}} \langle \Phi_{+}^{0} \vert \bigl[ \mathsf{V}(+) \bigr]^{N} G_{0}^{\dag}(-\beta_{1}) G_{0}^{\dag}(-\beta_{2}) \vert \Phi_{+}^{0} \rangle \, .
\end{aligned}
\end{equation}
The latter can be computed by recalling the similarity transformation induced by $\mathsf{V}$
\begin{equation}
\label{30052020_07}
\bigl[ \mathsf{V}(+) \bigr]^{-N} \mathbb{G}_{0}(\beta) \bigl[ \mathsf{V}(+) \bigr]^{N} = \mathsf{W}(\beta) \mathbb{G}_{0}(\beta) \, ,
\end{equation}
then, from the vacuum property $\langle \Phi_{+}^{0} \vert G_{0}^{\dag}(\beta)$ and the above definition it is simple to show that
\begin{equation}
\begin{aligned}
\kappa_{0}(\beta_{1},\beta_{2}) & = - \frac{ W_{21}(\beta_{1}) }{ W_{22}(\beta_{1}) } \delta_{\beta_{1},-\beta_{2}} \, ,
\end{aligned}
\end{equation}
where $W_{ij}(\beta)$ are entries of the $2\times2$ matrix $\mathsf{W}(\beta)$. The matrix $\mathsf{W}$ is computed as follows:
\begin{equation}
\label{03052021_1151}
\mathsf{W} = \mathsf{U}(-\delta^{*}/2) \Bigl[ \cosh(N\gamma) - \sinh(N\gamma) \tau^{z} \Bigr] \mathsf{U}(\delta^{*}/2) \, .
\end{equation}
By using the property
\begin{equation}
\mathsf{U}(-\theta) \tau^{z} \mathsf{U}(\theta) = \tau^{z} \mathsf{U}(2\theta) \, ,
\end{equation}
the calculation of (\ref{03052021_1151}) is immediate and gives
\begin{equation}
\label{03052021_1156}
\mathsf{W} = \cosh(N\gamma) - \sinh(N\gamma) \cos\delta^{*} \tau^{z} - \sinh(N\gamma) \sin\delta^{*} \tau^{y} \, ,
\end{equation}
or, more explicitly,
\begin{widetext}
\begin{equation}
\label{03052021_1157}
\mathsf{W} =
\left(\begin{array}{cc}
\cosh(N\gamma) - \sinh(N\gamma) \cos\delta^{*} &  \im \sinh(N\gamma) \sin\delta^{*} \\
-\im \sinh(N\gamma) \sin\delta^{*} & \cosh(N\gamma) + \sinh(N\gamma) \cos\delta^{*}
\end{array}\right) \, .
\end{equation}
\end{widetext}

Turing back to the calculation of the partition function ratio (\ref{30052020_04}), the above formalism allows us to find
\begin{equation}
\frac{Z_{+-}(M,N \vert s)}{Z_{++}(M,N)} = - M^{-1} \sum_{ \beta } \operatorname{e}^{\im s \beta} \frac{W_{21}(\beta)}{W_{22}(\beta)}.
\end{equation}
This result is obtained by using the fact that $\theta_{0}(\beta)$ is an odd function. By applying $M\to\infty$ limit, we get 
\begin{equation}
\zeta_{\mathrm{B}}(N,s) = \lim_{ M \rightarrow \infty} \frac{Z_{+-}}{Z_{++}} = \int_{-\pi}^{\pi} \frac{\textrm{d}\beta}{2\pi} \operatorname{e}^{\im s \beta} \frac{W_{21}(\beta)}{W_{22}(\beta)} \, ,
\end{equation}
which justifies the expression in Eq.~\eqref{03052021_0909}.

\begin{figure}[b]
\centering
\includegraphics[width=0.68\columnwidth]{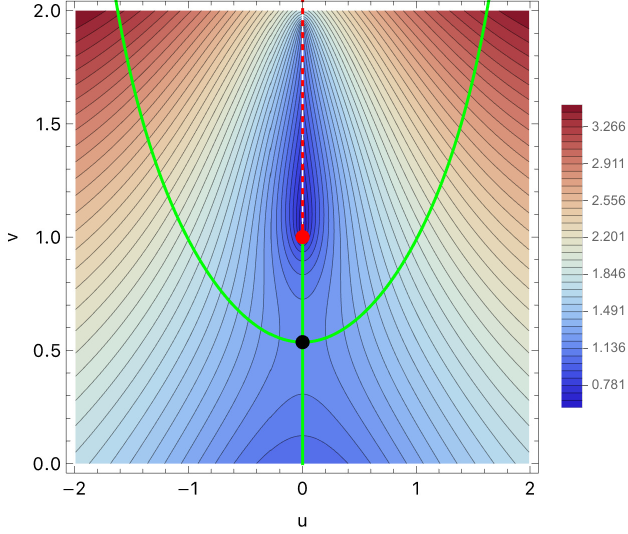}\\[4mm]
\includegraphics[width=0.68\columnwidth]{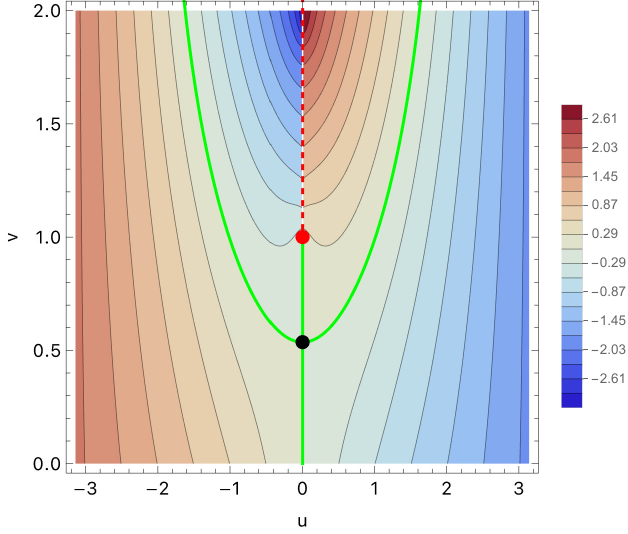}
\caption{The real (top panel) and imaginary (bottom panel) parts of the function $L(\omega) = \gamma(\omega)-\im \omega \tan\vartheta$. The locus $\textrm{Im}\left(L(\omega) \right)=0$ is indicated with green solid lines. The saddle point is indicated with a black bullet, the point $v=\hat{\gamma}(0)$ with a red dot, and the branch cut emanating from $\omega = \im \hat{\gamma}(0)$ is shown with a dashed red line. In these plots, $T=0.6\, T_{\rm c}$, $K_{1}=K_{2}$, $\vartheta=30^{\circ}$.}
\label{fig_sdp_densityplot}
\end{figure}

\subsection{Type-A interfaces}
The calculation for the lattice in Fig.~\ref{fig_3_new} can be performed by introducing the block rotation operator $R(1+s,1+s_{1})=\Gamma_{2s+1}\Gamma_{2s_{1}+1}$ for spins in the upper edge. We express this operator in terms of Fermi fields
\begin{multline} 
\label{02052021}
R(1+s, 1+s_{1})  = - M^{-1} \sum_{ \beta_{3}, \beta_{4} } \operatorname{e}^{-\im (1+s)\beta_{3}} \operatorname{e}^{-\im\left(1+s_{1}\right)\beta_{4}} \\ 
\times  \left( F\left(\beta_{3}\right) + F^{\dag}(-\beta_{3}) \right) \left( F\left(\beta_{4}\right) + F^{\dag}\left(-\beta_{4}\right) \right),
\end{multline}
and, successively, in terms of operators $G_{0}$. The ratio of partition functions we need to compute is
\begin{multline}
\frac{Z_{+-}(M,N \vert s, s_{1})}{Z_{++}(M,N)}= \\
  \langle + \vert \Gamma_{2s_{1}+1} \Gamma_{2s+1} \mathsf{V}^{N} \Gamma_{1} \Gamma_{2s_{1}+1} \vert + \rangle / \langle + \vert \mathsf{V}^{N} \vert + \rangle.
\end{multline}
We observe that each Kaufman spinor carries a combination of the form shown in (\ref{02052021}). After a simple manipulation, we obtain
\begin{equation}
\label{27032025_1612}
\frac{Z_{+-}(M,N \vert s, s_{1})}{Z_{++}(M,N)} = M^{-2} \sum_{ (\beta)_{4} } \mathcal{A}( (\beta)_{4} ) \mathbb{M}((\beta)_{4}) \, ,
\end{equation}
with
\begin{equation}
\mathcal{A}\left( \left(\beta\right)_{4} \right) = \operatorname{e}^{-\im \beta_{1}  -\im \left(1+s\right)\beta_{3} - \im (1+s_{1})(\beta_{2}+\beta_{4})} \prod_{j=1}^{4} \operatorname{e}^{ \im \theta_{0}\left(\beta_{j}\right)} \, ,
\end{equation}
and
\begin{equation}
\begin{aligned}
\mathbb{M}((\beta)_{4}) & = \langle \Phi_{+}^{0} \vert \bigl[ G_{0}(\beta_{4}) + G_{0}^{\dag}(-\beta_{4}) \bigr] \bigl[ G_{0}(\beta_{3}) + G_{0}^{\dag}(-\beta_{3}) \bigr]\\
& \times  \bigl[ \mathsf{V}(+) \bigr]^{N} \bigl[ G_{0}(\beta_{1}) + G_{0}^{\dag}(-\beta_{1}) \bigr] \\
& \times \bigl[ G_{0}(\beta_{2}) + G_{0}^{\dag}(-\beta_{2}) \bigr] \vert \Phi_{+}^{0} \rangle / Z_{0} \, .
\end{aligned}
\end{equation}
The above can be further simplified by using the anti-commutation relations satisfied by $G_{0}$ and $G_{0}^{\dag}$, the vacuum property $G_{0}(\beta) \vert \Phi_{+}^{0} \rangle$ and its adjoint. In addition, we use the following result \cite{Abraham_2012}:
\begin{equation}
\label{30052020_10}
\kappa_{0}( (\beta)_{2n} )  = \sum_{j=2}^{2n} (-1)^{j} \kappa_{0}(\beta_{1},\beta_{j}) \kappa_{0}( \Delta_{1,j}((\beta)_{2n}) ) \, ,
\end{equation}
that allows for the calculation of contraction functions of higher order. In the above, $\kappa_{0}(\emptyset)=1$ and $\Delta_{1,j}$ denotes the removal of two momenta $\beta$ with labels $1$ and $j$ from the sequence. In particular,
\begin{multline}
\kappa_{0}( (\beta)_{4} )  = \kappa_{0}( \beta_{1},\beta_{2} ) \kappa_{0}( \beta_{3},\beta_{4} ) - \kappa_{0}( \beta_{1},\beta_{3} ) \kappa_{0}( \beta_{2},\beta_{4} )\\
 + \kappa_{0}( \beta_{1},\beta_{4} ) \kappa_{0}( \beta_{2},\beta_{3} ).
\end{multline}
Leaving out the lengthy details, the final result of this calculation takes a very simple form which can be conveniently written as a sum of Wick's contractions of the indices of the four momenta $(\beta)_{4}$. With this in mind, we write the result as follows:
\begin{multline}
\label{20112024_1200}
\mathbb{M}((\beta)_{4})  = \mathbb{M}(\wick{ \c1 {\beta_{1},} \c1 {\beta_{2}} \vert \c1 {\beta_{4}}, \c1 {\beta_{3}} }) + \mathbb{M}(\wick{ \c1 {\beta_{1},} \c2 {\beta_{2}} \vert \c2 {\beta_{4}}, \c1 {\beta_{3}} }) \\
 + \mathbb{M}(\wick{ \c1 {\beta_{1},} \c2 {\beta_{2}} \vert \c1 {\beta_{4}}, \c2 {\beta_{3}} }),
\end{multline}
where
\begin{subequations}
\begin{align}
\mathbb{M}(\wick{ \c1 {\beta_{1},} \c2 {\beta_{2}} \vert \c2 {\beta_{4}}, \c1 {\beta_{3}} }) & = \frac{\delta_{\beta_{4},-\beta_{2}} \delta_{\beta_{3},-\beta_{1}}}{W_{22}(\beta_{4})W_{22}(\beta_{3})} \, , \\
\mathbb{M}(\wick{ \c1 {\beta_{1},} \c2 {\beta_{2}} \vert \c1 {\beta_{4}}, \c2 {\beta_{3}} }) & = - \frac{\delta_{\beta_{4},-\beta_{1}} \delta_{\beta_{3},-\beta_{2}}}{W_{22}(\beta_{4})W_{22}(\beta_{3})} \, , \\
\mathbb{M}(\wick{ \c1 {\beta_{1},} \c1 {\beta_{2}} \vert \c1 {\beta_{4}}, \c1 {\beta_{3}} }) & =  \biggl[ 1 - \frac{W_{21}(\beta_{1})}{W_{22}(\beta_{1})} \biggr] \biggl[ 1 - \frac{W_{21}(\beta_{3})}{W_{22}(\beta_{3})} \biggr] \nonumber\\
& \times \delta_{\beta_{1},-\beta_{2}} \delta_{\beta_{4},-\beta_{3}} \, .
\end{align}
\end{subequations}
Finally, we plug (\ref{20112024_1200}) into (\ref{27032025_1612}) and perform the summation over momenta. At this stage we observe that the summation of the terms involving a single Kronecker delta, as those in the contraction $(12)\vert(34)$, yield a vanishing contribution---this can be easily established by using the property
\begin{equation}
\sum_{\left(\beta\right)_{2}} \delta_{\beta_{1},-\beta_{2}} \operatorname{e}^{-\im s\beta} = \sum_{j=1}^{M} \operatorname{e}^{-\im s \left(2j-1\right)\pi/M} = 0,
\end{equation}
valid for any integer $s$.
Summarizing, the partition function ratio is given by
\begin{widetext}
\begin{equation}
\begin{aligned}
\frac{Z_{+-}(M,N \vert s, s_{1})}{Z_{++}(M,N)} & = \underbrace{ \biggl[ \frac{1}{M} \sum_{ \beta_{1}} \frac{\operatorname{e}^{-\im s\beta_{1}}}{W_{22}(\beta_{1})} \biggr] \biggl[ \frac{1}{M} \sum_{ \beta_{2}}\frac{1}{W_{22}(\beta_{2})} \biggr] }_{ \longrightarrow \pfig{bb} } + \underbrace{ (-1) \biggl[ \frac{1}{M} \sum_{ \beta_{1}} \frac{\operatorname{e}^{-\im s_{1}\beta_{1}}}{W_{22}(\beta_{1})} \biggr] \biggl[ \frac{1}{M} \sum_{ \beta_{2}}\frac{\operatorname{e}^{-\im (s-s_{1})\beta_{2}}}{W_{22}(\beta_{2})} \biggr] }_{ \longrightarrow \pfig{cc} } \\
& + \underbrace{ \biggl[ - \frac{1}{M} \sum_{ \beta_{1}} \frac{ W_{21}(\beta_{1}) }{ W_{22}(\beta_{1}) } \operatorname{e}^{-\im s_{1}\beta_{1}} \biggr] \biggl[ - \frac{1}{M} \sum_{ \beta_{3}} \frac{ W_{21}(\beta_{3}) }{ W_{22}(\beta_{3}) } \operatorname{e}^{-\im (s_{1}-s)\beta_{3}} \biggr] }_{ \longrightarrow \pfig{aa} }.
\end{aligned}
\end{equation}
\end{widetext}
By performing the limit $M \to \infty$, we finally obtain Eq.~\eqref{19112024_1446} with $\zeta_{\mathrm{A}}$ given by Eq.~\eqref{03052021_1247} and $\zeta_{\mathrm{B}}$ given by Eq.~\eqref{03052021_0909}.

\section{Function $L\left(\omega\right)$, angle-dependent surface tension and interface stiffness}
\label{app:B}

In Fig.~\ref{fig_sdp_densityplot}, we present a typical behaviour of the function $L\left(\omega\right)$ on the complex plane. The saddle point is denoted by black dot and there are two curves along which $\operatorname{Im} L\left(\omega\right)=0$. 
One of these curves is the SDP, the other one---along the imaginary axis---is the steepest \emph{ascent} path. Both paths intersect in the saddle point and the directions of the steepest descent are those which emanate from the saddle point in horizontal direction along the curved path indicated with green color. 

\begin{figure*}[t]
\centering
\includegraphics[height=80mm, trim=19 2 21 3]{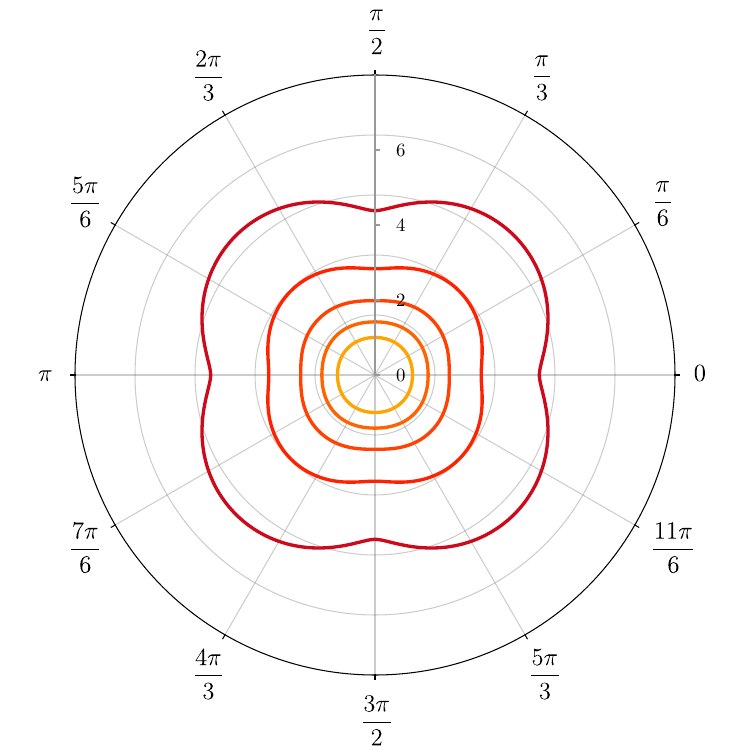}\hspace{1cm}
\includegraphics[height=80mm, trim=48 31 50 33]{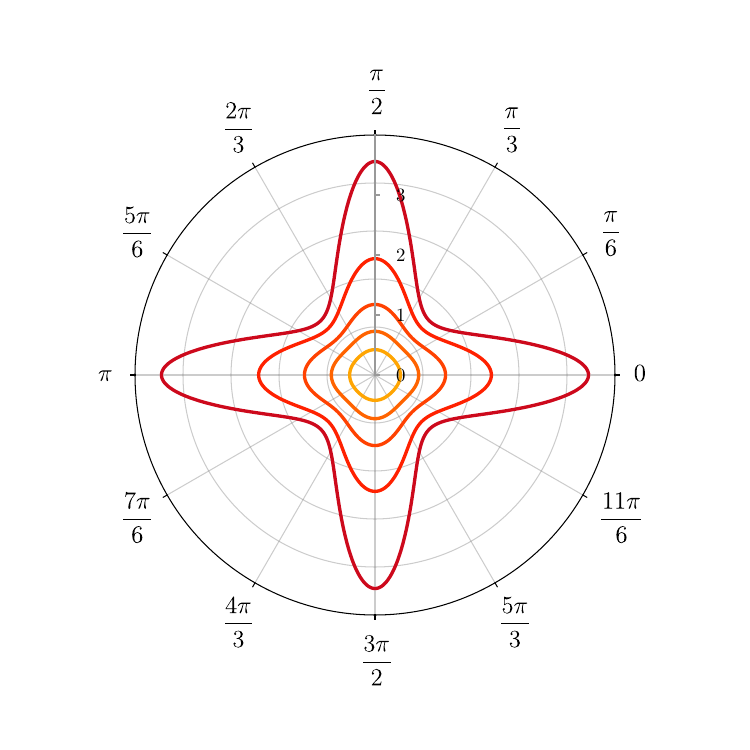}
\caption{Polar chart of the angle-dependent surface tension $\tau(\vartheta)$ (left panel) and interface stiffness $\Sigma(\vartheta)$ (right panel). In left panel $T/T_{\rm c}= 0.2, 0.3, 0.4, 0.5$, and $0.6$, respectively from the outer to the inner curve. In right panel $T/T_{\rm c}= 0.4, 0.5, 0.6, 0.7$, and $0.8$, respectively from the outer to the inner curve.}
\label{fig_polar}
\end{figure*}

Fig.~\ref{fig_polar} provides the polar charts showing the comparison between the angle-dependent surface tension $\tau(\vartheta)$ and interface stiffness, the latter can also be written in the form $\Sigma(\vartheta) = \tau(\vartheta) + \tau^{(2)}(\vartheta)$.

\section{Properties of point tension $\taufcp$}
\label{app:C}
In this appendix we study the properties of the point tension $\taufcp$ defined via relation \eqref{25112024_1119}
\begin{equation}\label{taufcp}
\exp\left[ -2\taufcp\left(\vartheta\right) \right] = - \tan^{2}\left[ \delta^{*}\left(\im v_{s}\left(\vartheta\right)\right)/2\right].
\end{equation}
We start from the trigonometric identity $\tan^{2}(\omega/2) = [1-\cos(\omega)]/[1+\cos(\omega)]$. Since $\cos[\delta^{*}(\im v)]>1$ for any $0 < v < \hat{\gamma}(0)$, it follows that
\begin{equation}
\label{25112024_1147}
-\tan^{2}[ \delta^{*}(\im v)/2] > 0,
\end{equation}
and therefore the point tension $\taufcp$ is real, as expected. 
The above result allows us to introduce parameter $\mathcal{T}\left(v\right)$
\begin{equation}
    \iu \mathcal{T}\left(v\right)=\tan\left[\delta^{*}(\iu v)/2\right],
\end{equation}
which is real for $0<v<\hat{\gamma}\left(0\right)$. From the hyperbolic sine relation \eqref{25112024_1235} for $\omega=\iu v$ it follows that
\begin{equation}
\label{25112024_1223}
\sin\delta^{*}(\im v) = \im s_{1}^{\star} \sinh v /\sinh\gamma(\im v).
\end{equation}
Since for $0<v<\hat{\gamma}\left(0\right)$ we have $\sinh\gamma(\im v)>0$, $\sin\delta^{*}(\im v)$ is purely imaginary with positive imaginary part. Using the trigonometric identity $\tan(\omega/2) = (\sin \omega)/(1+\cos \omega)$ for $\omega=\iu v$ this proves that $\mathcal{T}\left(v\right)>0$, and from Eq.~\eqref{taufcp} we get
\begin{equation}
\label{25112024_1226}
\taufcp(\vartheta) = - \ln\mathcal{T}(v_{s}(\vartheta)).
\end{equation}

Using the identities for Onsager hyperbolic triangle for $\omega=\iu v$, we derived
\begin{equation}
\mathcal{T}(v) = \left[ \frac{\sqrt{1+\mathcal{J}(v)}-1}{\sqrt{1+\mathcal{J}(v)}+1}\right]^{1/2},
\end{equation}
where
\begin{equation}
\label{25112024_1305}
\mathcal{J}\left(v\right) = \frac{1}{4} \frac{(A-B)^{2}(\operatorname{e}^{2v}-1)^{2}}{(A-\operatorname{e}^{v})(B-\operatorname{e}^{v})(A\operatorname{e}^{v}-1)(B\operatorname{e}^{v}-1)},
\end{equation}
and
\begin{equation}
\label{25112024_1304}
A = \operatorname{e}^{2K_{1}+2K_{2}^{\star}}, \qquad B = \operatorname{e}^{2K_{1}-2K_{2}^{\star}}.
\end{equation}
Since for $0<v<\hat{\gamma}(0)$ the function $\mathcal{J}(v)$ is positive, $\mathcal{T}\left(v\right)$ is bounded between zero and unity, and as a result, the point tension $\taufcp$ is always positive.

\unappendix

\balancecolsandclearpage

\cleardoublepage
\beginsupplement

\widetext
\begin{center}
\textbf{\large Supplementary Information for \\
\vspace{2mm}
\ourtitle}

\medskip

 Alessio~Squarcini$^{1,*}$, Piotr~Nowakowski$^{2,\dagger}$, Douglas~B.~Abraham$^{3}$, Anna~Macio{\l}ek$^{4,5}$\\[2mm]

\textit{\small{$^{1}$}Department of Physics and Materials Science, University of Luxembourg,\\ 30 Avenue des Hauts-Fourneaux, L-4362 Esch-sur-Alzette, Luxembourg}

\textit{\small{$^{2}$Group for Computational Life Sciences, Division of Physical Chemistry,\\ Ru\dj{}er Bo\v{s}kovi\'c Institute, Bijeni\v{c}ka 54, 10000, Zagreb, Croatia}}

\textit{\small{$^{3}$Rudolf Peierls Centre for Theoretical Physics, Clarendon Laboratory, Oxford OX1 3PU, UK}}

\textit{\small{$^{4}$}Institute of Physical Chemistry, Polish Academy of Sciences, Kasprzaka 44/52, PL-01-224 Warsaw, Poland}

\textit{\small{$^{5}$}Max Planck Institute for Intelligent Systems, Heisenbergstr.~3, D-70569, Stuttgart, Germany}

\vspace{10mm}

\texttt{\small{$^{*}$alessio.squarcini@uni.lu}}\\ \texttt{\small{$^{\dagger}$Piotr.Nowakowski@irb.hr}}

\end{center}


\section{Saddle point}
\label{sec_3}

This section provides a summary of properties of the saddle point of the function $L\left(\omega\right)=\gamma\left(\omega\right)-\iu \omega\tan\vartheta$  for arbitrary angle $0\leqslant\vartheta\leqslant \pi/2$ and arbitrary couplings $K_{1}$ and $K_{2}$. The interested reader can find previously existing results on Wulff construction and equilibrium shapes of Ising crystals and related models in Refs.~\cite{Abraham_1984_Surfaces_and_Roughening, Rottman_Wortis_PRB, ABSZ}, in the review articles \cite{Abraham.review, Rottman_Wortis, Pfister_2009, Pfister_2010}, and references therein.
The saddle point, given by the solution of Eq.~\eqref{09022021_0712}, is located on the imaginary axis $\omega_\mathrm{s}=\iu \vs\left(\vartheta\right)$,  where the function $\vs\left(\vartheta\right)$ fulfils the condition
\begin{subequations}\label{vs}
\begin{equation}
\label{01022021_1837}
F(\vs(\vartheta)) = \tan\vartheta,
\end{equation}
where the function $F\left(v\right)=-\iu \gamma^{\left(1\right)}\left(\iu v\right)$ is given by
\begin{equation}
\label{01022021_1838}
F(v) = s_{1}^{\star} s_{2}^{\vphantom{\star}} \frac{\sinh v}{\sinh\gamma(\im v)},\qquad\text {with }\cosh\gamma(\iu v) = c_{1}^{\star}c_{2}^{\vphantom{\star}} - s_{1}^{\star}s_{2}^{\vphantom{\star}} \cosh v,
\end{equation}
\end{subequations}
as can be calculated from Eq.~\eqref{01022021_1752}. For completeness, we recall the definition of the simplified notation that we use in this manuscript: $s_i=\sinh 2K_i$, $s^\star_i=\sinh 2K_i^\star$, $c_i=\cosh 2K_i$, $c^\star_i=\cosh 2K_i^\star$, for $i=1,2$.

\begin{lemma}[Existence, uniqueness and differentiability of $\vs$] \label{vs:properties}  For $\vartheta\in\left[0,\pi/2\right]$ the condition~\eqref{vs} has exactly one real solution and the defined function $\vs\left(\vartheta\right)$ is differentiable and it is strictly increasing upon increasing $\vartheta$.
\end{lemma}

\emph{Proof.} We start from analysing the properties of the Onsager $\gamma\left(\omega\right)$ function for imaginary arguments. As follows from Eq.~\eqref{01022021_1838}, upon increasing $v$ the value of $\gamma\left(\iu v\right)$ decreases from $\gamma\left(0\right)=2K_2^{\vphantom{\star}}-2K_1^\star$ for $v=0$ to $0$ for $v=\hat{\gamma}\left(0\right)=2K_2^\star-2K_1^{\vphantom{\star}}$, where  $\hat{\gamma}\left(\omega\right)$  is given by
\begin{equation}
\cosh\hat{\gamma}(\omega) = c_{1}^{\vphantom{\star}}c_{2}^{\star} - s_{1}^{\vphantom{\star}} s_{2}^{\star} \cos \omega,
\end{equation}
\textit{i.e.}, it is Onsager $\gamma\left(\omega\right)$ function with $K_{1}$ and $K_{2}$ interchanged. Moreover, the function $\gamma\left(\iu v\right)$ is differentiable and, therefore, also $F\left(v\right)$ is differentiable for $0\leqslant v<\hat{\gamma}\left(0\right)$.

Since $\gamma\left(\iu v\right)$ is a decreasing function of $v$, also $\sinh \gamma\left(\iu v\right)$ is a decreasing function of $v$, and from Eq.~\eqref{01022021_1838} it follows that the function $F\left(v\right)$ is a strictly increasing function of $v$. As we have calculated, $F\left(0\right)=0$ and $\lim_{v\to\hat{\gamma}\left(0\right)^-}F\left(v\right)=+\infty$, and therefore Eq.~\eqref{01022021_1838} has exactly one solution $\vs\left(\vartheta\right)$, for $0\leqslant \vartheta \leqslant \pi/2$.

Finally, we note that it can be easily checked that $\operatorname{d} F/\operatorname{d} v>0$, and from the implicit function theorem it follows that $\vs\left(\vartheta\right)$ is differentiable and it is an increasing function of its argument.\hfill $\square$

The above calculations allow us to determine the limiting values of $\vs$
\begin{equation}\label{vs:limits}
\vs\left(0\right)=0,\qquad \lim_{\vartheta\to\pi/2}\vs\left(\vartheta\right)=\hat{\gamma}\left(0\right).
\end{equation}



The formula for the saddle point $\vs(\vartheta)$ can be derived in a closed form for any value of $\vartheta$ and for arbitrary couplings; this is the result of the following
\begin{lemma}[Formula for $\vs$]\label{vs:formula}
\label{lemma_sp}
For $\vartheta = \pi/4$ the saddle point location is given by
\begin{subequations}
\begin{equation}
\label{01022021_1926}
\cosh \vs(\pi/4) = \frac{ (s_{1}^{\star}s_{2}^{\vphantom{\star}})^{2} + (c_{1}^{\star}c_{2}^{\vphantom{\star}})^{2} - 1 }{ 2 s_{1}^{\star}s_{2}^{\vphantom{\star}}c_{1}^{\star}c_{2}^{\vphantom{\star}}} \, .
\end{equation}
For $0\leqslant \vartheta<\pi/2$ and $\vartheta \neq \pi/4$ the saddle point location $\vs(\vartheta)$ is given by
\begin{equation}
\label{01022021_1857}
\cosh \vs(\vartheta) = \frac{ -V(\vartheta) + \sqrt{V^{2}(\vartheta)-U(\vartheta)W(\vartheta)}} { U(\vartheta) } \, ,
\end{equation}
\end{subequations}
where
\begin{subequations}
\label{01022021_1900}
\begin{align}
U(\vartheta) & = (s_{1}^{\star}s_{2}^{\vphantom{\star}})^{2} \bigl[ 1 - \tan^{2}\vartheta \bigr], \\
V(\vartheta) & = s_{1}^{\star}s_{2}^{\vphantom{\star}} c_{1}^{\star}c_{2}^{\vphantom{\star}} \tan^{2}\vartheta, \\
W(\vartheta) & = \tan^{2}\vartheta - (c_{1}^{\star}c_{2})^{2} \tan^{2}\vartheta - ( s_{1}^{\star}s_{2}^{\vphantom{\star}})^{2}.
\end{align}
\end{subequations}
\end{lemma}

\emph{Proof.} Eq.~\eqref{01022021_1857} can be obtained by taking the square of Eq.~\eqref{01022021_1837} and expressing $\sinh^{2}\vs(\vartheta)$, $\sinh^{2}\gamma( \im \vs(\vartheta) )$ in terms of $\cosh \vs(\vartheta)$ and its powers. The result is the quadratic equation in the variable $\cosh \vs(\vartheta)$
\begin{equation}
\label{09022021_1134}
U(\vartheta) \cosh^{2}\vs(\vartheta) + 2 V(\vartheta) \cosh \vs(\vartheta) + W(\vartheta) = 0,
\end{equation}
with $U(\vartheta)$, $V(\vartheta)$, and $W(\vartheta)$ given by \eqref{01022021_1900}. The sign ``$+$'' in front of the square root in Eq.~\eqref{01022021_1857} can be determined by requiring the consistency with values of $\vs$ for $\vartheta=0$ and $\vartheta\to\pi/2$ given in Eq.~\eqref{vs:limits}. Alternatively, one can consider the critical point for the symmetric lattice (for which $ss^{\star}=1$ and $cc^{\star}=2$). In this case Eq.~\eqref{01022021_1857} yields $\cosh \vs(\vartheta) = 1$, meaning that $\vs(\vartheta)=0$ for all possible values of $\vartheta$. 

In the special case of $\vartheta=\pi/4$, $U\left(\vartheta\right)=0$ and the quadratic equation \eqref{09022021_1134} reduces to a linear expression in $\cosh \vs(\vartheta)$ and Eq.~\eqref{01022021_1926} follows straightforwardly. \hfill $\square$

\section{Angle-dependent surface tension}
\label{SI:tau}

The angle-dependent surface tension $\tau(\vartheta)$, defined by Eqs.~\eqref{01022021_1750}, is given by
\begin{equation}\label{tau}
\tau(\vartheta) = \gamma( \im \vs(\vartheta))\cos\vartheta + \vs(\vartheta) \sin\vartheta.
\end{equation}
Formally, this definition is valid only for $0\leqslant \vartheta<\pi/2$, \ie, within domain of the function $\vs\left(\vartheta\right)$. Motivated by the physical interpretation of $\tau\left(\vartheta\right)$ as a surface tension of the interface forming an angle $\vartheta$ with the vertical direction of the lattice, we naturally extend this definition to an arbitrary angle by requiring that $\tau\left(\vartheta\right)$ is symmetric around $\vartheta=0$ and $\vartheta=\pi/2$, and by assuming that it is continues for $\vartheta=\pi/2$. This leads to the following conditions
\begin{equation}\label{tau:symmetry}
\tau\left(\vartheta\right)=\tau\left(-\vartheta\right), \qquad \tau\left(\pi/2+\vartheta\right)=\tau\left(\pi/2-\vartheta\right), \qquad \tau\left(\pi/2\right)=\hat{\gamma}\left(0\right).
\end{equation}
We note that the two symmetries of the surface tension follow from invariance of the lattice under reflections along vertical and horizontal directions.

\begin{lemma}[Differentiability of $\tau\left(\vartheta\right)$] The function $\tau\left(\vartheta\right)$ defined for arbitrary angle $\vartheta$ via Eqs.~\eqref{tau} and \eqref{tau:symmetry} is continuous and differentiable.
\end{lemma}

\emph{Proof}. For $\vartheta\in\left(0,\pi/2\right)$ the differentiability of $\tau\left(\vartheta\right)$ follows directly from the properties of $\vs\left(\vartheta\right)$, proved in Lemma~\ref{vs:properties}. The symmetries in Eq.~\eqref{tau:symmetry} guarantee that this function is also differentiable for $\vartheta$ located in mirror reflections of this segment: $\left(-\pi/2, 0\right)$, $\left(\pi/2, \pi\right)$, and $\left(\pi,3\pi/2\right)$.

The only points that need to be  checked separately are the points of symmetry $\vartheta=0$ and $\vartheta=\pi/2$. From the definition of $\tau$ we get
\begin{equation}
    \lim_{\vartheta\to 0^\pm}\tau\left(\vartheta\right)=\tau\left(0\right)=\gamma\left(0\right),\qquad \lim_{\vartheta\to \pi/2^\pm}\tau\left(\vartheta\right)=\tau\left(\pi/2\right)=\hat{\gamma}\left(0\right),
\end{equation}
and therefore the function is continuous in these points. From Eq.~\eqref{tau}, using the relation \eqref{01022021_1837}, for $0<\vartheta<\pi/2$ we derive
\begin{equation}
\label{01022021_1903}
\tau^{(1)}(\vartheta)=- \gamma( \im \vs(\vartheta))\sin\vartheta + \vs(\vartheta)\cos\vartheta,
\end{equation}
and thus
\begin{equation}
    \lim_{\vartheta\to 0^+} \tau^{(1)}\left(\vartheta\right)=0,\qquad \lim_{\vartheta\to \pi/2^-} \tau^{(1)}\left(\vartheta\right)=-\gamma\left(\iu \hat{\gamma}\left(0\right)\right)=0.
\end{equation}
The derivatives from opposite directions can be calculated using the symmetries \eqref{tau:symmetry}. We conclude that for $\vartheta=0$ and $\vartheta=\pi/2$ the function $\tau\left(\vartheta\right)$ is continuous and its left- and right-side derivatives are equal to $0$, and therefore
\begin{equation}
    \tau^{(1)}\left(0\right)=\tau^{(1)}\left(\pi/2\right)=0
\end{equation}
and the function is differentiable for arbitrary angle $\vartheta$. \hfill $\square$

\begin{lemma}[Behaviour of $\tau\left(\theta\right)$]\label{tau:lemma}
 For $T<T_\mathrm{c}$ the maximal and minimal values of the surface tension depend on the lattice couplings $K_1$ and $K_2$ in the following way:
 \begin{subequations}
\begin{enumerate}
    \item If 
    \begin{equation}\label{tau:cond1}
    s_1 \hat{\gamma}\left(0\right)\geqslant s_2 \sinh \hat{\gamma}\left(0\right)
    \end{equation}
    the maximal value of surface tension is $\tau\left(\pi/2\right)=\hat{\gamma}\left(0\right)$, the minimal value is $\tau\left(0\right)=\gamma\left(0\right)$, and for $0<\vartheta<\pi/2$ the function $\tau\left(\vartheta\right)$ is strictly increasing.
    \item If 
    \begin{equation}\label{tau:cond2}
    s_2 \gamma\left(0\right)\geqslant s_1 \sinh \gamma\left(0\right)
    \end{equation}
    the maximal value of surface tension is $\tau\left(0\right)=\gamma\left(0\right)$, the minimal value is $\tau\left(\pi/2\right)=\hat{\gamma}\left(0\right)$, and for $0<\vartheta<\pi/2$ the function $\tau\left(\vartheta\right)$ is strictly decreasing.
    \item If none of the above inequalities holds, the function $\tau\left(\vartheta\right)$ has two local minima, $\tau\left(0\right)=\gamma\left(0\right)$ and $\tau\left(\pi/2\right)=\hat{\gamma}\left(0\right)$, and the lower of them is the global minimum. Between these minima there is a single maximum (that is also a global maximum) located at $\vartheta=\vartheta_\mathrm{m}$, given by
    \begin{equation}\label{tau:thetam}
        \tan \vartheta_\mathrm{m}=\frac{\vs\left(\vartheta_\mathrm{m}\right)}{\gamma\left(\iu \vs\left(\vartheta_\mathrm{m}\right)\right)}, \qquad 0<\vartheta_\mathrm{m}<\pi/2,
    \end{equation}
where this equation has exactly one solution. The surface tension is strictly increasing function for $0<\vartheta<\vartheta_\mathrm{m}$ and strictly decreasing function for $\vartheta_\mathrm{m}<\vartheta<\pi/2$.
\end{enumerate}
\end{subequations}
The behaviour of $\tau\left(\vartheta\right)$ for angles outside of the range $\left[0,\pi/2\right]$ follows directly from the symmetries \eqref{tau:symmetry}. For completeness, we recall that for $T<T_\mathrm{c}$, $\gamma\left(0\right)=2K_2^{\vphantom{\star}}-2K_1^\star$ and $\hat{\gamma}\left(0\right)=2K_1^{\vphantom{\star}}-2K_2^\star$.
\end{lemma}
\emph{Proof}. We start from the condition $\tau^{(1)}\left(\vartheta\right)>0$, which guarantees that the surface tension is a strictly increasing function of $\vartheta$. For $0<\vartheta<\pi/2$, from Eq.~\eqref{01022021_1903} it follows that this condition is equivalent to
\begin{equation}
    \tan\vartheta<\frac{\vs\left(\vartheta\right)}{\gamma\left(\iu \vs\left(\vartheta\right)\right)}.
\end{equation}
Using Eq.~\eqref{vs}, after some algebra we get
\begin{equation}\label{tau:inequality}
    s_2 \frac{\sinh \vs\left(\vartheta\right)}{\vs\left(\vartheta\right)}<s_1 \frac{\sinh\gamma\left(\iu \vs\left(\vartheta\right)\right)}{\gamma\left(\iu \vs\left(\vartheta\right)\right)}.
\end{equation}
We now note that $\sinh x/x$ is a strictly increasing function for $x>0$. Since, as we have proven in Lemma~\ref{vs:properties}, $\vs\left(\vartheta\right)$ is also strictly increasing for $0<\vartheta<\pi/2$, left-hand-side of inequality \eqref{tau:inequality} is also a strictly increasing function of $\vartheta$. On the other hand, the function $\gamma\left(\iu v\right)$ is a strictly decreasing function of $v$ and, as a result, right-hand-side of inequality \eqref{tau:inequality} is a strictly decreasing function of $\vartheta$. Therefore, if inequality \eqref{tau:inequality} is satisfied for $\vartheta\to\pi/2$, it must be satisfied for all $\vartheta\in\left(0,\pi/2\right)$ and, thus, the surface tension is strictly increasing. By applying the limit $\vartheta\to\pi/2$, after short calculations (using identity $\lim_{x\to 0}\sinh x/x=1$) inequality \eqref{tau:inequality} simplifies to condition \eqref{tau:cond1}, and proves first point of the Lemma.

The above reasoning can be repeated for the condition $\tau^{(1)}\left(\vartheta\right)<0$, for which $\tau\left(\vartheta\right)$ is strictly decreasing function of $\vartheta$, giving inequality
\begin{equation}\label{tau:invinequality}
    s_2 \frac{\sinh \vs\left(\vartheta\right)}{\vs\left(\vartheta\right)}>s_1 \frac{\sinh\gamma\left(\iu \vs\left(\vartheta\right)\right)}{\gamma\left(\iu \vs\left(\vartheta\right)\right)}.
\end{equation}
Like in the previous case, left-hand-side is an increasing function and right-hand-side is a decreasing function  of $\vartheta$. Therefore, this inequality is satisfied for all $0<\vartheta<\pi/2$ if it is satisfied for $\vartheta\to 0$. Applying the limit $\vartheta \to 0$ to inequality \eqref{tau:invinequality} gives the condition \eqref{tau:cond2} and proves second point of the Lemma.

Finally, when inequality \eqref{tau:inequality} is not fulfilled for $\vartheta\to\pi/2$, and inequality \eqref{tau:invinequality} is not fulfilled for $\vartheta \to 0$, there must be points where both sides are equal, \ie, where $\tau^{(1)}\left(\vartheta\right)=0$. Since left-hand-side is strictly increasing and right-hand-side is strictly decreasing function of $\vartheta$, there is only one such point and we denote it by $\vartheta_\mathrm{m}$. The condition \eqref{tau:thetam} can be derived straightforwardly from \eqref{01022021_1903}. For $0<\vartheta<\vartheta_\mathrm{m}$ inequality \eqref{tau:inequality} is satisfied and $\tau\left(\vartheta\right)$ is strictly increasing, and for $\vartheta_\mathrm{m}<\vartheta<\pi/2$ inequality \eqref{tau:invinequality} is satisfied and the surface tension is strictly decreasing. This proves third point of the Lemma.\hfill $\square$

\begin{lemma}[$\tau\left(\vartheta\right)$ for symmetric lattice]\label{tau:symmetric}
In case of symmetric lattice ($K_1=K_2\equiv K$) the minimal value of the surface tension $\gamma\left(0\right)=\hat{\gamma}\left(0\right)=2K-2K^\star$ is attained for $\vartheta=0$ and $\vartheta=\pi/2$. The maximal value of surface tension is $\tau\left(\pi/4\right)$ and $\tau\left(\pi/4\right)<\sqrt{2} \tau\left(0\right)$. Moreover, the function $\tau\left(\vartheta\right)$ is symmetric around $\vartheta=\pi/4$ (which is in agreememt with an invariance of the symmetric lattice under reflections along diagonal lines).
\end{lemma}
\emph{Proof}. We use Lemma~\ref{tau:lemma} to find minimal and maximal value of surface tension for symmetric lattice. Since for this lattice $s_1=s_2=s$ and $\gamma\left(\omega\right)=\hat{\gamma}\left(\omega\right)$, the conditions \eqref{tau:cond1} and \eqref{tau:cond2} both reduce to $\sinh \gamma\left(0\right)/\gamma\left(0\right)\leqslant 1$,
a condition that cannot be satisfied for $T\neq T_\mathrm{c}$. Therefore, point 3 of Lemma~\ref{tau:lemma} applies to this case.

From Lemma~\ref{vs:formula} it follows that for symmetric lattice (for which $c_1=c_2=c$)
\begin{equation}\label{tau:iso1}
    \cosh \vs\left(\pi/4\right)=c c^\star/2.
\end{equation}
Moreover, from the definition of Onsager $\gamma$ function we get
\begin{equation}\label{tau:iso2}
    \cosh \gamma\left(\iu\vs\left(\pi/4\right)\right)=c c^\star/2.
\end{equation}
Using these results in Eq.~\eqref{tau:thetam} it is straightforward to check that $\vartheta_\mathrm{m}=\pi/4$. Using Lemma~\ref{tau:lemma}, we have therefore shown that the minima of $\tau\left(\vartheta\right)$ are located at $\vartheta=0$ and $\vartheta=\pi/2$, and maximum is located at $\vartheta=\pi/4$.

\begin{figure}[t]
\centering
\includegraphics[width=100mm]{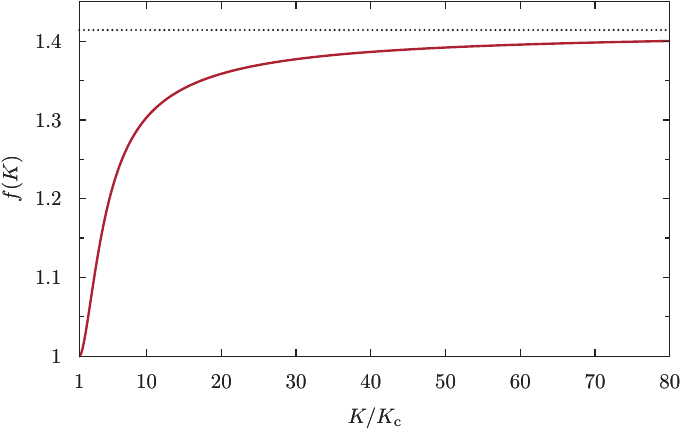}
\caption{The function $f\left(K\right)=\tau\left(\pi/4\right)/\tau\left(0\right)$. The dotted line denotes the asymptotic line $\sqrt{2}$ to which the function approaches in the limit $K\to\infty$.}
\label{fig_boundtension}
\end{figure}

In order to show that $\tau\left(\pi/4\right)<\sqrt{2} \tau\left(0\right)$, from Eqs.~\eqref{tau:iso1} and \eqref{tau:iso2} we derive
\begin{equation}
    \vs\left(\pi/4\right)=\gamma\left(\iu  \vs\left(\pi/4\right)\right)=\ln\sinh 2K,
\end{equation}
and therefore from \eqref{tau} we get
\begin{equation}
    \frac{\tau\left(\pi/4\right)}{\tau\left(0\right)}= \frac{ \sqrt{2} \log(\sinh(2K)) }{ 2K - \ln\coth(K) } \equiv f\left(K\right)<\sqrt{2},
\end{equation}
where we have introduced the function $f\left(K\right)$. The plot of this function is presented in Fig.~\ref{fig_boundtension}. We leave the proof of inequality $f\left(K\right)<\sqrt{2}$ for $T<T_\mathrm{c}$ (\ie, $K>K_\mathrm{c}=\frac{1}{2}\ln\left(1+\sqrt{2}\right)$) as a simple exercise.

We now show that the surface tension for symmetric lattice is symmetric around $\vartheta=\pi/4$, \ie, $\tau\left(\pi/4+\vartheta\right)=\tau\left(\pi/4-\vartheta\right)$, or equivalently $\tau\left(\vartheta\right)=\tau\left(\pi/2-\vartheta\right)$. To this end, we recall the function $F\left(v\right)$ defined in Eq.~\eqref{01022021_1838}. Simple calculation for symmetric lattice gives
\begin{equation}\label{tau:thetamirror}
    F\left(\gamma\left(\iu \vs\left(\vartheta\right)\right)\right)=\frac{\sinh \gamma\left(\iu \vs\left(\vartheta\right)\right)}{\sinh\gamma(\im \gamma\left(\iu \vs\left(\vartheta\right)\right))}=\frac{\sinh \gamma\left(\iu \vs\left(\vartheta\right)\right)}{\sinh \vs\left(\vartheta\right)}=1/F\left(\vs\left(\vartheta\right)\right)=1/\tan\vartheta=\tan\left(\pi/2-\vartheta\right),
\end{equation}
where we have used the definition \eqref{01022021_1837} of $\vs$ and identity $\gamma\left(\iu \gamma\left(\iu v\right)\right)=v$. Comparing Eq.~\eqref{tau:thetamirror} with Eq.~\eqref{01022021_1837} proofs that for symmetric lattice and $0<\vartheta<\pi/2$
\begin{equation}
    \vs\left(\pi/2-\vartheta\right)=\gamma\left(\iu \vs\left(\vartheta\right)\right),
\end{equation}
and therefore, from Eq.~\eqref{tau} we can derive
\begin{equation}
    \tau\left(\pi/2-\vartheta\right)= \gamma( \im \vs(\pi/2-\vartheta))\cos\left(\pi/2-\vartheta\right) + \vs(\pi/2-\vartheta) \sin\left(\pi/2-\vartheta\right)= \vs(\vartheta) \sin\vartheta+ \gamma( \im \vs(\vartheta))\cos\vartheta=\tau\left(\vartheta\right).
\end{equation}
This proofs the postulated symmetry of the surface tension.\hfill $\square$

\begin{figure}[t]
\centering
\includegraphics[height=6cm]{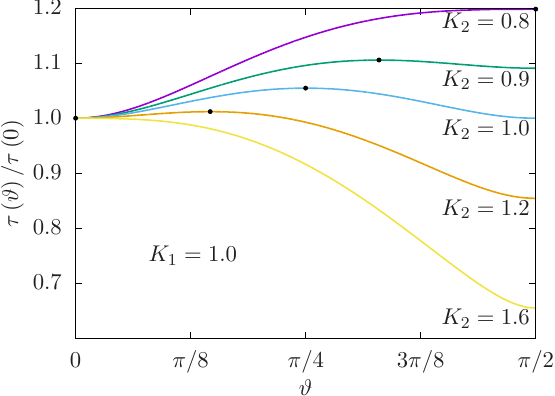}\hspace{0.5cm}
\includegraphics[height=6cm]{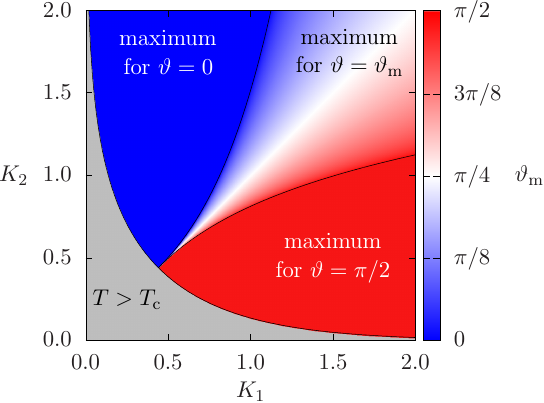}
\caption{The behaviour of angle-dependent surface tension $\tau\left(\vartheta\right)$ for the anisotropic lattice. In left panel we plot this function for $K_1=1.0$ and various values of $K_2$. Black dots denote the maxima of plotted curves. For $0\leqslant \vartheta\leqslant \pi/2$, when $K_2$ is small, the surface tension is monotonously increasing and its maximum is located for $\vartheta=\pi/2$. Upon increasing $K_2$ the surface tension becomes non-monotonic with a maximum that, upon increasing $K_2$, moves continuously from $\vartheta=\pi/2$ to $\vartheta=0$. Eventually, for $K_2$ large enough, $\tau\left(\vartheta\right)$ becomes monotonically decreasing function with a maximum located for $\vartheta=0$. Right panel presents the behaviour of surface tension for different values of $K_1$ and $K_2$. In blue region the maximum of $\tau\left(\vartheta\right)$ is located for $\vartheta=0$, and in red region this maximum is located for $\vartheta=\pi/2$. Between these two regions the value of $\vartheta$ for which maximum is attained varies continuously; we denote its position with the colour code. Finally, gray region denotes values of $K_1$ and $K_2$ for which the system is above its critical point.}
\label{tau_anisotropic}
\end{figure}

The above Lemma allows to understand the meaning of the three possible scenarios of behaviour of $\tau\left(\vartheta\right)$ presented in Lemma~\ref{tau:lemma}. Since $\sinh x/x\geqslant 1$, the condition \eqref{tau:cond1} in the first point of Lemma~\ref{tau:lemma} can only be satisfied when $K_1>K_2$. Similarly, condition \eqref{tau:cond2} in the second point of Lemma~\ref{tau:lemma} can only be satisfied when $K_2<K_1$. Lemma~\ref{tau:symmetric} shows than on the plane $\left(K_1,K_2\right)$ of possible lattices there is a region around line $K_1=K_2$ in which the surface tension has a non-trivial maximum (point 3 of Lemma~\ref{tau:lemma}). Outside of this region, for $K_1>K_2$ the surface tension $\tau\left(\vartheta\right)$ is an increasing function for $0<\vartheta<\pi/2$ (point 1 of Lemma~\ref{tau:lemma}), and for $K_2<K_1$ it is a decreasing function (point 2 of Lemma~\ref{tau:lemma}). We present the plots of the surface tension for different lattices in Fig.~\ref{tau_anisotropic}. 

\begin{lemma}[Bounds for $\tau\left(\vartheta\right)$] For $T<T_\mathrm{c}$ for any couplings $K_1$ and $K_2$, and any angle $\vartheta$ the surface tension satisfies inequalities
\begin{equation}
\min\left\{ \gamma\left(0\right), \hat{\gamma}\left(0\right)\right\}\leqslant \tau\left(\vartheta\right)<\left[\gamma^2\left(0\right)+\hat{\gamma}^2\left(0\right)\right]^{1/2}.
\end{equation}
\end{lemma}
\emph{Proof}. Due to the symmetries \eqref{tau:symmetry} of $\tau\left(\vartheta\right)$, it is enough to prove the inequalities for $0\leqslant \vartheta\leqslant \pi/2$. 

The lower bound follows directly from Lemma~\ref{tau:lemma}: no matter which of the three possible cases is applicable, minimal value of the surface tension is equal to the smaller of $\gamma\left(0\right)$ and $\hat{\gamma}\left(0\right)$.

To prove the upper bound, we first note that
\begin{equation}\label{tau:estimate}
    \vs\left(\vartheta\right)\leqslant \lim_{\vartheta\to\pi/2}\vs\left(\vartheta\right)=\hat{\gamma}\left(0\right), \qquad \gamma\left(\iu\vs\left(\vartheta\right)\right)\leqslant \gamma\left(\iu\vs\left(0\right)\right)=\gamma\left(0\right),
\end{equation}
where the inequalities follow from the fact that $\vs\left(\vartheta\right)$ is strictly increasing and  $\gamma\left(\iu\vs\left(\vartheta\right)\right)$ is strictly decreasing function of $\vartheta$. These inequalities allow us to estimate
\begin{equation}
\tau\left(\vartheta\right) = \gamma( \im \vs(\vartheta))\cos\vartheta + \vs(\vartheta) \sin\vartheta<\gamma( 0)\cos\vartheta +\hat{\gamma}\left(0\right) \sin\vartheta\leqslant \left[\gamma^2\left(0\right)+\hat{\gamma}^2\left(0\right)\right]^{1/2}.
\end{equation}
We note that $\tau\left(\vartheta\right)$ cannot reach the above upper bound because inequalities \eqref{tau:estimate} cannot be both satisfied for a single value of $\vartheta$. We leave the calculation of maximal value of $\gamma( 0)\cos\vartheta +\hat{\gamma}\left(0\right) \sin\vartheta$ as a simple mathematical exercise.\hfill $\square$



\section{Steepest descent path}
\label{sec_4}

In this section we prove basic properties of the steepest descent path (SDP) for $0\leqslant\vartheta<\pi/2$ for arbitrary $K_1$ and $K_2$. In the considerations below we use some well-known properties of SDPs, for completeness we discuss them in next section of this document.

In our problem, the SDP is defined as a curve on a complex plane of variable $\omega$ that goes horizontally through the saddle point $\omega=\iu \vs\left(\vartheta\right)$ and satisfies condition \eqref{00007}: 
\begin{equation}
    \operatorname{Im}\left[N \gamma\left(\omega\right)-\iu M\omega\right]=0.
\end{equation}
We parametrize the SDP with $\omega_\mathrm{sdp}\left(u\right)=u+\iu \vsdp\left(u\right)$, with the parameter $-\pi<u<\pi$ (the exact domain for $u$ is discussed below). As we have shown in the main text (see Eqs.~\eqref{00014} and \eqref{00011}), the function $\vsdp\left(u\right)$ is given implicitly by
\begin{subequations}\label{sdp:def}
\begin{equation}\label{sdp:def:eq}
\frac{\mathcal{A}^{2}\left(u,\vsdp\left(u\right)\right) }{ \cos^{2}\left(u\tan\vartheta\right) } - \frac{ \mathcal{B}^{2}\left(u,\vsdp\left(u\right)\right) }{ \sin^{2}\left(u\tan\vartheta\right) } = 1,
\end{equation}
with
\begin{align}
\mathcal{A}\left(u,v\right) & =  c_{1}^{\star} c_{2}^{\vphantom{\star}} - s_{1}^{\star} s_{2}^{\vphantom{\star}} \cos u \cosh v, \\
\mathcal{B}(u,v) & =  s_{1}^{\star} s_{2}^{\vphantom{\star}} \sin u \sinh v.    
\end{align}
\end{subequations}
\begin{proposition}[Symmetry of SDP]
\label{property_09022021_0811}
The SDP is symmetric under reflection around the imaginary axis.
\end{proposition}
\emph{Proof.} The SDP is symmetric under reflections around imaginary axis when $\vsdp\left(u\right)=\vsdp\left(-u\right)$. It is straightforward to check that if $\vsdp\left(u\right)=v_0$ is a solution of Eq.~\eqref{sdp:def}, also $\vsdp\left(-u\right)=v_0$ solves this equation. \hfill $\square$\\

\begin{proposition}[Range of $u$]
\label{property_09022021_0813}
The SDP does not touch the side lines $u= \pm \pi$.
\end{proposition}
\emph{Proof.} In order to prove this property it is enough to check that for $u=\pi$ Eq.~\eqref{sdp:def} cannot be satisfied. Since $\mathcal{B}(\pi,v)=0$, Eq.~\eqref{sdp:def:eq} gives
\begin{equation}
\left( c_{1}^{\star}c_{2}^{\vphantom{\star}} + s_{1}^{\star} s_{2}^{\vphantom{\star}} \cosh \vsdp\left(\pi\right) \right)^{2} = \cos^{2} \left(\pi \tan\vartheta\right),
\end{equation}
or equivalently
\begin{equation}\label{sdp:pieq}
\left[ \cosh{\gamma}(\pi) + s_{1}^{\star} s_{2}^{\vphantom{\star}} \left(\cosh \vsdp\left(\pi\right)-1\right) \right]^{2} = \cos^{2}\left(\pi \tan\theta\right) .
\end{equation}
Since $\cosh\gamma\left(\pi\right)>1$ and $(\cosh \vsdp\left(\pi\right)-1)>0$, the left-hand-side of Eq.~\eqref{sdp:pieq} is bigger than one. Since, the right-hand-side of Eq.~\eqref{sdp:pieq} does not exceed one, there is no solution for $\vsdp\left(\pi\right)$. \hfill $\square$\\

\begin{proposition}[Asymptotes of SDP]
\label{property_09022021_0818}
The SDP approaches the vertical asymptotes $\omega=\pm\iu \uex(\vartheta)$ with $\uex\left(\vartheta\right) = \pi/\left(1+\tan\vartheta\right)$, \ie, $\vsdp\left(u\right)$ diverges to infinity for $u\to\uex\left(\vartheta\right)$.
\end{proposition}
\emph{Proof.} Due to the symmetry of SDP, it is enough to consider only the case of $u>0$. For $v \to + \infty$ the functions $\mathcal{A}$ and $\mathcal{B}$ exhibit the following asymptotic behaviour
\begin{subequations} \label{00015}
\begin{align}
\mathcal{A}\left(u,v\right) & = - \frac{1}{2} s_{1}^{\star} s_{2}^{\vphantom{\star}} \left(\cos u\right) \operatorname{e}^{v}+ \operatorname{O}\left(\operatorname{e}^{-v}\right), \\
\mathcal{B}\left(u,v\right) & = \frac{1}{2} s_{1}^{\star} s_{2}^{\vphantom{\star}} \left(\sin u\right) \operatorname{e}^{v}+ \operatorname{O}\left(\operatorname{e}^{-v}\right).
\end{align}
\end{subequations}
By inserting the above into Eq.~\eqref{sdp:def:eq}, we find
\begin{equation}
\left( s_{1}^{\star} s_{2}^{\vphantom{\star}} \right)^{2} \sin\left[u\left(-1+\tan\vartheta\right)\right] \sin\left[u\left(1+\tan\vartheta\right)\right]  = \sin^{2}(2u\tan\vartheta) \operatorname{e}^{-2v}. 
\end{equation}
Since for $v\to\infty$ right-hand-side vanishes, the asymptote of the SDP is located at the smallest positive $u$ for which the left-hand-side also vanishes. This proves that
\begin{equation}
\label{00016}
\uex\left(\vartheta\right) = \frac{\pi}{1+\tan\vartheta},
\end{equation}
and when $u$ approaches $\uex\left(\vartheta\right)$ from below, the SDP approaches asymptotically the vertical asymptote $\omega=\iu\uex\left(\vartheta\right)$, with  $\uex\left(\vartheta\right)< \pi$. \hfill $\square$\\

\begin{proposition}[SDP for $\vartheta=\pi/4$ and symmetric lattice]
\label{property_09022021_0823}
The SDP for $\vartheta=\pi/4$ and symmetric lattice is given by
\begin{equation}
\label{00019}
\vsdp\left(u\right) = \operatorname{arccosh}\left(\frac{ \cosh \vs\left(\pi/4\right) }{ \cos u }\right).
\end{equation}
For $u \in (0,\pi/2)$ the above function $\vsdp\left(u\right)$ is convex, increases monotonically and diverges for $u \to \pi/2$. 
\end{proposition}
\emph{Proof.} For symmetric lattice $K_1=K_2$, and therefore $s_{1}^{\star}s_{2}^{\vphantom{\star}}=s^{\star}s=1$, $c_{1}^{\star}c_{2}^{\vphantom{\star}}=cc^{\star}$. In this case, for $\vartheta=\pi/4$ Eq.~\eqref{sdp:def:eq} simplifies to
\begin{equation}
\label{00017}
\cos u \cosh \vsdp\left(u\right) = cc^{\star}/2.
\end{equation}
Since $\cosh v_{s}(\pi/4)=cc^{\star}/2$, the above equation is equivalent to Eq.~\eqref{00019}. The first and second derivatives of $\vsdp$ with respect to $u$ can be computed straightforwardly:
\begin{subequations}
\begin{align}
\label{00018}
\frac{\operatorname{d}}{\operatorname{d}u} \vsdp\left(u\right) &= \tan u \coth \vsdp\left(u\right),\\
\label{00020}
\frac{\operatorname{d}^{2}}{\operatorname{d}u^{2}} \vsdp\left(u\right) &= \frac{\coth v_{\rm sdp}(u)}{\cos^2 u} .
\end{align}
\end{subequations}
Both functions are positive and diverge for $u \rightarrow \pi/2$ (in agreement with Property \ref{property_09022021_0818}), which proves the postulated properties of $\vsdp$. \hfill $\square$\\

\begin{theorem}[Analytic expression for SDP]
\label{lemma_09022021_0906}
The SDP $\omega=u+\iu \vsdp\left(u\right)$ for $-\uex\left(\vartheta\right)<u<\uex\left(\vartheta\right)$, with $\uex$ given by Eq.~\eqref{00016}, can be calculated from the following analytic formulae:\\
For $\vartheta = \pi/4$ the steepest descent path $v_{\rm sdp}(u)$ is given by
\begin{equation}
\label{09022021_0921}
\cosh \vsdp\left(u\right) = \frac{ \left( c_{1}^{\star}c_{2}^{\vphantom{\star}} \right)^{2} + \left( s_{1}^{\star}s_{2}^{\vphantom{\star}} \right)^{2} \cos^{2}u - \cos^{2}u }{ 2 s_{1}^{\star} s_{2}^{\vphantom{\star}} c_{1}^{\star} c_{2}^{\vphantom{\star}} \cos u };
\end{equation}
for $\vartheta \neq \pi/4$ the steepest descent path is given by
\begin{equation}
\label{09022021_1005}
\cosh \vsdp\left(u\right) = \frac{ - \mathcal{Q}\left(u,\vartheta\right) \pm \sqrt{\mathcal{Q}^{2}\left(u,\vartheta\right) - \mathcal{P}\left(u,\vartheta\right)\mathcal{R}\left(u,\vartheta\right)}} { \mathcal{P}\left(u,\vartheta\right)},
\end{equation}
with
\begin{subequations}\label{27012021_0958}
\begin{align}
\mathcal{P}\left(u,\vartheta\right) 
& = (s_{1}^{\star} s_{2}^{\vphantom{}})^{2} \sin\left[u\left(1-\tan\vartheta\right)\right] \sin\left[u\left(1+\tan\vartheta\right)\right],\\
\mathcal{Q}\left(u,\vartheta\right) & = s_{1}^{\star} s_{2}^{\vphantom{}} c_{1}^{\star} c_{2}^{\vphantom{}} \cos u\; \sin^{2}\left(u\tan\vartheta\right), \\
\mathcal{R}\left(u,\vartheta\right)  & = - \left( c_{1}^{\star} c_{2}^{\vphantom{}} \right)^{2} \sin^{2}\left(u\tan\vartheta\right) - \left( s_{1}^{\star} s_{2}^{\vphantom{}} \right)^{2} \sin^{2}u\; \cos^{2}\left(u\tan\vartheta\right) + \sin^{2}\left(u\tan\vartheta\right) \cos^{2}\left(u\tan\vartheta\right).
\end{align}
\end{subequations}
The sign before square root in Eq.~\eqref{09022021_1005} is chosen as follows: for $0 < \vartheta <\pi/4$ it is ``$+$'', for $\pi/4 < \vartheta <\pi/2$ and $0 < u \leqslant  u_{0}\left(\vartheta\right)$ it is also ``$+$'', and for $\pi/4 < \vartheta <\pi/2$ and $u_{0}\left(\vartheta\right) < u < \uex(\vartheta)$ it is ``$-$''. Here  $u_{0}\left(\vartheta\right)=\pi/\left(2\tan\vartheta\right)$.
\end{theorem}

\emph{Proof}. Equation \eqref{sdp:def:eq} can be transformed to the following quadratic equation in the variable $\cosh \vsdp\left(u\right)$
\begin{equation}
\label{27012021_0957}
\mathcal{P}\left(u,\vartheta\right) \cosh^{2}\vsdp\left(u\right) + 2 \mathcal{Q}\left(u,\vartheta\right) \cosh \vsdp\left(u\right) + \mathcal{R}\left(u,\vartheta\right)=0,
\end{equation}
with the functions $\mathcal{P}\left(u,\vartheta\right)$, $\mathcal{Q}\left(u,\vartheta\right)$, and $\mathcal{R}\left(u,\vartheta\right)$ as defined in Eq.~\eqref{27012021_0958}. The expression for the SDP for $\vartheta = \pi/4$, given in Eq.~\eqref{09022021_0921}, follows by noting that $\mathcal{P}\left(u,\pi/4\right)=0$. 

In order to establish the sign in front of the square root, we consider various limiting cases. We begin by examining the limit $u \rightarrow 0$, in which we must retrieve the formula for the saddle point. It is straightforward to check that 
\begin{equation}
\lim_{u \rightarrow 0} \frac{\mathcal{P}\left(u,\vartheta\right)}{u^{2}} = U\left(\vartheta\right), \qquad
\lim_{u \rightarrow 0} \frac{\mathcal{Q}\left(u,\vartheta\right)}{u^{2}} = V\left(\vartheta\right), \qquad
\lim_{u \rightarrow 0} \frac{\mathcal{R}\left(u,\vartheta\right)}{u^{2}} = W\left(\vartheta\right),
\end{equation}
and thus, Eq.~\eqref{27012021_0957} reduces to the equation for $\vs$ (Eq.~\eqref{09022021_1134}) in the limit $u \to 0$. The steepest descent path $\vsdp\left(u=0\right)$ yields the saddle point $\vs\left(\vartheta\right)$ provided the sign ``$+$'' is chosen in Eq.~\eqref{09022021_1005}.

The above analysis is valid on the imaginary axis $u=0$. In the following, we examine what happens away from the imaginary axis. We proceed by examining separately two cases: $0 < \vartheta < \pi/4$, and $\pi/4 < \theta < \pi/2$.

\textbf{Case 1.} For $0<\vartheta<\pi/4$ we look at Eq.~\eqref{09022021_1005} in the limit $u\to \uex\left(\vartheta\right)=2/\left(1+\tan\left(\vartheta\right)\right)$. Since for $0<\vartheta<\pi/4$ we have $\cos\uex\left(\vartheta\right)<\cos\left(\pi/2\right)=0$, and in this limit $\mathcal P\to 0$ and $\mathcal Q<0$. Using this result in Eq.~\eqref{09022021_1005} we get $\cosh \vsdp\to \infty$ for ``$+$'' sign, and $\cosh \vsdp\to -\mathcal{R}/2\mathcal{Q}$ for ``$-$'' sign. Since for $u\to\uex\left(\vartheta\right)$ SDP approaches to its vertical asymptote, only the result with ``+'' sign is correct.

For $0<u<\uex\left(\vartheta\right)$, in order to establish the correct sign in Eq.~\eqref{09022021_1005}, we use the properties of SDP. Since $\vsdp\left(u\right)$ must be continuous, the sign can change only when the discriminant $\Delta=\mathcal{Q}^2-\mathcal{P}\mathcal{R}=0$. In Lemma \ref{lemma_09022021_0929} below, we show that $\Delta$ is non-negative, and it is equal to 0 only for $u=u_n\left(\vartheta\right)=\left(2n+1\right)\pi/\left(2\tan\vartheta\right)$, where $n=0,1,2,\ldots$. For $0<\vartheta<\pi/4$ we have $\uex\left(\vartheta\right)<u_n\left(\vartheta\right)$ and, therefore, the discriminant $\Delta$ is strictly positive for $0<u<\uex\left(\vartheta\right)$ and the sign cannot change. This shows that in this case ``+'' is the correct sign in Eq.~\eqref{09022021_1005}.

\textbf{Case 2.} For $\pi/4<\vartheta<\pi/2$ from the Lemma \ref{lemma_09022021_0929} below it follows that the first zero of the discriminant $0<u_0\left(\vartheta\right)<\uex\left(\vartheta\right)$, and therefore, the sign in Eq.~\eqref{09022021_1005} can change at $u=u_0\left(\vartheta\right)$. We note that for $n\geqslant 1$,$u_n\left(\vartheta\right)>\uex\left(\vartheta\right)$ and thus the other zeros of discriminant are still not relevant.

When $0<u<u_0\left(\vartheta\right)$ the continuity of SDP requires that the sign in Eq.~\eqref{09022021_1005} is not changing. Therefore it must be the same as for $u=0$, i.e., ``+''.

When $u_0\left(\vartheta\right)<u<\uex\left(\vartheta\right)$ the sign can be establish by considering the limit $u\to\uex\left(\vartheta\right)$. Since for $\pi/4<\vartheta<\pi/2$ we have $\cos\uex\left(\vartheta\right)>\cos\left(\pi/2\right)=0$, in this limit $\mathcal{P}\to 0$ and $\mathcal Q>0$. Using this result in Eq.~\eqref{09022021_1005} we get $\cosh \vsdp\to -\mathcal{R}/2\mathcal{Q}$ for ``$+$'' sign, and $\cosh \vsdp\to \infty$ for ``$-$'' sign. Therefore, in this case the sign ``$-$'' gives the correct result in Eq.~\eqref{09022021_1005}.

Finally, we note that when $u=u_0\left(\vartheta\right)$, since the discriminant is zero, both posible signs in Eq.~\eqref{09022021_1005} give the same, correct result. \hfill $\square$\\

In order to prove Theorem \ref{lemma_09022021_0906} it is useful to introduce the following lemma.
\begin{lemma}[Non-negativity and zeros of the discriminant]
\label{lemma_09022021_0929}
For $0<\vartheta<\pi/2$ the discriminant
\begin{equation}
\Delta\left(u,\vartheta\right) = \mathcal{Q}^{2}\left(u,\vartheta\right) - \mathcal{P}\left(u,\vartheta\right) \mathcal{R}\left(u,\vartheta\right)
\end{equation}
is non-negative for any $u \in (0,\pi)$ and vanishes only for $u = u_{n}\left(\vartheta\right)$ with 
\begin{equation}
\label{09022021_0939}
u_{n}\left(\vartheta\right) = \frac{2n+1}{2} \frac{\pi}{\tan\vartheta}, \qquad \text{for }n \in \mathbb{Z}.
\end{equation}
The functions $\mathcal{P}\left(u,\vartheta\right)$, $\mathcal{Q}\left(u,\vartheta\right)$, and $\mathcal{R}\left(u,\vartheta\right)$ are defined in Eq.~\eqref{27012021_0958}.
\end{lemma}
\emph{Proof.} After some algebra, we get the formula for the discriminant
\begin{equation}
\label{09022021_0936}
\Delta\left(u,\vartheta\right) = \frac{\left( s_{1}^{\star}s_{2} \right)^{2}}{4} \cos^{2}\left(u\tan\vartheta\right) \left[ X_{1}\left(u,\vartheta\right) + X_{2}\left(u,\vartheta\right) \sin^{2}u \right],
\end{equation}
where
\begin{subequations}\label{09022021_0937}
\begin{align}
X_{1}\left(u,\vartheta\right) & = \left[ \cos\left(2u\right) - \cos\left(2u\tan\vartheta\right) \right]^{2}, \\
X_{2}\left(u,\vartheta\right) & = 4 \left( c_{1}^{\star}c_{2}^{\vphantom{\star}} \right)^{2} \sin^{2}\left[u\tan\vartheta\right]  + 2\left[ \left( s_{1}^{\star}s_{2}^{\vphantom{\star}} \right)^{2}-1 \right] \left[ \cos\left(2u\tan\vartheta\right) - \cos\left(2u\right) \right].
\end{align}
\end{subequations}

We start our analysis from the factor in square brackets in Eq.~\eqref{09022021_0936}. We transform the formula to
\begin{equation}\label{sdp:factor}
   X_{1}\left(u,\vartheta\right) + X_{2}\left(u,\vartheta\right) \sin^{2}u = \left( s_{1}^{\star}s_{2}^{\vphantom{\star}} \right)^{2} x^2+y^2+xy\left[\left(s_1^\star\right)^2+s_2^2\right],
\end{equation}
where we have introduced
\begin{equation}
    x  = 1-\cos\left(2u\right), \qquad y  = 1-\cos\left(2u\tan\vartheta\right), 
\end{equation}
and used identity
\begin{equation}
  \left( c_{1}^{\star}c_{2}^{\vphantom{\star}} \right)^{2}-  \left( s_{1}^{\star}s_{2}^{\vphantom{\star}} \right)^{2}-1=\left(s_1^\star\right)^2+s_2^2.
\end{equation}
Since for $0<u<\pi$, $x>0$ and $y\geqslant 0$, the factor~\eqref{sdp:factor} is strictly positive and, therefore, the discriminant~\eqref{09022021_0936} is non-negative.

Finally, from Eq.~\eqref{09022021_0936} it follows that the zeros of the discriminant are given by solutions of $\cos\left(u\tan\vartheta\right)=0$, \ie, $u=u_n\left(\vartheta\right)$ given by Eq.~\eqref{09022021_0939}.\hfill $\square$


\section{General properties of the steepest descent path}
\label{sec_6}

In this section we discuss some basic properties of the SDP which were necessary to prove the theorems in the previous section of this document. Even though they are well-known and even considered obvious, we were not able to find their systematic proofs. For completeness of our manuscript we provide these proofs.

In the following we assume that $f\left(z\right)$ is a holomorphic function defined on an open subset $D$ of the complex plane and we will look at the steepest descent line defined by the condition
\begin{equation}
    \operatorname{Im}f\left(z\right)=0.
\end{equation}

\begin{theorem}[smoothness of SDP]
    If there is no saddle point in $D$, the steepest descent line is continuous and differentiable. Also the line cannot have endpoint inside $D$.
\end{theorem}
\emph{Proof.} Let $z_0$ be any point on the steepest descent path. Since it is not a saddle point, $f^\prime\left(z_0\right)\neq 0$ and from holomorphic inverse function theorem [N.~Bleistein and R.~A.~Handelsman, ``Asymptotic Expansions of Integrals''], there exists an open set $D_1\subset D$ around $z_0$ on which $f$ can be inverted, and the inverse function $f^{-1}\left(z\right)$ is also holomorphic. Let us now consider the domain of $f^{-1}\left(z\right)$, the open set $A=f\left(D_1\right)$. Of course, $f\left(z_0\right)\in A\cap\mathbb{R}$. Let $t_1<f\left(z_0\right)<t_2$ be two real numbers, such that the segment $B=\left(t_1,t_2\right)\subset A$. The steepest descent path around $z_0$ can be parametrized $\gamma\left(t\right)=f^{-1}\left(t\right)$ for $t_1<t<t_2$. This proves that steepest descent path is continuous, differentiable, and $z_0$ is not its endpoint.\hfill $\square$\\

We note that the continuity and smoothness of SDP was crucial in the proof of Theorem \ref{lemma_09022021_0906}, where we used these properties to determine the correct sign in Eq.~\eqref{09022021_1005}.

\begin{theorem}[intersection of SDP]
    Two steepest descent paths can intersect only in the saddle point.
\end{theorem}
\emph{Proof.} Let as assume that two steepest descent lines intersect in $z_0$ and it is not a saddle point. Since $f^\prime\left(z_0\right)\neq 0$ from inverse function theorem, in the vicinity of $z_0$ there exists inverse function $f^{-1}\left(z\right)$. 

At the same time, in any open neighbourhood of $z_0$ we have four distinct curves starting in $z_0$, along which $f\left(z\right)$ is real (a pair for each intersecting steepest descent path).
If along any of these curves real $f\left(z\right)$ is non-monotonic, we can find two points $z_1$ and $z_2$ for which $f\left(z_1\right)=f\left(z_2\right)$. If, on the other hand, $f\left(z\right)$ is monotonic along all these curves, we can still find two of them along which $f\left(z\right)$ has the same monotonicity (is strictly increasing or decreasing); and along these two paths we can find two points $z_1$ and $z_2$ for which $f\left(z_1\right)=f\left(z_2\right)$. This shows that $f\left(z\right)$ is not injective in the vicinity of $z_0$, and therefore, the inverse function does not exist, which is in contradiction with the inverse function theorem.\hfill $\square$\\

The two above theorems draw a picture of steepest descent paths connecting, across the complex plane, saddle points or points in infinity (and possibly ending in non-analyticity points). However, they do not exclude cases when infinitely long SDP occupies a finite open subset of the complex plane, like for example a logarithmic spiral. Impossibility of such a scenario follows from the theorem below.

\begin{theorem}[no ending of SDP]
    Let $\gamma\left(t\right)$ be a non-intersecting differentiable contour inside $D$. If there is no saddle point in $D$, and the steepest descent path crosses $\gamma\left(t\right)$ and enters the region inside the contour, there exists a second point where the steepest descent path crosses $\gamma\left(t\right)$ and leaves this region.
\end{theorem}
\emph{Proof.} Let $z_0=\gamma\left(t_0\right)$ be the point in $D$ where the steepest descent path and the contour $\gamma\left(t\right)$ intersect. We now look at the function $\operatorname{Im}\left[f\left(\gamma\left(t\right)\right)\right]$. Clearly this continuous function has a zero for $t=t_0$. If there is another zero for some $t=t_1$, the point $\gamma\left(t_1\right)$ is where the steepest descent path leaves the region inside the contour and the theorem is proven. We therefore assume that $\operatorname{Im}\left[f\left(\gamma\left(t\right)\right)\right]$ does not have any other zeros. 

If $\operatorname{Im}\left[f\left(\gamma\left(t\right)\right)\right]$ has only one zero, since the function is defined on a closed contour, it must have a maximum or a minimum at $t=t_0$. This means that the derivative of $\operatorname{Im}\left[f\left(z\right)\right]$ at $z=z_0$ in the direction tangent to the contour is zero. At the same time, derivative of $\operatorname{Im}\left[f\left(z\right)\right]$ at $z=z_0$ in the direction tangent to steepest descent line is also zero. If these two directions are different, the derivative in any direction must be zero and via Cauchy--Riemann equations it is straightforward to show that $z_0$ is a saddle point---a contradiction with the assumptions of the theorem.

If, on the other hand, the steepest descent path enters the interior of the contour in $z_0$ being tangent to the contour (like $y=x$ and $y=\tan x$ in the origin), we slightly modify the contour to make these two curves not tangent in $z_0$. This can be done using holomorphic inverse function theorem: We first use $f\left(z\right)$ to map both curves. Now, the steepest descent path lies on the real axis and the contour is some curve intersecting it in a tangent manner. Second, we modify the contour keeping it continuous and differentiable, such that is intersects real axis in $f\left(z_0\right)$ with non-zero angle. Third, we map both curves back using $f^{-1}\left(z\right)$. This way, the modified contour intersects with steepest descent path with non-zero angle, in which case the theorem has already been proven. Here, we skip the details of this construction. \hfill $\square$\\

\begin{theorem}[intersection of SDP in saddle point]
    If at some saddle point $z_0$ inside $D$, $f^{\prime\prime}\left(z_0\right)\neq 0$, two steepest descent paths intersect at right angles in $z_0$.
\end{theorem}

\emph{Proof.} We prove this theorem using the Morse lemma for holomorphic functions \cite{Zoladek}. In $z_0$ we have $\operatorname{Im}f\left(z_0\right)=0$, $f^\prime\left(z_0\right)=0$, and $f^{\prime\prime}\left(z_0\right)\neq 0$, and therefore, there exist holomorphic map $\varphi\left(z\right)$ such that
\begin{equation}
    \varphi\left(z_0\right)=0,\quad f\left(\varphi^{-1}\left(w\right)\right)=f\left(z_0\right)+w^2,
\end{equation}
for $w$ from some open neighbourhood of $0$. By taking $w=\varphi\left(z\right)$, we get
\begin{equation}
    \operatorname{Im}f\left(z\right)=\operatorname{Im}\varphi^2\left(z\right),
\end{equation}
for $z$ in the open neighbourhood of $z_0$. The steepest descent paths $\operatorname{Im}f\left(z\right)=0$ are equivalent to  
\begin{equation}
    \operatorname{Im}\varphi\left(z\right)=0,\qquad \text{or }\operatorname{Re}\varphi\left(z\right)=0,
\end{equation}
so from Cauchy--Riemann equations it follows that there are two perpendicular steepest descent lines intersecting in $z_0$. \hfill $\square$\\

The above theorem proves that in the saddle point $\omega=\iu \nus\left(\vartheta\right)$ we have intersection of two  perpendicular SDPs. Since one of them is located along the imaginary axis, the second SDP must go from the saddle point in the direction parallel to the real axis. This justifies (at least for small $u$) the parametrisation of SDP $\omega=u+\iu \vsdp\left(u\right)$ used in Section~\ref{sec_4}.

%

\begin{thebibliography}{35}%
\makeatletter
\providecommand \@ifxundefined [1]{%
 \@ifx{#1\undefined}
}%
\providecommand \@ifnum [1]{%
 \ifnum #1\expandafter \@firstoftwo
 \else \expandafter \@secondoftwo
 \fi
}%
\providecommand \@ifx [1]{%
 \ifx #1\expandafter \@firstoftwo
 \else \expandafter \@secondoftwo
 \fi
}%
\providecommand \natexlab [1]{#1}%
\providecommand \enquote  [1]{``#1''}%
\providecommand \bibnamefont  [1]{#1}%
\providecommand \bibfnamefont [1]{#1}%
\providecommand \citenamefont [1]{#1}%
\providecommand \href@noop [0]{\@secondoftwo}%
\providecommand \href [0]{\begingroup \@sanitize@url \@href}%
\providecommand \@href[1]{\@@startlink{#1}\@@href}%
\providecommand \@@href[1]{\endgroup#1\@@endlink}%
\providecommand \@sanitize@url [0]{\catcode `\\12\catcode `\$12\catcode `\&12\catcode `\#12\catcode `\^12\catcode `\_12\catcode `\%12\relax}%
\providecommand \@@startlink[1]{}%
\providecommand \@@endlink[0]{}%
\providecommand \url  [0]{\begingroup\@sanitize@url \@url }%
\providecommand \@url [1]{\endgroup\@href {#1}{\urlprefix }}%
\providecommand \urlprefix  [0]{URL }%
\providecommand \Eprint [0]{\href }%
\providecommand \doibase [0]{https://doi.org/}%
\providecommand \selectlanguage [0]{\@gobble}%
\providecommand \bibinfo  [0]{\@secondoftwo}%
\providecommand \bibfield  [0]{\@secondoftwo}%
\providecommand \translation [1]{[#1]}%
\providecommand \BibitemOpen [0]{}%
\providecommand \bibitemStop [0]{}%
\providecommand \bibitemNoStop [0]{.\EOS\space}%
\providecommand \EOS [0]{\spacefactor3000\relax}%
\providecommand \BibitemShut  [1]{\csname bibitem#1\endcsname}%
\let\auto@bib@innerbib\@empty
\bibitem [{\citenamefont {Peierls}(1936)}]{peierls1936ising}%
  \BibitemOpen
  \bibfield  {author} {\bibinfo {author} {\bibfnamefont {R.}~\bibnamefont {Peierls}},\ }\href {https://doi.org/doi:10.1017/S0305004100019174} {\bibfield  {journal} {\bibinfo  {journal} {Math. Proc. Camb. Philos. Soc.}\ }\textbf {\bibinfo {volume} {32}},\ \bibinfo {pages} {477} (\bibinfo {year} {1936})}\BibitemShut {NoStop}%
\bibitem [{\citenamefont {Dobrushin}(1968)}]{Dobrushin_1968}%
  \BibitemOpen
  \bibfield  {author} {\bibinfo {author} {\bibfnamefont {R.~L.}\ \bibnamefont {Dobrushin}},\ }\href {https://doi.org/10.1007/BF01075681} {\bibfield  {journal} {\bibinfo  {journal} {Funct. Anal. Its Appl.}\ }\textbf {\bibinfo {volume} {2}},\ \bibinfo {pages} {292} (\bibinfo {year} {1968})}\BibitemShut {NoStop}%
\bibitem [{\citenamefont {Griffiths}(1964)}]{Griffiths}%
  \BibitemOpen
  \bibfield  {author} {\bibinfo {author} {\bibfnamefont {R.~B.}\ \bibnamefont {Griffiths}},\ }\href {https://doi.org/https://doi.org/10.1103/PhysRev.136.A437} {\bibfield  {journal} {\bibinfo  {journal} {Phys. Rev.}\ }\textbf {\bibinfo {volume} {136}},\ \bibinfo {pages} {A437} (\bibinfo {year} {1964})}\BibitemShut {NoStop}%
\bibitem [{\citenamefont {Griffiths}(1971)}]{Griffiths_1970}%
  \BibitemOpen
  \bibfield  {author} {\bibinfo {author} {\bibfnamefont {R.~B.}\ \bibnamefont {Griffiths}},\ }in\ \href@noop {} {\emph {\bibinfo {booktitle} {Statistical Mechanics and Quantum Field Theory (Les Houches 1970)}}},\ \bibinfo {series and number} {Summer School of Theroetical Physics},\ \bibinfo {editor} {edited by\ \bibinfo {editor} {\bibfnamefont {C.}~\bibnamefont {{De Witt}}}\ and\ \bibinfo {editor} {\bibfnamefont {R.}~\bibnamefont {Stora}}}\ (\bibinfo  {publisher} {Gordon and Breach Science Publishers},\ \bibinfo {address} {New York, London, Paris},\ \bibinfo {year} {1971})\ p.\ \bibinfo {pages} {241}\BibitemShut {NoStop}%
\bibitem [{\citenamefont {Onsager}(1944)}]{Onsager_44}%
  \BibitemOpen
  \bibfield  {author} {\bibinfo {author} {\bibfnamefont {L.}~\bibnamefont {Onsager}},\ }\href {https://doi.org/doi.org/10.1103/PhysRev.65.117} {\bibfield  {journal} {\bibinfo  {journal} {Phys. Rev.}\ }\textbf {\bibinfo {volume} {65}},\ \bibinfo {pages} {117} (\bibinfo {year} {1944})}\BibitemShut {NoStop}%
\bibitem [{\citenamefont {Gallavotti}\ \emph {et~al.}(1973)\citenamefont {Gallavotti}, \citenamefont {Martin-L{\"o}f},\ and\ \citenamefont {Miracle-Sol{\'e}}}]{GMM72}%
  \BibitemOpen
  \bibfield  {author} {\bibinfo {author} {\bibfnamefont {G.}~\bibnamefont {Gallavotti}}, \bibinfo {author} {\bibfnamefont {A.}~\bibnamefont {Martin-L{\"o}f}},\ and\ \bibinfo {author} {\bibfnamefont {S.}~\bibnamefont {Miracle-Sol{\'e}}},\ }in\ \href@noop {} {\emph {\bibinfo {booktitle} {Statistical Mechanics and Mathematical Problems}}},\ \bibinfo {series} {Lecture Notes in Physics}, Vol.~\bibinfo {volume} {20},\ \bibinfo {editor} {edited by\ \bibinfo {editor} {\bibfnamefont {A.}~\bibnamefont {Lenard}}}\ (\bibinfo  {publisher} {Springer},\ \bibinfo {address} {Berlin, Heidelberg},\ \bibinfo {year} {1973})\ p.\ \bibinfo {pages} {162}\BibitemShut {NoStop}%
\bibitem [{\citenamefont {Gallavotti}(1972)}]{Gallavotti_1972}%
  \BibitemOpen
  \bibfield  {author} {\bibinfo {author} {\bibfnamefont {G.}~\bibnamefont {Gallavotti}},\ }\href {https://doi.org/10.1007/BF01645615} {\bibfield  {journal} {\bibinfo  {journal} {Commun. Math. Phys.}\ }\textbf {\bibinfo {volume} {27}},\ \bibinfo {pages} {103} (\bibinfo {year} {1972})}\BibitemShut {NoStop}%
\bibitem [{\citenamefont {Gallavotti}(1999)}]{Gallavotti}%
  \BibitemOpen
  \bibfield  {author} {\bibinfo {author} {\bibfnamefont {G.}~\bibnamefont {Gallavotti}},\ }\href {https://books.google.hr/books?id=2jDuMESyj6MC} {\emph {\bibinfo {title} {{Statistical Mechanics: A Short Treatise}}}},\ Theoretical and Mathematical Physics\ (\bibinfo  {publisher} {Springer},\ \bibinfo {address} {Berlin, Heidelberg},\ \bibinfo {year} {1999})\BibitemShut {NoStop}%
\bibitem [{\citenamefont {Abraham}\ \emph {et~al.}(1988)\citenamefont {Abraham}, \citenamefont {Ko},\ and\ \citenamefont {$\check{\textrm{S}}$vraki\'c}}]{AKS_1988}%
  \BibitemOpen
  \bibfield  {author} {\bibinfo {author} {\bibfnamefont {D.~B.}\ \bibnamefont {Abraham}}, \bibinfo {author} {\bibfnamefont {L.~F.}\ \bibnamefont {Ko}},\ and\ \bibinfo {author} {\bibfnamefont {N.~M.}\ \bibnamefont {$\check{\textrm{S}}$vraki\'c}},\ }\href {https://doi.org/10.1103/PhysRevLett.61.2393} {\bibfield  {journal} {\bibinfo  {journal} {Phys. Rev. Lett.}\ }\textbf {\bibinfo {volume} {61}},\ \bibinfo {pages} {2393} (\bibinfo {year} {1988})}\BibitemShut {NoStop}%
\bibitem [{\citenamefont {Abraham}\ \emph {et~al.}(1989)\citenamefont {Abraham}, \citenamefont {Ko},\ and\ \citenamefont {$\check{\textrm{S}}$vraki\'c}}]{AKS_1989}%
  \BibitemOpen
  \bibfield  {author} {\bibinfo {author} {\bibfnamefont {D.~B.}\ \bibnamefont {Abraham}}, \bibinfo {author} {\bibfnamefont {L.~F.}\ \bibnamefont {Ko}},\ and\ \bibinfo {author} {\bibfnamefont {N.~M.}\ \bibnamefont {$\check{\textrm{S}}$vraki\'c}},\ }\href {https://doi.org/10.1007/BF01016767} {\bibfield  {journal} {\bibinfo  {journal} {J. Stat. Phys.}\ }\textbf {\bibinfo {volume} {56}},\ \bibinfo {pages} {563} (\bibinfo {year} {1989})}\BibitemShut {NoStop}%
\bibitem [{\citenamefont {Abraham}\ \emph {et~al.}(2005)\citenamefont {Abraham}, \citenamefont {Mustonen},\ and\ \citenamefont {Wood}}]{AMW_2005}%
  \BibitemOpen
  \bibfield  {author} {\bibinfo {author} {\bibfnamefont {D.~B.}\ \bibnamefont {Abraham}}, \bibinfo {author} {\bibfnamefont {V.}~\bibnamefont {Mustonen}},\ and\ \bibinfo {author} {\bibfnamefont {A.~J.}\ \bibnamefont {Wood}},\ }\href {https://doi.org/10.1103/PhysRevE.71.036106} {\bibfield  {journal} {\bibinfo  {journal} {Phys. Rev. E}\ }\textbf {\bibinfo {volume} {71}},\ \bibinfo {pages} {036106} (\bibinfo {year} {2005})}\BibitemShut {NoStop}%
\bibitem [{\citenamefont {Abraham}\ \emph {et~al.}(2004)\citenamefont {Abraham}, \citenamefont {Mustonen},\ and\ \citenamefont {Wood}}]{AMW_2004}%
  \BibitemOpen
  \bibfield  {author} {\bibinfo {author} {\bibfnamefont {D.~B.}\ \bibnamefont {Abraham}}, \bibinfo {author} {\bibfnamefont {V.}~\bibnamefont {Mustonen}},\ and\ \bibinfo {author} {\bibfnamefont {A.~J.}\ \bibnamefont {Wood}},\ }\href {https://doi.org/10.1103/PhysRevLett.93.076101} {\bibfield  {journal} {\bibinfo  {journal} {Phys. Rev. Lett.}\ }\textbf {\bibinfo {volume} {93}},\ \bibinfo {pages} {076101} (\bibinfo {year} {2004})}\BibitemShut {NoStop}%
\bibitem [{\citenamefont {Abraham}\ \emph {et~al.}(2017)\citenamefont {Abraham}, \citenamefont {Macio\l{}ek}, \citenamefont {Squarcini},\ and\ \citenamefont {Vasilyev}}]{AMSV_2017}%
  \BibitemOpen
  \bibfield  {author} {\bibinfo {author} {\bibfnamefont {D.~B.}\ \bibnamefont {Abraham}}, \bibinfo {author} {\bibfnamefont {A.}~\bibnamefont {Macio\l{}ek}}, \bibinfo {author} {\bibfnamefont {A.}~\bibnamefont {Squarcini}},\ and\ \bibinfo {author} {\bibfnamefont {O.}~\bibnamefont {Vasilyev}},\ }\href {https://doi.org/10.1103/PhysRevE.96.042154} {\bibfield  {journal} {\bibinfo  {journal} {Phys. Rev. E}\ }\textbf {\bibinfo {volume} {96}},\ \bibinfo {pages} {042154} (\bibinfo {year} {2017})}\BibitemShut {NoStop}%
\bibitem [{\citenamefont {Abraham}\ \emph {et~al.}(2014)\citenamefont {Abraham}, \citenamefont {Macio\l{}ek},\ and\ \citenamefont {Vasilyev}}]{AMV_2014}%
  \BibitemOpen
  \bibfield  {author} {\bibinfo {author} {\bibfnamefont {D.~B.}\ \bibnamefont {Abraham}}, \bibinfo {author} {\bibfnamefont {A.}~\bibnamefont {Macio\l{}ek}},\ and\ \bibinfo {author} {\bibfnamefont {O.}~\bibnamefont {Vasilyev}},\ }\href {https://doi.org/10.1103/PhysRevLett.113.077204} {\bibfield  {journal} {\bibinfo  {journal} {Phys. Rev. Lett.}\ }\textbf {\bibinfo {volume} {113}},\ \bibinfo {pages} {077204} (\bibinfo {year} {2014})}\BibitemShut {NoStop}%
\bibitem [{\citenamefont {Abraham}\ \emph {et~al.}(2024)\citenamefont {Abraham}, \citenamefont {Macio\l{}ek},\ and\ \citenamefont {Squarcini}}]{AMS_2024}%
  \BibitemOpen
  \bibfield  {author} {\bibinfo {author} {\bibfnamefont {D.~B.}\ \bibnamefont {Abraham}}, \bibinfo {author} {\bibfnamefont {A.}~\bibnamefont {Macio\l{}ek}},\ and\ \bibinfo {author} {\bibfnamefont {A.}~\bibnamefont {Squarcini}},\ }\href {https://doi.org/10.1103/PhysRevE.109.054121} {\bibfield  {journal} {\bibinfo  {journal} {Phys. Rev. E}\ }\textbf {\bibinfo {volume} {109}},\ \bibinfo {pages} {054121} (\bibinfo {year} {2024})}\BibitemShut {NoStop}%
\bibitem [{\citenamefont {Abraham}(2012)}]{Abraham_2012}%
  \BibitemOpen
  \bibfield  {author} {\bibinfo {author} {\bibfnamefont {D.~B.}\ \bibnamefont {Abraham}},\ }\href {https://doi.org/10.1063/1.4752460} {\bibfield  {journal} {\bibinfo  {journal} {J. Math. Phys.}\ }\textbf {\bibinfo {volume} {53}},\ \bibinfo {pages} {095224} (\bibinfo {year} {2012})}\BibitemShut {NoStop}%
\bibitem [{\citenamefont {Abraham}\ and\ \citenamefont {Macio\l{}ek}(2010)}]{AM_2010}%
  \BibitemOpen
  \bibfield  {author} {\bibinfo {author} {\bibfnamefont {D.~B.}\ \bibnamefont {Abraham}}\ and\ \bibinfo {author} {\bibfnamefont {A.}~\bibnamefont {Macio\l{}ek}},\ }\href {https://doi.org/10.1103/PhysRevLett.105.055701} {\bibfield  {journal} {\bibinfo  {journal} {Phys. Rev. Lett.}\ }\textbf {\bibinfo {volume} {105}},\ \bibinfo {pages} {055701} (\bibinfo {year} {2010})}\BibitemShut {NoStop}%
\bibitem [{\citenamefont {Abraham}\ and\ \citenamefont {Macio\l{}ek}(2013)}]{AM_2013}%
  \BibitemOpen
  \bibfield  {author} {\bibinfo {author} {\bibfnamefont {D.~B.}\ \bibnamefont {Abraham}}\ and\ \bibinfo {author} {\bibfnamefont {A.}~\bibnamefont {Macio\l{}ek}},\ }\href {https://doi.org/10.1209/0295-5075/101/20006} {\bibfield  {journal} {\bibinfo  {journal} {EPL}\ }\textbf {\bibinfo {volume} {101}},\ \bibinfo {pages} {20006} (\bibinfo {year} {2013})}\BibitemShut {NoStop}%
\bibitem [{\citenamefont {Abraham}(1971)}]{Abraham_SAM_1971}%
  \BibitemOpen
  \bibfield  {author} {\bibinfo {author} {\bibfnamefont {D.~B.}\ \bibnamefont {Abraham}},\ }\href {https://doi.org/10.1002/sapm197150171} {\bibfield  {journal} {\bibinfo  {journal} {Stud. Appl. Math.}\ }\textbf {\bibinfo {volume} {50}},\ \bibinfo {pages} {71} (\bibinfo {year} {1971})}\BibitemShut {NoStop}%
\bibitem [{\citenamefont {Schultz}\ \emph {et~al.}(1964)\citenamefont {Schultz}, \citenamefont {Mattis},\ and\ \citenamefont {Lieb}}]{SML}%
  \BibitemOpen
  \bibfield  {author} {\bibinfo {author} {\bibfnamefont {T.~D.}\ \bibnamefont {Schultz}}, \bibinfo {author} {\bibfnamefont {D.~C.}\ \bibnamefont {Mattis}},\ and\ \bibinfo {author} {\bibfnamefont {E.~H.}\ \bibnamefont {Lieb}},\ }\href@noop {} {\bibfield  {journal} {\bibinfo  {journal} {Rev. Mod. Phys.}\ }\textbf {\bibinfo {volume} {36}},\ \bibinfo {pages} {856} (\bibinfo {year} {1964})}\BibitemShut {NoStop}%
\bibitem [{\citenamefont {Kaufman}(1949)}]{Kaufman_49}%
  \BibitemOpen
  \bibfield  {author} {\bibinfo {author} {\bibfnamefont {B.}~\bibnamefont {Kaufman}},\ }\href {https://doi.org/10.1103/PhysRev.76.1232} {\bibfield  {journal} {\bibinfo  {journal} {Phys. Rev.}\ }\textbf {\bibinfo {volume} {76}},\ \bibinfo {pages} {1232} (\bibinfo {year} {1949})}\BibitemShut {NoStop}%
\bibitem [{\citenamefont {Bleistein}\ and\ \citenamefont {Handelsman}(1986)}]{Bleistein1987}%
  \BibitemOpen
  \bibfield  {author} {\bibinfo {author} {\bibfnamefont {N.}~\bibnamefont {Bleistein}}\ and\ \bibinfo {author} {\bibfnamefont {R.}~\bibnamefont {Handelsman}},\ }\href {https://books.google.hr/books?id=3GZf-bCLFxcC} {\emph {\bibinfo {title} {{Asymptotic Expansions of Integrals}}}},\ Dover Books on Mathematics\ (\bibinfo  {publisher} {Dover Publications},\ \bibinfo {year} {1986})\BibitemShut {NoStop}%
\bibitem [{\citenamefont {Abraham}\ and\ \citenamefont {Reed}(1977)}]{AR74}%
  \BibitemOpen
  \bibfield  {author} {\bibinfo {author} {\bibfnamefont {D.~B.}\ \bibnamefont {Abraham}}\ and\ \bibinfo {author} {\bibfnamefont {P.}~\bibnamefont {Reed}},\ }\href {https://doi.org/10.1088/0305-4470/10/6/006} {\bibfield  {journal} {\bibinfo  {journal} {J. Phys. A: Math. Gen.}\ }\textbf {\bibinfo {volume} {10}},\ \bibinfo {pages} {L121} (\bibinfo {year} {1977})}\BibitemShut {NoStop}%
\bibitem [{\citenamefont {Abraham}\ and\ \citenamefont {Upton}(1988)}]{AU88}%
  \BibitemOpen
  \bibfield  {author} {\bibinfo {author} {\bibfnamefont {D.~B.}\ \bibnamefont {Abraham}}\ and\ \bibinfo {author} {\bibfnamefont {P.~J.}\ \bibnamefont {Upton}},\ }\href {https://doi.org/10.1103/PhysRevB.37.3835} {\bibfield  {journal} {\bibinfo  {journal} {Phys. Rev. B}\ }\textbf {\bibinfo {volume} {37}},\ \bibinfo {pages} {3835(R)} (\bibinfo {year} {1988})}\BibitemShut {NoStop}%
\bibitem [{\citenamefont {Fisher}\ \emph {et~al.}(1982)\citenamefont {Fisher}, \citenamefont {Fisher},\ and\ \citenamefont {Weeks}}]{Fisher_Fisher_Weeks}%
  \BibitemOpen
  \bibfield  {author} {\bibinfo {author} {\bibfnamefont {M.~P.~A.}\ \bibnamefont {Fisher}}, \bibinfo {author} {\bibfnamefont {D.~S.}\ \bibnamefont {Fisher}},\ and\ \bibinfo {author} {\bibfnamefont {J.~D.}\ \bibnamefont {Weeks}},\ }\href {https://doi.org/10.1103/PhysRevLett.48.368} {\bibfield  {journal} {\bibinfo  {journal} {Phys. Rev. Lett.}\ }\textbf {\bibinfo {volume} {48}},\ \bibinfo {pages} {368} (\bibinfo {year} {1982})}\BibitemShut {NoStop}%
\bibitem [{\citenamefont {Abraham}\ and\ \citenamefont {$\check{\textrm{S}}$vraki$\acute{\textrm{c}}$}(1991)}]{Abraham_Svrakic_1991}%
  \BibitemOpen
  \bibfield  {author} {\bibinfo {author} {\bibfnamefont {D.~B.}\ \bibnamefont {Abraham}}\ and\ \bibinfo {author} {\bibfnamefont {N.~M.}\ \bibnamefont {$\check{\textrm{S}}$vraki$\acute{\textrm{c}}$}},\ }\href {https://doi.org/10.1007/BF01030000} {\bibfield  {journal} {\bibinfo  {journal} {J. Stat. Phys.}\ }\textbf {\bibinfo {volume} {63}},\ \bibinfo {pages} {1077} (\bibinfo {year} {1991})}\BibitemShut {NoStop}%
\bibitem [{\citenamefont {Abraham}\ and\ \citenamefont {Macio\l{}ek}(2002)}]{AM_2002}%
  \BibitemOpen
  \bibfield  {author} {\bibinfo {author} {\bibfnamefont {D.~B.}\ \bibnamefont {Abraham}}\ and\ \bibinfo {author} {\bibfnamefont {A.}~\bibnamefont {Macio\l{}ek}},\ }\href {https://doi.org/10.1103/PhysRevLett.89.286101} {\bibfield  {journal} {\bibinfo  {journal} {Phys. Rev. Lett.}\ }\textbf {\bibinfo {volume} {89}},\ \bibinfo {pages} {286101} (\bibinfo {year} {2002})}\BibitemShut {NoStop}%
\bibitem [{\citenamefont {Abraham}\ and\ \citenamefont {Macio\l{}ek}(2005)}]{AM_2005}%
  \BibitemOpen
  \bibfield  {author} {\bibinfo {author} {\bibfnamefont {D.~B.}\ \bibnamefont {Abraham}}\ and\ \bibinfo {author} {\bibfnamefont {A.}~\bibnamefont {Macio\l{}ek}},\ }\href {https://doi.org/10.1103/PhysRevE.72.031601} {\bibfield  {journal} {\bibinfo  {journal} {Phys. Rev. E}\ }\textbf {\bibinfo {volume} {72}},\ \bibinfo {pages} {031601} (\bibinfo {year} {2005})}\BibitemShut {NoStop}%
\bibitem [{\citenamefont {Squarcini}\ and\ \citenamefont {Tinti}(2023)}]{Squarcini_Tinti_2023}%
  \BibitemOpen
  \bibfield  {author} {\bibinfo {author} {\bibfnamefont {A.}~\bibnamefont {Squarcini}}\ and\ \bibinfo {author} {\bibfnamefont {A.}~\bibnamefont {Tinti}},\ }\href@noop {} {\bibfield  {journal} {\bibinfo  {journal} {SciPost Phys.}\ }\textbf {\bibinfo {volume} {15}},\ \bibinfo {pages} {164} (\bibinfo {year} {2023})}\BibitemShut {NoStop}%
\bibitem [{\citenamefont {Abraham}\ and\ \citenamefont {Owczarek}(1990)}]{Abraham_Owczarek}%
  \BibitemOpen
  \bibfield  {author} {\bibinfo {author} {\bibfnamefont {D.~B.}\ \bibnamefont {Abraham}}\ and\ \bibinfo {author} {\bibfnamefont {A.~L.}\ \bibnamefont {Owczarek}},\ }\href@noop {} {\bibfield  {journal} {\bibinfo  {journal} {Phys. Rev. Lett.}\ }\textbf {\bibinfo {volume} {64}},\ \bibinfo {pages} {2595} (\bibinfo {year} {1990})}\BibitemShut {NoStop}%
\bibitem [{Note1()}]{Note1}%
  \BibitemOpen
  \bibinfo {note} {See \cite {DS_wetting} and \cite {DV, Squarcini_Multipoint}, respectively, for a derivation of \protect \eqref {10062025_2217} and the latter result in field theory.}\BibitemShut {Stop}%
\bibitem [{\citenamefont {Abraham}\ and\ \citenamefont {Gates}(1974)}]{DBA_Gates}%
  \BibitemOpen
  \bibfield  {author} {\bibinfo {author} {\bibfnamefont {D.~B.}\ \bibnamefont {Abraham}}\ and\ \bibinfo {author} {\bibfnamefont {D.~J.}\ \bibnamefont {Gates}},\ }\href@noop {} {\bibfield  {journal} {\bibinfo  {journal} {Physica}\ }\textbf {\bibinfo {volume} {72}},\ \bibinfo {pages} {73} (\bibinfo {year} {1974})}\BibitemShut {NoStop}%
\bibitem [{\citenamefont {Delfino}\ and\ \citenamefont {Squarcini}(2013)}]{DS_wetting}%
  \BibitemOpen
  \bibfield  {author} {\bibinfo {author} {\bibfnamefont {G.}~\bibnamefont {Delfino}}\ and\ \bibinfo {author} {\bibfnamefont {A.}~\bibnamefont {Squarcini}},\ }\href@noop {} {\bibfield  {journal} {\bibinfo  {journal} {J. Stat. Mech.: Theory Exp.}\ }\textbf {\bibinfo {volume} {2013}},\ \bibinfo {pages} {P05010}}\BibitemShut {NoStop}%
\bibitem [{\citenamefont {Delfino}\ and\ \citenamefont {Viti}(2012)}]{DV}%
  \BibitemOpen
  \bibfield  {author} {\bibinfo {author} {\bibfnamefont {G.}~\bibnamefont {Delfino}}\ and\ \bibinfo {author} {\bibfnamefont {J.}~\bibnamefont {Viti}},\ }\href@noop {} {\bibfield  {journal} {\bibinfo  {journal} {J. Stat. Mech.: Theory Exp.}\ }\textbf {\bibinfo {volume} {2012}},\ \bibinfo {pages} {P10009}}\BibitemShut {NoStop}%
\bibitem [{\citenamefont {Squarcini}(2021)}]{Squarcini_Multipoint}%
  \BibitemOpen
  \bibfield  {author} {\bibinfo {author} {\bibfnamefont {A.}~\bibnamefont {Squarcini}},\ }\href {https://doi.org/https://doi.org/10.1007/JHEP11(2021)096} {\bibfield  {journal} {\bibinfo  {journal} {J. High Energ. Phys.}\ }\textbf {\bibinfo {volume} {2021}}\bibinfo  {number} { (11)},\ \bibinfo {pages} {96}}\BibitemShut {NoStop}%
\end{thebibliography}

\begin{thebibliography}{8}%
\makeatletter
\providecommand \@ifxundefined [1]{%
 \@ifx{#1\undefined}
}%
\providecommand \@ifnum [1]{%
 \ifnum #1\expandafter \@firstoftwo
 \else \expandafter \@secondoftwo
 \fi
}%
\providecommand \@ifx [1]{%
 \ifx #1\expandafter \@firstoftwo
 \else \expandafter \@secondoftwo
 \fi
}%
\providecommand \natexlab [1]{#1}%
\providecommand \enquote  [1]{``#1''}%
\providecommand \bibnamefont  [1]{#1}%
\providecommand \bibfnamefont [1]{#1}%
\providecommand \citenamefont [1]{#1}%
\providecommand \href@noop [0]{\@secondoftwo}%
\providecommand \href [0]{\begingroup \@sanitize@url \@href}%
\providecommand \@href[1]{\@@startlink{#1}\@@href}%
\providecommand \@@href[1]{\endgroup#1\@@endlink}%
\providecommand \@sanitize@url [0]{\catcode `\\12\catcode `\$12\catcode `\&12\catcode `\#12\catcode `\^12\catcode `\_12\catcode `\%12\relax}%
\providecommand \@@startlink[1]{}%
\providecommand \@@endlink[0]{}%
\providecommand \url  [0]{\begingroup\@sanitize@url \@url }%
\providecommand \@url [1]{\endgroup\@href {#1}{\urlprefix }}%
\providecommand \urlprefix  [0]{URL }%
\providecommand \Eprint [0]{\href }%
\providecommand \doibase [0]{https://doi.org/}%
\providecommand \selectlanguage [0]{\@gobble}%
\providecommand \bibinfo  [0]{\@secondoftwo}%
\providecommand \bibfield  [0]{\@secondoftwo}%
\providecommand \translation [1]{[#1]}%
\providecommand \BibitemOpen [0]{}%
\providecommand \bibitemStop [0]{}%
\providecommand \bibitemNoStop [0]{.\EOS\space}%
\providecommand \EOS [0]{\spacefactor3000\relax}%
\providecommand \BibitemShut  [1]{\csname bibitem#1\endcsname}%
\let\auto@bib@innerbib\@empty
\bibitem [{\citenamefont {Abraham}(1984)}]{Abraham_1984_Surfaces_and_Roughening}%
  \BibitemOpen
  \bibfield  {author} {\bibinfo {author} {\bibfnamefont {D.~B.}\ \bibnamefont {Abraham}},\ }\href@noop {} {\bibfield  {journal} {\bibinfo  {journal} {J. Stat. Phys.}\ }\textbf {\bibinfo {volume} {34}},\ \bibinfo {pages} {793} (\bibinfo {year} {1984})}\BibitemShut {NoStop}%
\bibitem [{\citenamefont {Rottman}\ and\ \citenamefont {Wortis}(1981)}]{Rottman_Wortis_PRB}%
  \BibitemOpen
  \bibfield  {author} {\bibinfo {author} {\bibfnamefont {C.}~\bibnamefont {Rottman}}\ and\ \bibinfo {author} {\bibfnamefont {M.}~\bibnamefont {Wortis}},\ }\href@noop {} {\bibfield  {journal} {\bibinfo  {journal} {Phys. Rev. B}\ }\textbf {\bibinfo {volume} {24}},\ \bibinfo {pages} {6274} (\bibinfo {year} {1981})}\BibitemShut {NoStop}%
\bibitem [{\citenamefont {Avron}\ \emph {et~al.}(1982)\citenamefont {Avron}, \citenamefont {van Beijeren}, \citenamefont {Schulman},\ and\ \citenamefont {Zia}}]{ABSZ}%
  \BibitemOpen
  \bibfield  {author} {\bibinfo {author} {\bibfnamefont {J.~E.}\ \bibnamefont {Avron}}, \bibinfo {author} {\bibfnamefont {H.}~\bibnamefont {van Beijeren}}, \bibinfo {author} {\bibfnamefont {L.~S.}\ \bibnamefont {Schulman}},\ and\ \bibinfo {author} {\bibfnamefont {R.~K.~P.}\ \bibnamefont {Zia}},\ }\href@noop {} {\bibfield  {journal} {\bibinfo  {journal} {J. Phys. A: Math. Gen.}\ }\textbf {\bibinfo {volume} {15}},\ \bibinfo {pages} {L81} (\bibinfo {year} {1982})}\BibitemShut {NoStop}%
\bibitem [{\citenamefont {Abraham}(1986)}]{Abraham.review}%
  \BibitemOpen
  \bibfield  {author} {\bibinfo {author} {\bibfnamefont {D.~B.}\ \bibnamefont {Abraham}},\ }\href@noop {} {\emph {\bibinfo {title} {Surface Structures and Phase Transitions--Exact Results}}},\ edited by\ \bibinfo {editor} {\bibfnamefont {C.}~\bibnamefont {Domb}}\ and\ \bibinfo {editor} {\bibfnamefont {J.~L.}\ \bibnamefont {Lebowitz}},\ \bibinfo {series} {Phase Transitions and Critical Phenomena}, Vol.~\bibinfo {volume} {10}\ (\bibinfo  {publisher} {Academic Press},\ \bibinfo {address} {London},\ \bibinfo {year} {1986})\ pp.\ \bibinfo {pages} {1--74}\BibitemShut {NoStop}%
\bibitem [{\citenamefont {Rottman}\ and\ \citenamefont {Wortis}(1984)}]{Rottman_Wortis}%
  \BibitemOpen
  \bibfield  {author} {\bibinfo {author} {\bibfnamefont {C.}~\bibnamefont {Rottman}}\ and\ \bibinfo {author} {\bibfnamefont {M.}~\bibnamefont {Wortis}},\ }\href@noop {} {\bibfield  {journal} {\bibinfo  {journal} {Physics Reports}\ }\textbf {\bibinfo {volume} {103}},\ \bibinfo {pages} {59} (\bibinfo {year} {1984})}\BibitemShut {NoStop}%
\bibitem [{\citenamefont {Pfister}(2009)}]{Pfister_2009}%
  \BibitemOpen
  \bibfield  {author} {\bibinfo {author} {\bibfnamefont {C.-E.}\ \bibnamefont {Pfister}},\ }\href@noop {} {\bibfield  {journal} {\bibinfo  {journal} {arXiv:0911.5232}\ } (\bibinfo {year} {2009})}\BibitemShut {NoStop}%
\bibitem [{\citenamefont {Pfister}(2010)}]{Pfister_2010}%
  \BibitemOpen
  \bibfield  {author} {\bibinfo {author} {\bibfnamefont {C.-E.}\ \bibnamefont {Pfister}},\ }\href {http://www.scholarpedia.org/article/Interface_free_energy} {\bibfield  {journal} {\bibinfo  {journal} {Scholarpedia}\ }\textbf {\bibinfo {volume} {5}},\ \bibinfo {pages} {9218} (\bibinfo {year} {2010})}\BibitemShut {NoStop}%
\bibitem [{\citenamefont {{\.Z}o{\l}{\k a}dek}(2025)}]{Zoladek}%
  \BibitemOpen
  \bibfield  {author} {\bibinfo {author} {\bibfnamefont {H.}~\bibnamefont {{\.Z}o{\l}{\k a}dek}},\ }\href@noop {} {\emph {\bibinfo {title} {{The monodromy group}}}}\ (\bibinfo  {publisher} {Birk\"auser},\ \bibinfo {year} {2025})\BibitemShut {NoStop}%
\end{thebibliography}
\end{document}